\documentclass[rmp,twocolumn,superscriptaddress,showpacs,floatfix]{revtex4}
\usepackage{amsmath}
\usepackage{graphicx}

\begin{document}
\title{Artificial Brownian motors: Controlling transport on the nanoscale}

\author{Peter H\"anggi}
\email{peter.hanggi@physik.uni-augsburg.de}
\address{Institut
f\"ur  Physik, Universit\"at Augsburg, Universit\"atsstr. 1, D-86135 Augsburg,
Germany}
\address{Department of Physics and Centre for Computational Science and
Engineering, National University of Singapore, Singapore 117542}
\author{Fabio Marchesoni}
\email{fabio.marchesoni@pg.infn.it}
\address{Dipartimento di
Fisica, Universit\`a di Camerino, I-62032 Camerino, Italy}
\address{School of Physics, Korea Institute for Advanced Study, Seoul 130-722, Korea}

\begin{abstract}
In systems possessing spatial or dynamical symmetry breaking,
Brownian motion combined with
unbiased
external input signals, deterministic or random, alike, can assist
directed motion of particles at the submicron scales. In such cases,
one speaks of ``Brownian motors". In this review the constructive
role of Brownian motion is exemplified for various physical and
technological setups, which are inspired by the cell molecular
machinery: working principles and characteristics of stylized
devices are discussed to show how fluctuations, either thermal or
extrinsic, can be used to control diffusive particle transport.
Recent experimental demonstrations of this concept are surveyed with
particular attention to transport in artificial, i.e. non-biological
nanopores and optical traps, where single particle currents have
been first measured. Much emphasis is given to two- and
three-dimensional devices containing many interacting particles of
one or more species; for this class of artificial motors, noise
rectification results also from the interplay of particle Brownian
motion and geometric constraints. Recently, selective control and
optimization of the transport of interacting colloidal particles and
magnetic vortices have been successfully achieved, thus leading to
the new generation of microfluidic and superconducting devices
presented hereby.  The field has recently been enriched with
impressive experimental achievements in building artificial Brownian
motor devices  that even operate within the quantum domain by
harvesting quantum Brownian motion. Sundry akin topics include
activities aimed at noise-assisted shuttling other degrees of
freedom such as charge, spin or even heat and the assembly of
chemical synthetic molecular motors.  Our survey ends with a
perspective for future roadways  and potential new applications.
\end{abstract}
\date{\today}
\maketitle
\
\tableofcontents

\section{INTRODUCTION} \label{introduction}

Over the past two decades advances in microscopy and microscale
control have allowed scientists and engineers to delve into the
workings of the biological matter. One century after Kelvin's death,
today's researchers aim to explain how the engines of life actually
work by stretching thermodynamics beyond its 19th-century limits
\cite{haw2007}.

Carnot realized that all engines transform energy from one form into
another with a maximum possible efficiency that does not depend on
the technology developed or on the fuel utilized, but rather on
fundamental quantities such as heat and temperature. Kelvin and
Clausius came up with two ``rules of the engine", later known as the
first two laws of thermodynamics. The first law states that energy
cannot be destroyed or created but only transformed; the second law
sets fundamental limitations to what energy transformation can
achieve in practical terms. Just as the first law was centered on
the notion of energy (from Greek for ``work capability"), the second
law revolved around the new concept of entropy (a term coined by
Clausius also from Greek for ``change capability"). \footnote{Rudolf
Julius Emanuel  Clausius throughout his life allegedly preferred the
German word ``Verwandlungswert'' rather than ``entropy'', although
his colleagues suggested him to choose a name for his new
thermodynamic quantity ``S'' (his chosen symbol for labeling
``entropy'', possibly in honor of ``S''adi Carnot) that sounded as
close as possible to the word ``energy''.} When expressed in such
terms, the second law states that entropy cannot decrease during any
spontaneous or natural process. Notably, within the whole, virtual
manufactory of all natural processes the first law takes on the role
of an account clerk,  keeping track of all energy changes, while the
second law takes on the role of the director, determining the
direction and action of all processes.

The fathers of thermodynamics developed their laws having in mind
macroscopic systems that they could describe in terms of state
(i.e., average) quantities such as pressure and temperature, a
reasonable assumption when dealing with the monster steam engines of
the Victorian industry. A typical protein, however, is a few
nanometers in size and consists of just a few tens of thousands of
atoms. As a consequence the movements of a protein engine are
swamped by the fluctuations resulting from the Brownian motion of
its surroundings, which causes the energy of any its part to
fluctuate continually in units of $kT$, with $k$ denoting the
Boltzmann constant and $T$ the temperature. The effects of such
energy fluctuations were brilliantly demonstrated by Yanagida's
group \cite{nishiyama2001NatureCell3,nishiyama2002NatureCell4}, who
observed kinesin molecules climbing the cytoskeletal track in a
juddering motion made up of random hesitations, jumps and even
backward steps. Similar results have been reported for a host of
protein engines. The key question in modern thermodynamics is
therefore how far energy fluctuations drive micro- and nano-engines
beyond the limits of macroscopic laws.

A revealing experiment was performed recently by
\textcite{bustamante2005PhysToday58}, who first stretched a single
RNA molecule, by optically tugging at a tiny plastic bead attached
to one end, and then released the bead to study the effect of random
energy fluctuations on the molecule recovery process. By repeating
many identical stretching cycles, these authors found that the
molecule ``relaxation path" was different every time. In fact, the
bead was drawing useful energy from the thermal motion of the
suspension fluid and transforming it into motion. However, by
averaging over increasingly longer bead trajectories that is,
approaching a macroscopic situation,
\textcite{bustamante2005PhysToday58} were able to reconcile their
findings with the second law. These results lead to the conclusion
that the straightforward extension of the second law to microscopic
systems was ungranted; individual small systems do evolve under {\it
inherently nonequilibrium conditions}.

However, a decade ago \textcite{jarzynski1997PRL78} showed that the
demarcation line between equilibrium and non-equilibrium processes
is not always as clear cut as we were used to think. Imagining a
microscopic single molecule process, Jarzynski evaluated not the
simple average of the change of the (random) work of the underlying
perturbed nanosystem, as it was pulled away from equilibrium
according to an arbitrary protocol of forcing, but rather the
average of the {\it exponential} of that tailored nonequilibrium
work.
Remarkably, such a quantity turned out to be the same as for an
adiabatically slow version of the same process, and most remarkably,
equals the exponential of the system's equilibrium free energy
change. This result, also experimentally demonstrated by
\textcite{bustamante2005PhysToday58}, came as much of a surprise
because it meant that information about macroscopic equilibrium was
somehow buried inside individual, randomly fluctuating microscopic
systems far from equilibrium, see also
\cite{gallavotti1995PRL74,crooks1999PRE60,jarzynski2007CRP8,talkner2007PRE75,talkner2008PRE77}.

There is an additional limitation of 19th-century thermodynamics
that is potentially even more significant in the design and
operation of engines at the sub-micron scales. Kelvin's
thermodynamics was based on the simplifying notion of an isolated
system. The laws of macroscopic thermodynamics therefore apply only
to systems that are either separated from their environment or
coupled to it under controlled operating conditions, that is,
measured in terms of the state variables of the system itself.
However, at variance with a cylinder inside a steam engine, protein
engines {\it do not (cannot) work in isolation}.

Very much on the footsteps of the 19th-century scientists and
engineers, modern experimenters have probed the proteins that play a
crucial role in the cell individually by feeding them with energy by
injecting some chemical fuel ``by hand" (e.g., ATP molecules) or
exerting mechanical actions of some sort
\cite{astumian1997Science276,bustamante2005PhysToday58}. In their
natural setting, however, life engines are just parts of a closely
interconnected functional web that keeps a cell alive. The great
challenge of systems biology is therefore to put our understanding
of isolated life engines back into the real world of the cell.

\subsubsection{Artificial nanodevices}

Nanotechnology has been intricately linked with biological systems
since its inception. Fascinated by the complexity and smallness of
the cell, \textcite{feynman1960} challenged the scientific community
to ``make a thing very small which does what we want". In his
visionary response, \textcite{drexler1992} proposed to focus on
protein synthesis as a pathway for creating nanoscale devices. Both
Feynman and Drexler's propositions have been met with much
skepticism as accurate manipulations at the nanoscale were deemed
impossible. However, in view of the recent advances in systems
biology \cite{gross1999}, cellular mechanisms are now being cited as
the key proof of the nanotechnological viability of devices with
atomic precision. In spite of their established complementarity, a
fundamental difference between systems biology and nanotechnology is
their ultimate goal. Systems biology aims to uncover the fundamental
operation of the cell in an effort to predict the exact response to
specific stimuli and genetic variations, whereas nanotechnology is
chiefly concerned with useful design.

Manufacturing nanodevices through positional assembly and
self-assembly of biological components available at the cellular
level is the goal of the so-called biomimetic approach -- as opposed
to the inorganic approach aimed at fabricating nanomechanical
devices in hard, inorganic materials (e.g., using modern
lithographic techniques, atomic force and scanning tunneling
microscopy, etc). Nature has already proven that it is possible to
engineer complex machines on the nanoscale; there is an existing
framework of working components manufactured by Nature than can be
used as a guide to develop our own {\it biology inspired}
nanodevices. It is also true that the molecular machinery still
outperforms anything that can be artificially manufactured by many
orders of magnitude. Nevertheless, inorganic nanodevices are
attracting growing interest as a viable option due to their
potential simplicity and robustness, without forgetting that
inorganic nanodevices may provide additional experimental access to
the molecular machinery itself.

With this review the authors intend to pursue further the inorganic
approach to nanodevices, based on three main assumptions: (1) That,
in view of the most recent developments on nonequilibrium
thermodynamics, the science of nanodevices, regardless of the
fabrication technique, is inseparable from the thermodynamics of
microscopic engines \cite{hanggi2005AnPhys14}; (2) That the
fabrication techniques on the nanoscales become more and more
performing following the trend of the last two decades; (3) That a
better understanding of the molecular machinery can help devise and
implement new transport and control mechanisms for biology inspired
nanodevices. In other words, we bet on a two-way cross-fertilization
between the biomimetic and  the inorganic approach.

\subsubsection{Brownian motors}

Nature provided microorganisms, with characteristic sizes of
 about 10$^{-5}$m, with a variety of self-propulsion mechanisms, all of
which pertain to motion induced by cyclic shape changes. During one
such cycle the configuration changes follow an asymmetric sequence,
where one half cycle does not simply retrace the other half, in
order to circumvent the absence of hydrodynamic inertia at the
microscales, i.e., for low Reynolds numbers
\cite{purcell1977AmJPhys45}. A typical example are motile bacteria
that ``swim" in a suspension fluid by rotating or waving their
flagella \cite{astumian2002PhysToday55,
astumian2007Physchemchemphys9}. As anticipated above, a further
complication arises when the moving parts of a (sub)micron-engine
have  molecular dimensions of 10$^{-8}$m or so. In that case,
diffusion caused by Brownian motion competes with self-propelled
motion.
For example, a molecular motor mechanism becomes superior if at room
temperature, and in a medium with viscosity close to that of water,
a bacterium needs more time to diffuse a body length than it does to
swim the same distance. A passive diffusive mechanism operating
alone simply becomes inefficient.

A solution common to most cell machinery is to have molecular motors
operating on a track that constrains the motion to essentially one
dimension along a periodic sequence of wells and barriers. The
energy barriers significantly suppress the diffusion, while thermal
noise plays a constructive role by providing a mechanism, thermal
activation \cite{hanggi1990RMP62}, by which motors can escape over
the barriers. The energy necessary for directed motion is supplied
by {\it asymmetrically} raising and lowering the barriers and wells,
either via an external time-dependent modulation (e.g., due to the
coupling with other motors) or by energy input from a nonequilibrium
source such as a chemical reaction, like the ATP hydrolysis.
Thus,  in agreement with the reasoning underlying the analysis
\footnote{Note in this context also the insightful examination of
Feynman's analysis by \textcite{parrondo1996AJP64}.} of the
Smoluchowski-Feynman stylized ratchet engine
\cite{smoluchowski1912ZP13,feynman1963}, under appropriate
nonequilibrium conditions, structural anisotropy can  sustain
directed motion. Such a device clearly does not violate the second
law of thermodynamics because the very presence of nonequilibrium
renders inoperative those limiting (thermal equilibrium)
restrictions.

In the case of a bacterium, as for any ordinary heat engine, the
relevant state variables, namely, its position and the phase of the
flagellum stroke, always cycle through one and the same periodic
time sequence; the two variables are tightly coupled and almost
synchronized. In clear contrast to this familiar scenario, the state
variables of molecular motors are often loosely coupled due to the
prominent action of fluctuations, a salient feature nicely captured
by H\"anggi who originally coined the term  {\it Brownian motors} in
the feature \cite{bartussek1995PhysBl51}. \footnote{ The notion of
``molecular motor" is reserved within this review for motors
specifying biological, intracelluar transport. Likewise, the notion
of ``Brownian ratchet" or ``thermal ratchet" is reserved for the
operating principle of protein translocation processes.  The latter
term seemingly has been introduced by \textcite{simon1992PNAS89} to
describe isothermal trapping of Brownian particles to drive protein
translocation, see also in \textcite{wang2002ApplPhysA75}.}

Important hallmarks of any genuine Brownian motor are
\cite{astumian2002PhysToday55,hanggi2005AnPhys14}: (i) The presence
of some amount of (not necessarily thermal) noise. The intricate
interplay among nonlinearity, noise-activated escape dynamics and
non-equilibrium driving implies that, generally, not even the
direction of transport is a priori predictable; (ii) Some sort of
symmetry-breaking supplemented by temporal periodicity (typically
via an unbiased, non-equilibrium forcing), if a cyclically operating
device is involved. Therefore, not every small ratchet device falls
under the category of Brownian motors. This holds true especially if
the governing transport principle is deterministic, like in
mechanical ratchet devices of macro- or mesoscopic size.

The following prescriptions should be observed when designing an
efficient Brownian motor: (a) Spatial and temporal periodicity
critically affect rectification; (b) All acting forces and gradients
must vanish after averaging over space, time, and statistical
ensembles; (c) Random forces (of thermal, non-thermal, or even
deterministic origin) assume a prominent role; (d) Detailed balance
symmetry, ruling thermal equilibrium dynamics, must be broken by
operating the device away from thermal equilibrium; (e) A
symmetry-breaking mechanism must apply. There exist several
possibilities to induce symmetry-breaking. First, the spatial
inversion symmetry of the periodic system itself may be broken
intrinsically; that is, already in the absence of non-equilibrium
perturbations. This is the most common situation and typically
involves a type of periodic, asymmetric ratchet potential. A second
option consists in the use of an unbiased driving force
(deterministic or stochastic, alike) possessing non-vanishing,
higher order odd time-correlations. Yet a third possibility arises
via collective effects in coupled, perfectly symmetric
non-equilibrium systems, namely in the form of spontaneous symmetry
breaking. Note that in the latter two cases we speak of a Brownian
motor dynamics even though a ratchet potential is not necessarily
involved.

The reasoning of using unbiased thermal fluctuations and/or unbiased
nonequilibrium  perturbations to drive directed motion of particles
and alike has seen several rediscoveries since the visionary works
by Marian von \textcite{smoluchowski1912ZP13} and Richard P.
\textcite{feynman1963}. From a historic perspective, the theme of
directed transport and Brownian motors was put into the limelight of
the statistical and biological physics research with the appearance
of several ground-breaking works, both theoretical and experimental,
which appeared during the (1992-1994) period. Most notably,
\textcite{ajdari1992ComptRend315,magnasco1993PRL71,astumian1994PRL72,bartussek1994EPL28,
chauwin1994EPL27,doering1994PRL72,millonas1994PLA185,rousselet1994Nature370,
prost1994PRL72} helped ignite a tumultuous activity in this topical
area which kept growing until the present days. The readers may
deepen their historical insight by consulting earlier introductory
reports such as those published by
\textcite{hanggi1996LNP476,julicher1997RMP69,astumian1997Science276,
astumian2002PhysToday55,reimann2002ApplPhysA75,reimann2002PhysRep361,
hanggi2005AnPhys14}.

This review focuses on the recent advances in the science of
non-biological, {\it artificial} Brownian motors.

In contrast to those reviews and feature articles mentioned above,
which beautifully cover the rich variety of possible Brownian motor
scenarios and working principles, our focus in this review is on
non-biological, {\it artificial}, mostly, solid state based Brownian
motors. In this spirit the authors have attempted to present a
comprehensive overview of the present status of this field,
including newest theory developments, most compelling experimental
demonstrations, and first successful technological applications.
Some closely related topics, such as the engineering of synthetic
molecular motors and nanomachines based on chemical species, are
only briefly discussed herein because comprehensive, up-to-date
reviews have been published recently by research groups very active
in that area
\cite{kottas2005ChemRev105,balzani2006CSR35,kay2007AngewChem46}.


\section{SINGLE-PARTICLE TRANSPORT} \label{singleparticle}

Signal rectification schemes in the absence of noise have been known
since long ago, especially in the electrical engineering literature.
However, rectification in a nanodevices cannot ignore fluctuations
and Brownian motion, in particular. New experiments on both
biological and artificial devices showed how noise rectification can
actually be utilized to effectively control particle transport on
the small scales. By now, noise rectification has become one of the
most promising techniques for powering micro- and nanodevices.

In order to set the stage, in the next subsection we first consider
the case of systems where rectification cannot occur. In the
following subsections we then single out all ingredients that do
make rectification possible.

Let us consider a Brownian particle with mass $m$, coordinate
$x(t)$, and friction coefficient $\gamma$ in one dimension,
subjected to an external static force $F$ and thermal noise
$\xi(t)$. The corresponding stochastic dynamics is described by the
inertial Langevin equation
\begin{equation}
\label{xLE} m\ddot x = - V'(x) - m\gamma\,\dot x + F + \xi(t),
\end{equation}
where $V(x)$ is a periodic potential with period $L$, namely
$V(x+L)=V(x)$, and $'$ indicates the differentiation with respect to
$x$ and $\dot x$  the differentiation with respect to time $t$.
Thermal fluctuations are modeled by a stationary Gaussian noise of
vanishing mean, $\langle\xi(t)\rangle = 0$, satisfying the
fluctuation-dissipation relation
\begin{equation}
\label{noise} \langle\xi(t)\xi(0)\rangle = 2 D_p \delta(t),
\end{equation}
where the momentum-diffusion strength reads $D_p = m\gamma kT$, with
$k$ denoting the Boltzmann constant, and $T$ is the temperature of
an equilibrium heat bath.

In extremely small systems, particle dynamics and fluctuations
occurring in biological  and liquid environments  are often
described well by the {\it overdamped} limit of Eq. (\ref{xLE}), --
for an illustrative discussion of this fact see the account
 given by  \textcite{purcell1977AmJPhys45}--, in terms of a the massless Langevin
equation, which is driven by a position-diffusion $D_x =
kT/(m\gamma) \equiv D$, i.e.,
\begin{equation}
\dot x =  - V'(x) + F + \xi (t)\;, \label{oLE}
\end{equation}
with a corresponding noise correlation
\begin{equation}
\langle\xi(t)\xi(0)\rangle = 2 D \delta(t) \;. \label{noiseoLE}
\end{equation}
In this overdamped regime, the inertia term $m\ddot x$ can be
dropped altogether with respect to the friction term $- m\gamma \dot
x$ (Smoluchowski approximation). Here $m\gamma$ has been scaled to
unity for convenience, i.e., $D\equiv kT$.

In the absence of an external bias, i.e. $F=0$, the equilibrium
stochastic mechanism in Eq. (\ref{xLE}) cannot sustain a non-zero
stationary current, i.e., $\langle \dot x(t) \rangle = 0$, no matter
what $V(x)$. This can be readily proven upon solving the
corresponding Fokker-Planck equation with periodic boundary
conditions \cite{risken1984,melnikov1991PhysRep209}.

\subsection{Symmetric substrates}\label{symmetricsubstrates}

\begin{figure}[btp]

\begin{center}
\hspace*{-0.5cm}
\includegraphics*[width=6.0cm]{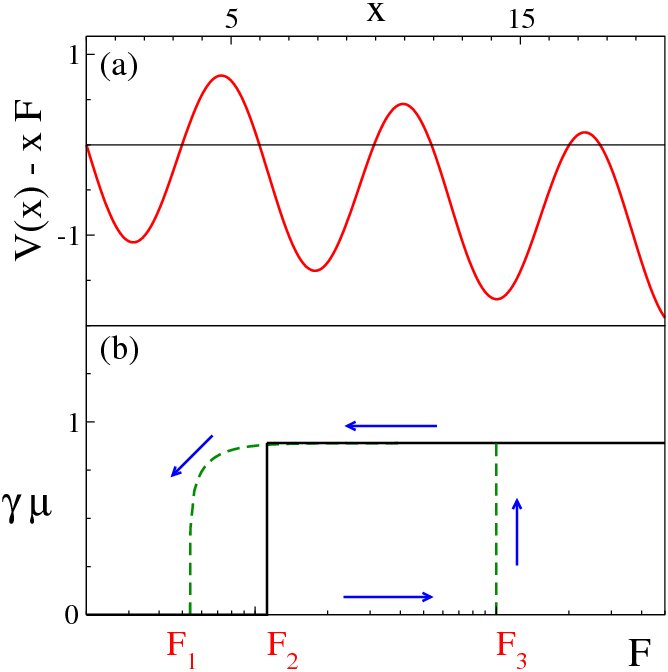}
\end{center}
\caption{(Color online) (a) Tilted periodic potential of Eq.
(\ref{cosine}), $V(x)-xF$, with $F=0.1$; (b) Locked-to-running
transitions. The thresholds $F_1$ and $F_3$ of the hysteretic loop
(dashed curves) and the zero temperature step at $F_2$ (solid curve)
are marked explicitly. Parameter values: $m = V_0 = 1$, i.e.
$F_3=1$, and $\gamma = 0.03$.} \label{Fwashboard}
\end{figure}

Let us consider first the case when the periodic substrate with the
potential $V(x)$ is symmetric under reflection, that is
$V(x-x_0)=V(-x+x_0)$ for certain $x_0$ with $x_0 \leq x_0 <L$. The
most studied example  is the symmetric washboard potential
\cite{risken1984}
\begin{equation}
V(x) = -V_0\sin(2\pi x/L), \label{cosine}
\end{equation}
displayed in presence of a static tilt-force $F$ in Fig.
\ref{Fwashboard}a. The particle mobility $\mu(F) \equiv \langle \dot
x\rangle/F$ is symmetric for $F\rightarrow -F$, namely,
$\mu(F)=\mu(-F)$. For this reason in this section we restrict
ourselves to $F\geq 0$.

\subsubsection{dc drive}
\label{dcdrive}

The Brownian motion, Eq. (\ref{xLE}), in the washboard potential,
Eq. (\ref{cosine}), is detailed in \textcite{risken1984} textbook.
To make contact with Risken's notation one must rewrite Eq.
(\ref{xLE}) in terms of the rescaled quantities $x \to
\frac{2\pi}{L}x$,
$F\rightarrow \frac{2\pi}{L}\frac{F}{m}$, and $T\rightarrow
(\frac{2\pi}{L})^2 \frac{T}{m}$, so that
\begin{equation}
\label{yLE} \ddot x = - \gamma\,\dot x + \omega_0^2\cos x + F +
\xi(t).
\end{equation}
The angular frequency $\omega_0 = (2\pi/L)\sqrt{V_0/m}$
characterizes the oscillating motion of the particle at the bottom
of the potential wells. To reconcile Eq. (\ref{yLE}) with Eq.
(\ref{xLE}) for the potential in Eq. (\ref{cosine}), it suffices to
scale
$m=1$ and $L=2\pi$, as assumed throughout this section, unless
stated
otherwise.

The dynamics of Eq. (\ref{yLE}) is characterized by random switches
occurring either between two locked states, dwelling a well minimum,
or switches between a locked state with zero average velocity and a
(downhill) running state with a finite,  average asymptotic velocity
$\langle \dot x\rangle = F/\gamma$. In terms of mobility, locked and
running states correspond to $\gamma\mu = 0$ and $\gamma\mu = 1$,
respectively.

In the overdamped regime,  Eq. (\ref{oLE}), the particle is trapped
most of the time at a local minimum of the tilted substrate as long
as $F \leq F_3 = \omega_0^2$; for $F>F_3$ there exist no such minima
and the particle runs in the $F$ direction with average speed
approaching $F/\gamma$. This behavior is described quantitatively by
the mobility formula \cite{risken1984}:
\begin{equation}
\label{mobility} \mu(F)=\frac{L}{\langle t(L,F) \rangle F},
\end{equation}
where $\langle t(L,F) \rangle$, the mean first-passage time of the
particle across a substrate unit cell in the $F$ direction, can be
computed explicitly for any choice of $V(x)$ \cite{hanggi1990RMP62}.

In the underdamped regime $\gamma \ll \omega_0$ the
locked-to-running transition depends crucially on the presence of
noise, no matter how weak. In the {\it noiseless} case, $\xi(t)
\equiv 0$, the average speed of a Brownian particle with coordinate
$x(t)$ depends on the initial conditions according to the hysteretic
cycle illustrated in Fig. \ref{Fwashboard}b: The transition from the
locked to the running state occurs on raising $F$ above $F_3$
(depinning threshold), while the opposite transition takes place on
lowering $F$ below $F_1 = (4/\pi)\gamma \omega_0$ (re-pinning
threshold). Of course, for sufficiently large values of $\gamma$,
say, in the damped regime with $\gamma > \gamma_0=(\pi/4) \omega_0$,
the distinction between $F_1$ and $F_3$ becomes meaningless; the
locked-to-running transition occurs at $F = F_3$, no matter what the
initial conditions.

At {\it zero temperature}, $T=0+$, we have a totally different
scenario, as the {\it stationary} dynamics of $x(t)$ is controlled
by one threshold $F_2=3.3576\dots \gamma \omega_0$, only
\cite{risken1984}: For $F < F_2$ the Brownian particle remains
trapped in one potential well; for $F > F_2$ it falls down the
tilted washboard potential with speed close to $F/\gamma$. At the
threshold $F_2$ the quantity $\gamma \mu$ jumps from zero to (very
close to) one, stepwise (Fig. \ref{Fwashboard}b). Note that, at
variance with $F_3$, $F_2$ indicates a dynamical transition
occurring in the presence of a relatively small tilt. Nevertheless,
at low damping, switches between locked and running states
correspond to long forward particle jumps, which can span over many
substrate unit cells $L$; the distributions of the relevant jump
lengths exhibit persistent non-exponential tails
\cite{pollak1993PRL70,ferrando1995PRE51,costantini1999EPL48,borromeo2000PRL84,shushin2002EPL60}.

For finite but low temperatures, $kT \ll V_0$, the transition from
the locked to the running state is continuous but still confined
within a narrow neighborhood of $F_2$; the relevant
locked-to-running transition threshold $F_{\rm th}$ is numerically
identifiable with high accuracy as a convex function of $\gamma$;
$F_2$ and $F_3$ are respectively the $\gamma \to 0$ and $\gamma \to
\infty$ asymptotes of the numerical curve  $F_{\rm th}(\gamma)$
\cite{risken1984}.

\subsubsection{ac drive} \label{acdrive}

Suppose now that the external force $F(t)$ acting on the unit-mass
Brownian particle is periodic in time.
The simplest case possible is represented by a {\it noiseless}
particle, $\xi(t)\equiv 0$,  moving on a sinusoidal potential, Eq.
(\ref{cosine}), under the action of a harmonic force
\begin{equation} \label{forceac}
F(t) = A_1 \cos(\Omega_1 t +\phi_1).
\end{equation}

The wide class of devices thus modeled may be assimilated to a
damped-driven pendulum operated at zero noise level, a chaotic
system investigated at depth in the 1980s \cite{baker1990}. The
dynamics of a massive particle in a sinusoidal potential was
reproduced in terms of a climbing-sine map
\cite{grossmann1982PRA26,geisel1982PRL48}: Running orbits, leading
to a spontaneous symmetry breaking, can be either periodic or
diffusive, depending on the value of the map control parameter (viz.
the amplitude of the sine term). The phase-space portrait of the
actual damped-driven pendulum was computed by
\textcite{huberman1980APL37}, who revealed the existence of
delocalized strange attractors with an intricate structure on all
scales, later recognized to be fractal objects \cite{baker1990}.
This means that, despite the global reflection symmetry of the
dynamics in Eqs. (\ref{xLE}) and (\ref{forceac}), for sufficiently
small $\gamma$ the particle drifts either to the right or the to
left, depending on the initial conditions, but with equal
probability in phase-space.

Coupling the particle to a heat bath, no matter how low the
temperature, changes this scenario completely. The action of the
noise source $\xi(t)$ amounts to scrambling the initial conditions,
which therefore must be averaged on. As a consequence, trajectories
to the right and to the left compensate one another and the symmetry
is restored: No Brownian drift is expected in the zero temperature
limit.

A system symmetry can be broken by adding a dc component $A_0$ to
the external force,
\begin{equation} \label{forcedcac}
F(t) = A_0+ A_1 \cos(\Omega_1 t +\phi_1).
\end{equation}

The most evident effect of such a periodic drive is the appearance
of hysteresis loops \cite{borromeo1999PRL82} in the parametric
curves of the mobility $\mu(t)$ versus $F(t)$. For
$\Omega_1\ll\omega_0$ and $A_0\ll F_{th}$, the mobility hysteresis
loop is centered on the static mobility curve $\mu(A_0)$ and
traversed counterclockwise; with decreasing $\Omega_1$, its major
axis approaches the tangent to the curve $\mu(A_0)$. Hysteresis
loops have been observed even for $\Omega_1$ much smaller than the
relevant Kramers rate, the smallest relaxation rate in the
unperturbed stationary process of Eq. (\ref{yLE}) with $F=0$. The
area encircled by the hysteretic loops is maximum for $A_0 \simeq
F_{\rm th}$, that is close to the transition jump. Even more
interestingly, it exhibits a resonant dependence on both the forcing
frequency and the temperature, thus pointing to a dynamical
stochastic resonance \cite{gammaitoni1998RMP70} mechanism between
locked and running states \cite{borromeo2000PRL84}.

Finally, \textcite{machura2007PRL98} showed that under special
conditions, involving small Brownian motion at temperature $T$ and
appropriate relaxation constants $\gamma \lesssim \Omega_1 \lesssim
\omega_0$, the damped process in Eq. (\ref{yLE}) occasionally
exhibits the phenomenon of absolute negative mobility: The ac cycle
averaged drift velocity $\langle \dot x \rangle$ may be oriented
against the dc bias $A_0$, as the result of a delicate interplay of
chaotic and stochastic dynamics. This observation was corroborated
theoretically by \textcite{speer2007PRE76} and
\textcite{kostur2008PRB77} and experimentally by
\textcite{nagel2008PRL100}. Yet, another artificial Brownian motor
system where this phenomenon can likely be validated experimentally
are cold atoms dwelling in periodic optical lattices.

\subsubsection{Diffusion peak} \label{diffusionpeak}

Diffusive transport of particles or, more generally, small objects
is a ubiquitous feature of physical and chemical reaction systems
\cite{burada2009chemphyschem}. Directed Brownian motor transport is
typically  controlled both by the fluctuation statistics of the
jittering objects and the phase space available to their dynamics.
For example, as the particle in Eq. (\ref{yLE}) drifts with average
speed $\langle \dot x \rangle$ in the direction of the external
force $F$, the random switches between locked and running state
cause a spatial dispersion of the particle around its average
position
The corresponding {\it normal} diffusion constant,
\begin{equation}
\label{diff.const} D(F) = \lim_{t\rightarrow \infty} \frac {\langle
x(t)^2\rangle - \langle x(t)\rangle^2} {2t},
\end{equation}
was computed numerically as a function of $F$ at constant
temperature by \textcite{costantini1999EPL48}. A peak in the curves
of $D$ versus $F$ is detectable in the vicinity of the transition
threshold $F_{\rm th}$, irrespective of the value of $\gamma$. In
particular, at low damping, where $F_{\rm th} \simeq F_2$, and low
temperature, $kT \ll \omega_0^2$, the peak of $D$ at $F = F_2$ is
very pronounced; on increasing the damping, the diffusion peak
eventually shrinks down to just a bump corresponding to the
threshold $F_{\rm th} \simeq F_3$. In any case, in an appropriate
neighborhood of $F_{\rm th}$ the diffusion constant can grow larger
than Einstein's diffusion constant for the free Brownian motion in
one dimension, $D_0 = kT/\gamma$. On increasing $T$ the diffusion
peak eventually disappears, no matter what $\gamma$.

%
%
A refined analytical formula for the diffusion peak was obtained by
Reimann and collaborators who regarded the locked-to-running
transition in the overdamped regime as a renewal process, that is
\cite{reimann2001PRL87,reimann2002PRE65}:
\begin{equation}\label{diff.const2}
D(F)= \frac{L^2}{2}\,\frac{\langle t^2(L,F)\rangle - \langle
t(L,F)\rangle^2} {\langle t(L,F)\rangle ^3},
\end{equation}
where the $n$-th moments of the first passage time, $\langle
t^n(L,F)\rangle$, can be computed explicitly. For $F\to 0$, Eq.
(\ref{diff.const2}) reproduces the zero-bias identity
$D/D_0=\gamma\mu$ \cite{festa1978PhysicaA90}.

\subsubsection{Single-file diffusion}\label{singlefile}

When a gas of particles is confined to a linear path, an individual
particle cannot diffuse past its neighbors. This constrained 1D
geometry is often called a ``single-file", or Jepsen gas
\cite{jepsen1965JMP6,harris1974AnProb2}. Let us consider a file of
$N$ indistinguishable, unit-mass Brownian particles moving on a
tilted sinusoidal substrate in Eq. (\ref{yLE}) of length $L$. If the
particle interaction is hard-core (zero radius), the file
constituents can be labeled according to a given ordered sequence
and the long-time diffusion of an individual particle gets strongly
suppressed. In early studies
\cite{jepsen1965JMP6,lebowitz1967PR155,levitt1973PRA8,harris1974AnProb2,marchesoni2006PRL97}
the mean square displacement of a single particle in the
thermodynamic limit ($L,N \to \infty$ with constant density
$\rho=N/L$) was calculated analytically. Those results were
generalized to the diffusion of a single-file of driven Brownian
particles on a periodic substrate by \textcite{taloni2006PRL96} who
derived the subdiffusive law
\begin{equation} \lim_{t \to \infty} [\label{sf1}
\langle x(t)^2\rangle - \langle
x(t)\rangle^2]=\frac{2}{\rho}\sqrt{\frac{D(F)t}{\pi}},
\end{equation}
with $D(F)$ given in Eq. (\ref{diff.const2}). This result applies to
any choice of the substrate potential $V(x)$
\cite{burada2009chemphyschem} and to the transport of composite
objects \cite{heinsalu2008PRE77}, as well. Excess diffusion peaks
have been obtained experimentally in the context of particle
transport in quasi-1D systems (Sec. \ref{nanopore}).

\subsection{Rectification of asymmetric processes} \label{rectification}

Stochastic transport across a device surely can be induced by
applying a macroscopic gradient, like a force or a temperature
difference. However, under many practical circumstances this is no
viable option: (a) Current induced by macroscopic gradients are
rarely selective; (b) A target particle that carries no charge or
dipole, can hardly be manipulated by means of an external field of
force; (c) External controls, including powering, critically
overburden the design of a small device. Ideally, the optimal
solution would be a self-propelled device that operates by
rectifying environmental signals. In the quest for rectification
mechanisms of easy implementation, we will start from the symmetric
dynamics of Eq. (\ref{xLE}) and add the minimal ingredients needed
to induce and control particle transport.

If $V(x)$ is symmetric under reflection, the only way to induce a
drift of the Brownian particle consists in driving it by means of a
non-symmetric force $F(t)$, either deterministic or random
\cite{luczka1995EPL31,hanggi1996EPL35,chialvo1997PRL78,astumian2002PhysToday55,reimann2002ApplPhysA75}.
Here, ``symmetric'' means that all force moments are invariant under
sign reversal $F \rightarrow -F$. Note that the condition of a
vanishing dc component, $\lim_{t\rightarrow
\infty}\frac{1}{t}\int_0^tF(s)ds=0$, would not be sufficient. For
instance, a bi-harmonic signal with commensurate frequencies and
arbitrary phase constants, although zero-mean valued, is in general
nonsymmetric.

On the contrary, particles in an asymmetric potential can drift on
average in one direction even when all perturbing forces or
gradients are symmetric. However, as pointed out in Sec.
\ref{molecularmotor}, to achieve directed transport in such a class
of devices,
the external perturbation is required to be at least
time-correlated, like in the presence of a non-Markovian noise
source (correlation ratchets) or a time periodic drive (rocked and
pulsated ratchets).

The interplay of time and space asymmetry in the operation of a
Brownian motor has been established on firmer mathematical grounds
by \textcite{yevtushenko2001EPL54} and by
\textcite{reimann2001PRL86}. Let us slightly generalize the
overdamped dynamics in Eq. (\ref{oLE}) to incorporate the case of
time dependent substrates, that is
\begin{equation}
\label{xtLE} \dot x = -V'[x,F_2(t)]+F_1(t) +\xi(t).
\end{equation}
The potential $V[x,F_2(t)]$ is termed supersymmetric
\cite{marchesoni1988PRL61,jung1991PRA44} if, for an appropriate
choice of the $x$ and $t$ origins,
\begin{equation}
\label{susy1} -V[x,F_2(t)] = V[x+L/2,F_2(-t)].
\end{equation}
Analogously, the additive drive $F_1(t)$ is supersymmetric if for an
appropriate $t$ origin,
\begin{equation}
\label{susy2} -F_1(t) = F_1(-t).
\end{equation}
Should $F_i(t)$, with $i=1,2$, be stationary noises, clearly no
restriction can be set on the $t$ origin; the equalities in Eq.
(\ref{susy1}) and in Eq. (\ref{susy2}) must then hold in a
statistical sense, meaning that the two terms of each equality must
be statistically indistinguishable.

Let us consider now the time reversed process $z(t)=x(-t)+L/2$. By
definition, $\langle \dot z \rangle = - \langle \dot x \rangle$,
whereas, on simultaneously imposing the supersymmetry conditions
(\ref{susy1}) and (\ref{susy2}), the Langevin equation (\ref{xtLE})
yields $\langle \dot z \rangle = \langle \dot x \rangle$; hence,
$\langle \dot x \rangle = 0$.  As a consequence, a nonzero
rectification current requires that either the substrate or the
additive drive (or both) are {\it non-supersymmetric}.

We remark that the above theorem has been proven only for zero-mass
particles, that is, when the Smoluchowski regime $m\ddot x=0$
applies. In the presence of inertia, instead, rectification may
occur, under very special conditions, also in fully supersymmetric
devices, as shown lately by \textcite{machura2007PRL98} for a rocked
cosine potential.

\subsection{Nonlinear mechanisms} \label{nonlinearmechanism}

In this Section we review transport on symmetric substrates driven
by asymmetric forces. The rectification mechanisms outlined below
can be traced back to the {\it nonlinear} nature of the substrate;
for this reason, at variance with Brownian motors, they work also,
if not more effectively, in the absence of noise.

We remind the reader that these mechanisms have been introduced and
demonstrated experimentally in the most diverse fields of physics
and engineering. Direct applications to various categories of
artificial nanodevices will be discussed in the subsequent sections.

\subsubsection{Harmonic mixing} \label{HM}

A charged particle spatially confined by a nonlinear force is
capable of mixing two alternating input electric fields of angular
frequencies $\Omega_1$ and $\Omega_2$, its response containing all
possible higher harmonics of $\Omega_1$ and $\Omega_2$. For
commensurate input frequencies, i.e., $\Omega_1/\Omega_2=n/m$ with
$n$ and $m$ coprime integer numbers, the output contains a dc
component, too \cite{schneider1966APL8}; harmonic mixing thus
induces a rectification effect of the $(n+m)$-th order in the
dynamical parameters of the system \cite{marchesoni1986PLA119,
goychuk1998EPL43}.

Let us consider the overdamped stochastic dynamics of Eq.
(\ref{oLE}) in the potential of Eq. (\ref{cosine}), driven by the
bi-harmonic force
\begin{equation}
\label{2freq} F(t)=A_1\cos(\Omega_1 t+\phi_1) +A_2\cos(\Omega_2
t+\phi_2)
\end{equation}
Let the two harmonic components of $F(t)$ be commensurate with one
another, meaning that $\Omega_1$, $\Omega_2$ are integer-valued
multiples of a fundamental frequency $\Omega_0$, i.e.,
$\Omega_1=n\Omega_0$ and $\Omega_2=m\Omega_0$. For small amplitudes
and low frequencies of the drive in Eq. (\ref{2freq}), a simple
expansion of the mobility function, Eq. (\ref{mobility}), in powers
of $F(t)$, yields, after time averaging, the following approximate
expression for the non-vanishing dc component of the particle
velocity:
\begin{eqnarray}
\label{HM1} \langle \dot x \rangle = 2{\overline
\mu}~~\frac{m+n}{m!~n!} \left ( \frac{A_1}{2}\right)^m \left (
\frac{A_2}{2}\right)^n \cos (\Delta_{m,n}),
\end{eqnarray}
where ${\overline \mu}$ is the positive $(n+m-1)$-th derivative of
$\mu(F)$ at $F=0$, $\Delta_{m,n}=n\phi_2- m\phi_1$, and $m+n$ is an
odd number. Harmonic mixing currents  $j=\langle \dot x \rangle/L$
clearly result from a spontaneous symmetry breaking mechanism:
indeed, averaging $j$ over $\phi_1$ or $\phi_2$ would eliminate the
effect completely.  This is true under any regime of temperature and
forcing, as proven by means of standard perturbation techniques
\cite{breymayer1981ZPB43,marchesoni1986PLA119}. In particular,
rectification of two small commensurate driving signals, $A_1,A_2
\ll \omega_0^2$, is a noise assisted process and, therefore, is
strongly suppressed for $kT\ll \omega_0^2$, when ${\overline \mu}$
decays exponentially to zero \cite{risken1984}. Moreover, we
anticipate that accounting for finite inertia effects requires
introducing in Eq. (\ref{HM1}) an additional damping and frequency
dependent phase-lag, as discussed in Sec. \ref{biharmonic}.

In all calculations of $\langle \dot x \rangle$, including Eq.
(\ref{HM1}), the reflection symmetry of $V(x)$ plays no significant
role; rectification via harmonic mixing is caused solely by the {\it
nonlinearity} of the substrate. However, it must be noticed, that a
symmetric nonlinear device cannot mix rectangular waveforms, like
\begin{equation}
\label{HM5}F(t)=A_1\hbox{sgn}[\cos(\Omega_1 t+\phi_1)]
+A_2\hbox{sgn}[\cos(\Omega_2 t+\phi_2)],
\end{equation}
with $A_1,A_2 \geq 0$ and $\hbox{sgn}[\dots]$ denoting the sign of
$[\dots]$ \cite{savelev2004EPL67,savelev2004PRE70b}. As shown in
Sec. \ref{asymmetry}, {\it asymmetric} devices do not exhibit this
peculiarity. Moreover, we underscore that for incommensurate
frequencies, $\Omega_1/\Omega_2$ not a rational number, harmonic
mixing takes place only in the presence of spatial asymmetry, as
reported in Sec. \ref{asymmetry}.

\begin{figure}[btp]
\begin{center}
\hspace*{-0.5cm}\includegraphics*[width=8.0cm]{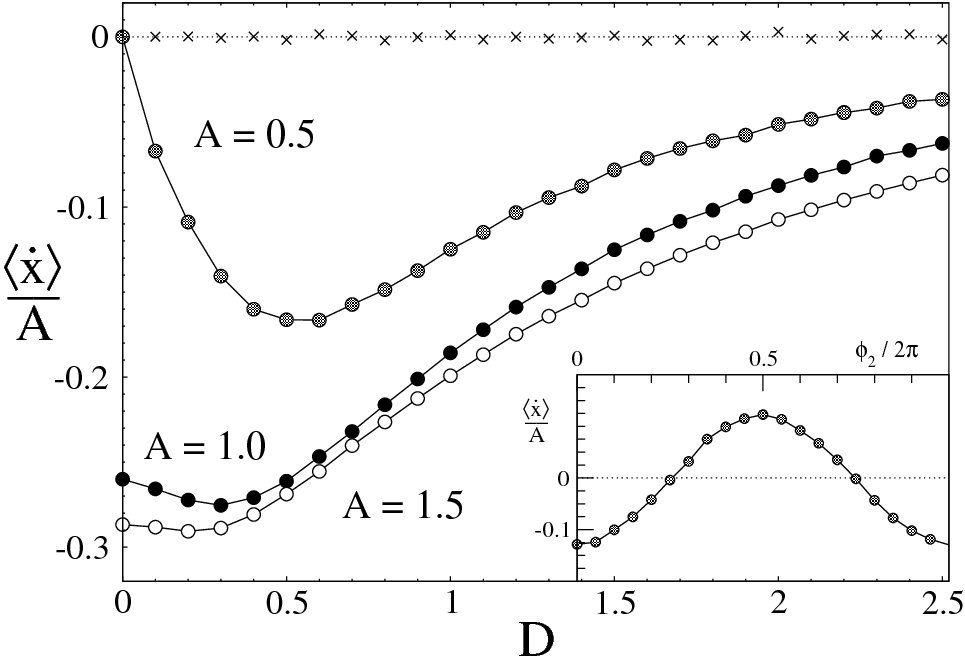}
\end{center}
\caption{Gating mechanism. Equation (\ref{G1}) has been simulated
numerically for $\Omega_1=\Omega_2=0.01$, $\phi_1=\phi_2=0$,
$A_1=A_2=A$, and $V(x) =\omega_0^2(1-\cos x)$ with $\omega_0=1$; the
net velocity $\langle \dot x\rangle$ has been plotted versus $D=kT$
for different amplitudes $A$. For a comparison we show  $\langle
\dot x\rangle$ versus $D$ for $\Omega_1=0.01$, $\Omega_2/\Omega_1=
\sqrt{2}$ and $A=0.5$ (crosses). Inset: $\langle \dot x\rangle$
versus $\phi_2$ for $\phi_1=0$, $A=0.5$, and $D=1$. Note the
resonant behavior of $\langle \dot x \rangle$ for subthreshold
drives, $A<1$, where a low noise level enhances rectification;
increasing $D$ for $A>1$ degrades the rectification effect.
}\label{Fgating}
\end{figure}

For practical applications, any commensuration condition, like
$\Omega_1/\Omega_2=n/m$ for harmonic mixing (but see also Secs.
\ref{gating} and \ref{asymmetry}) is affected by an uncertainty that
is inverse proportional to the observation time, i.e., the code
running time for numerical simulations or the data recording time
for real experiments. As such time is necessarily finite, only a
limited number of commensuration spikes can be actually detected.

We conclude by reminding that the notion of harmonic mixing has been
introduced long ago, e.g., to interpret the output of certain charge
wave density experiments \cite{schneider1966APL8} and to design
laser ring gyroscopes \cite{chow1985RMP57} and annular Josephson
junctions \cite{ustinov2004PRL93}; applications in the context of
nano-particle transport is more recent (see Secs. \ref{noiseinduced}
and \ref{coldatom}).

\subsubsection{Gating} \label{gating}

A periodically-driven Brownian motion can also be rectified by
modulating the amplitude of the substrate potential $V(x)$. Let us
specialize the overdamped dynamics in Eq. (\ref{xtLE}) as follows:
\begin{equation}
\label{G1} \dot x=-V'(x)[1+F_2(t)] +F_1(t) +\xi(t)
\end{equation}
To avoid interference with possible harmonic mixing effects we
follow the prescription of Sec. \ref{HM}, namely we take $F_i(t)=A_i
\hbox{sgn}[\cos(\Omega_it+\phi_i)]$, $i=1,2$, with $A_1 > 0$ and
$0<A_2<1$. Mixing of the additive $F_1(t)$ and the multiplicative
signal $F_2(t)$ provides a control mechanism of potential interest
in device design.

In the adiabatic limit, the Brownian particle moves cyclicly back
and forth subjected to opposite dc forces with amplitude $A_1$; the
substrate potential $V(x,t)=V(x)[1+F_2(t)]$ switches, in turn,
periodically between the two symmetric configurations
$V_{\pm}(x)=V(x)(1 \pm A_2)$. The relevant mobilities
$\mu_{\pm}(A_1)$ can be easily related to the static mobility
function $\mu(A)$ for the tilted potential $V(x)$ studied in Fig.
\ref{Fwashboard}, namely \cite{savelev2004EPL67,savelev2004PRE70b}
\begin{equation}
\mu_{\pm}(A_1)= (1\pm A_2)\, \mu\left[\frac{A_1}{1\pm A_2}\right],
\end{equation}
with $T\to T/(1\pm A_2)$.

In particular, for any pair of commensurate frequencies $\Omega_1$
and $\Omega_2$ such that $\Omega_2/\Omega_1 = (2m-1)/(2n-1)$ (with
$m$, $n$ positive integers), the net particle velocity mediated over
an integer number of cycles of both $F_1(t)$ and $F_2(t)$ can be
cast in the form
\begin{equation}
\label{gate.mob} \langle \dot x \rangle =
-\frac{{(-1)^{m+n}A_1}}{{(2m-1)(2n-1)}}~\Delta
\mu(A_1,A_2)~p(\Delta_{n,m}),
\end{equation}
where $\Delta \mu(A_1,A_2)=\mu_-(A_1)-\mu_+(A_1)$
and $p(\Delta_{n,m})\equiv {|\pi-\Delta_{n,m}|}/{\pi}-0.5$ is a
modulation factor with $\Delta_{n,m}= (2n-1)\phi_2-(2m-1)\phi_1$,
mod($2\pi$). Note that for different choices of the ratio
$\Omega_2/\Omega_1$, no induced drift is predicted (see Fig.
\ref{Fgating}).

As a consequence, a relatively small modulation of the sinusoidal
potential amplitude at low temperatures results in a net transport
current as illustrated in Fig.~\ref{Fgating}. Let us consider the
simplest possible case, $\Omega_1=\Omega_2$ and $\phi_1=\phi_2$: As
the ac drive points to the right, the amplitude of $V(x,t)$ is set
at its maximum value $\omega_0^2(1+A_2)$; at low temperatures the
Brownian particle can hardly overcome the substrate barriers within
a half ac-drive period $\pi/\Omega_1$. In the subsequent half period
$F_1(t)$ switches to the left, while the potential amplitude drops
to its minimum value $\omega_0^2(1-A_2)$: The particle has a better
chance to escape a potential well to the left than to the right,
thus inducing a negative current with maximum intensity for
$\Omega_1$ of the order of the Kramers rate (resonant activation
\cite{borromeo2004EPL68}). Of course, amplitude and sign of the net
current may be controlled via the modulation parameters $A_2$ and
$\phi_2$, too (see inset of Fig.~\ref{Fgating}).

\subsubsection{Noise induced transport} \label{noiseinduced}

Induced transport in the symmetric dynamics can be achieved also by
employing two {\it correlated} noisy signals.

{\it (i) Noise mixing.} \textcite{borromeo2004EPL68} showed that the
gating process in Eq. (\ref{G1}) can be driven also by two
stationary, zero-mean valued Gaussian noises, $F_i(t)=\eta_i(t)$
with $i=1,2$. The two random drives may be cross- and
auto-correlated with
\begin{equation}
\label{G4} \langle \eta_{i}(t) \eta_{j}(0)\rangle = \sqrt{Q_i Q_j}
\,\,\,\frac{\lambda_{ij}}{\tau_{ij}} \exp\left (
-\frac{|t|}{\tau_{ij}}\right ), ~~~~i,j=1,2.
\end{equation}
Without loss of generality, one sets $\lambda_{11}=\lambda_{22}=1$
and $\lambda_{12}=\lambda_{21}=\lambda$, and to avoid technical
complications, $\tau_{ij} \equiv \tau$. Of course $\tau \rightarrow
0$ corresponds to taking the white noise limit of $\eta_i(t)$, Eq.
(\ref{noiseoLE}). The parameter $\lambda$ characterizes the
cross-correlation of the two signals; in particular, $\lambda=0$:
independent signals; $\lambda=1$: identical signals, $\eta_2(t)
\equiv \eta_1(t)$; $\lambda=-1$: signals with reversed sign,
$\eta_2(t) \equiv -\eta_1(t)$. Two such signals may have been
generated by a unique noise source and then partially de-correlated
through different transmission or coupling mechanisms.

In the white noise limit, $\tau \rightarrow 0$, the Fokker-Planck
equation associated with the process of Eqs. (\ref{G1}) and
(\ref{G4}) admits of a stationary solution in closed form.
Non-vanishing values of $\langle \dot x \rangle$ for $\lambda \neq
0$ are the signature of a stochastic symmetry breaking due to {\it
nonlinear noise mixing}; the sign of $\lambda$, similarly to the
relative phase $\Delta_{m,n}$ in the gating set-up of Sec.
\ref{gating}, determines the direction of the particle drift. The
interpretation of this result is straightforward. Let us consider,
for instance, the case of the symmetric potential in Eq.
(\ref{cosine}), rocked and pulsated by the same signal, i.e.,
$\eta_1(t)=\eta_2(t)$: For $\lambda=1$, when pushed to the left
($\eta_i<0$), the Brownian particle encounters lower substrate
barriers than when pushed to the right ($\eta_i>0$), hence the
negative average drift current detected by means of numerical
simulation \cite{borromeo2005CHAOS15,borromeo2006PRE74}.

\begin{figure}[ht]
\includegraphics*[width=6.5cm]{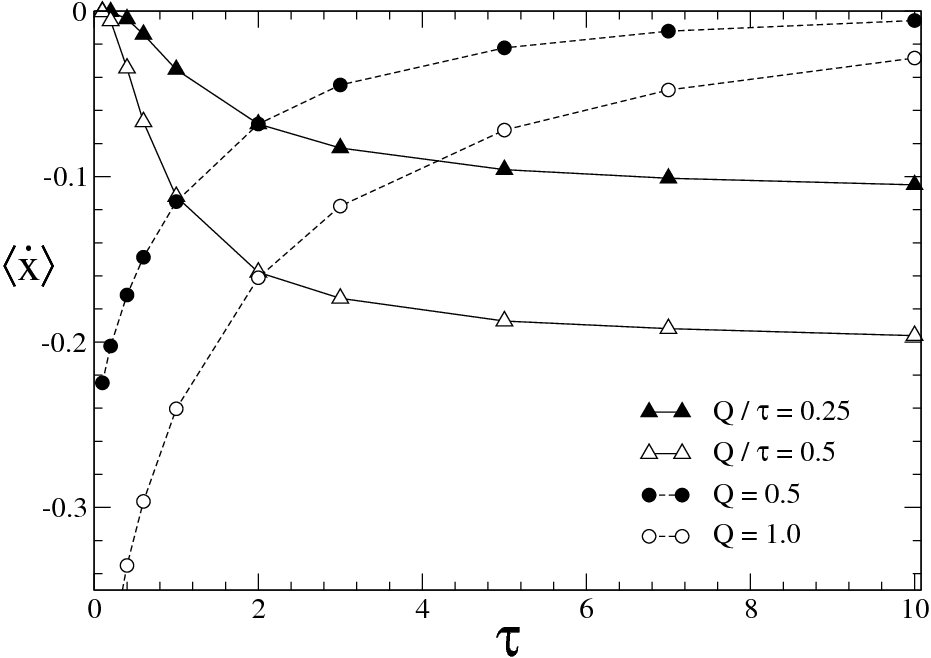}
\includegraphics*[width=6.5cm]{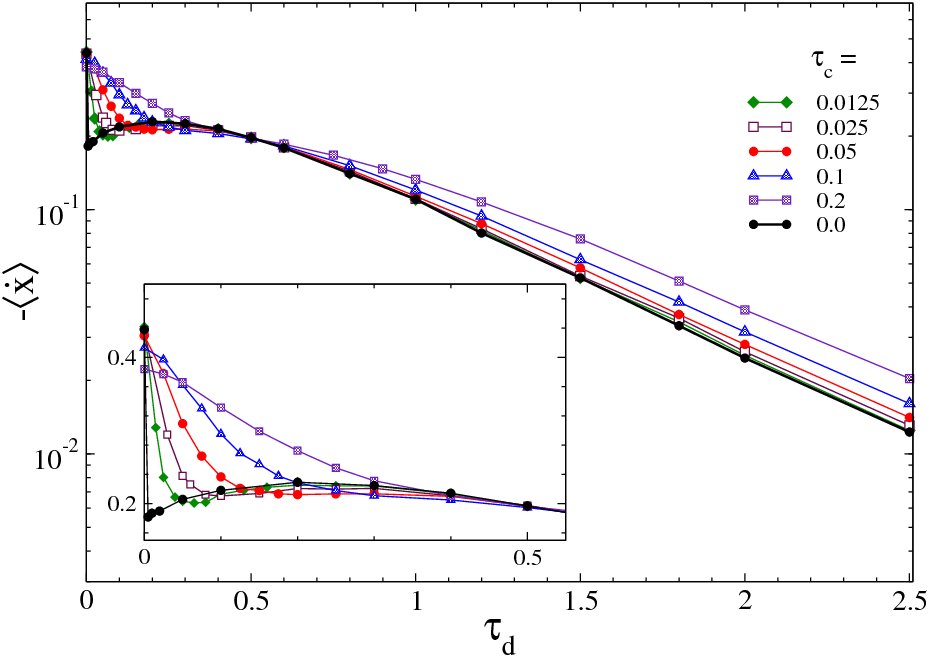}
\caption{\label{Fdelay} (Color online) Equation (\ref{G1}), with
$F_i(t)= \eta_i(t)$, $i=1,2$, and $V(x) =-\sin x$, has been
simulated numerically; the noises $\eta_i(t)$, Eq. (\ref{G4}), have
same strength, $Q_1=Q_2=Q$. After \textcite{borromeo2005CHAOS15}:
(a) Nonlinear noise mixing: Characteristics curve $\langle \dot
x\rangle$-$\tau$ for different intensities $Q$ of the noises and
$\lambda=1$; $D=0.1$. (b) Noise recycling:  Characteristics curve
$v$-$\tau_d$, where $\tau_d$ is the relative time delay of $\eta_i$
(see text). Data are for different $\tau$ and $Q=1$; $D=0$,
see in \textcite{borromeo2006PRE74}.}
\end{figure}

The magnitude of the induced current is controlled by the
correlation time $\tau$ \cite{borromeo2005CHAOS15}. While one can
easily estimate $\langle \dot x \rangle$ for $\lambda \neq 0$ and
$\tau=0$ (white noise \cite{risken1984}) or $\tau \rightarrow
\infty$ (strongly correlated noise \cite{hanggi1989JSP54}), the
intermediate $\tau$ values are accessible solely through numerical
simulation. In Fig. \ref{Fdelay}(a) we report $\langle \dot x
\rangle$ versus $\tau$ for different noise intensities: the two sets
of curves at $Q$ and $Q/\tau$ fixed, illustrate well the noise
mixing effect for $\tau \to 0$ and $\tau \to \infty$, respectively. \\

{\it (ii) Noise recycling.} If the noises $\eta_i(t)$ are generated
by the same source and then coupled to the diffusing particle
through different paths, it may well happen that they are simply
delayed in time. Under stationary conditions, we can assume that
$\eta_2(t)=\eta_1(t-\tau_d)$, with $\eta_i(t)$ given in Eq.
(\ref{G4}) and, for simplicity, $\lambda=0$. In the notation of
Refs. \cite{borromeo2006PRE74,borromeo2007PRE75}, $\eta_1(t)$
represents the primary noise source and $\eta_2(t)$ a recycled noise
to be used as a control signal.

By the same argument as for noise mixing, we expect that the
Brownian dynamics gets rectified to the left with negative velocity
$\langle \dot x (\tau_d) \rangle$. Note that: $\langle \dot x
(-\tau_d) \rangle=\langle \dot x (\tau_d) \rangle$ and $\langle \dot
x (\tau_d) \rangle \to -\langle \dot x (\tau_d) \rangle$, upon
changing the relative signs of $\eta_i$. The dependence of the
characteristic curve $\langle \dot x (\tau_d) \rangle$ on the time
constants $\tau_d$ and $\tau$ displayed in Fig. \ref{Fdelay}(b), is
important in view of practical applications. Indeed, in many
circumstances, it would be extremely difficult to recycle a control
signal $\eta_2(t)$ so that $\tau_d \ll \tau$; stated otherwise,
measuring $\langle \dot x (0) \rangle$ requires a certain degree of
experimental sophistication. On the contrary, if we agree to work on
the resonant tail of its response curve $|\langle \dot x (\tau_d)
\rangle|$, a noise controlled rectification device can be operated
with less effort; its net output current may be not the highest for
$\tau_d>\tau$, but is still appreciable and, more importantly,
stable against the accidental floating of the time constant
$\tau_d$.

In this sense both schemes discussed in this section are a
simple-minded attempt at implementing the operation of a Maxwell's
daemon: The ideal device we set up is intended to gauge the primary
random signal $\eta_1$ at the sampling time $t$ and, depending on
the sign of each reading, to lower or raise the gate barriers
accordingly at a later time $t+\tau_d$, i.e., open or close the trap
door. The rectifying power of such a daemon is far from optimal;
lacking the dexterity of Maxwell's ``gate-keeper''
\cite{leff2003,maruyama2008RMP}, it only works ``in average" like an
automaton.


\subsection{Brownian motors} \label{molecularmotor}

As detailed in Sec. \ref{rectification}, a necessary condition for
the rectification of symmetric signals, random or periodic in time,
alike, is the spatial asymmetry of the substrate. Rectification
devices involving asymmetric substrates are termed {\it ratchets}.
In most such devices, however, noise (no matter what its source,
i.e. stochastic, or chaotic, or thermal) plays a non-negligible, or
even the dominant role. Under such conditions one speaks of Brownian
motors
\cite{astumian1997Science276,astumian2002PhysToday55,bartussek1995PhysBl51,hanggi1996LNP476,reimann1996PLA215,
linke2002ApplPhysA75,hanggi2002ChemPhys281,hanggi2005AnPhys14,
kay2007AngewChem46}. The label Brownian motor should not be abused
to refer to all small ratchet-like devices. For instance, the rocked
ratchets of Sec. \ref{rockedratchet} work quite differently in the
presence or in the absence of noise, whereas the pulsated and
correlation ratchets (sometimes also referred to as ``stochastic"
ratchets) of Secs. \ref{pulsatedratchet} and \ref{thermalratchet}
work only in the presence of noise.

The hallmarks of genuine Brownian motors are listed in Sec.
\ref{introduction}. In this Section we discuss in detail noise
rectification and directed transport on asymmetric periodic
substrates and potentials. We caution the reader that: (1) Strict
periodicity is not a requirement for the operation of a Brownian
motor. The ratchet system may contain small amounts of disorder
\cite{harms1997PRL79,popescu2000PRL85,kafri2005JOP17,
martinez2008PRL100} or even be non-periodic
 \cite{marchesoni1997PRE56}; (2)
Spatial asymmetry can also result as a collective effect, for
instance, in the extended systems consisting of interacting,
symmetric dynamical components, introduced in Sec.
\ref{collectivetransport}.

\begin{figure}[ht]
\includegraphics[width=7.0cm]{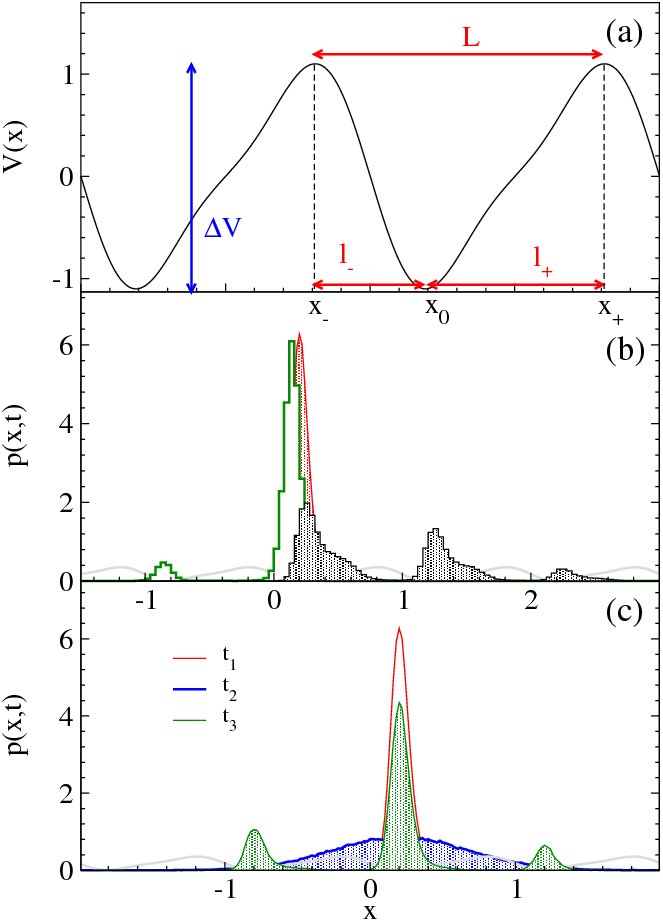}
\caption{(Color online) Ratchet mechanism. (a) Sketch of the
potential in Eq. (\ref{doublesine}) with $L=1$ and $V_0=L/2\pi$. The
three consecutive extremal points, $x_-=-0.19$, $x_{0}=0.19$ and
$x_{+}=0.81$, define a potential well; the barrier height is $\Delta
V=V(x_{\pm})-V(x_0)\simeq 0.35$ and its asymmetry is quantified by
the difference $l_+-l_-$ with $l_{\pm}=|x_{\pm}-x_0|$. The curvature
of the potential at the bottom of the well is $\overline
{\omega}_0^2=(3\sqrt{3}/2)\omega_0^2\simeq 10.1$. (b) Rocket
ratchet. Probability density $p(x,t)$ for a Brownian particle
initially centered around $x=0$ (red) and then driven for $t=10$ by
a dc force $A_0=-0.5$ (green) and $0.5$ (grey). The backward
displacement is strongly suppressed, hence the positive natural
orientation of this ratchet. (c) Pulsated ratchet. The ratchet
potential is switched ``on" and ``off" periodically with $t^{\rm
on}=1$, $t^{\rm off}=3$ and $T_2=t^{\rm on}+t^{\rm off}$ (see text).
The particle, initially set at $x=x_0$, relaxes first in the
starting well during $t^{\rm on}$ (curve 1, red, $t_1=t^{\rm on}$),
then diffuses symmetrically only driven by noise for $t^{\rm off}$
(curve 2, blue, $t_2=T_2$), and finally gets retrapped in the
neighboring wells as the next cycle begins (curve 3, green,
$t_3=T_2+t^{\rm on}$). As the left side peak of $p(x,t_3)$ is more
pronounced than the right peak, the natural orientation of this
ratchet is negative. In the simulations of (b) and (c) the ratchet
potential $V(x)$ is the same as in (a) (also drawn to guide the eye)
and the intensity of the noise in Eq. (\ref{noiseoLE}) is
$D=0.1\Delta V$. Courtesy of Marcello Borromeo.}\label{Frratchet1}
\end{figure}

The archetypal model of ratchet substrates in 1D is the double-sine
potential proposed by \textcite{bartussek1994EPL28}
\begin{equation} \label{doublesine}
V(x)=-V_0\left [\sin\left(\frac{2\pi x}{L}\right)
+\frac{1}{4}\sin\left(\frac{4\pi x}{L}\right)\right ],
\end{equation}
or, in the rescaled units of Eq. (\ref{yLE}), $V(x)=-\omega_0^2[\sin
x +\frac{1}{4}\sin(2x)]$. In the sketch of Fig. \ref{Frratchet1},
the barriers are skewed to the right with ratchet length $l_+>l_-$.
A Brownian particle with Langevin equation (\ref{xLE}) moving on
such a substrate is characterized by an asymmetric mobility
function, $\mu(F)\neq \mu(-F)$; as the particle mobility depends now
not only on the amplitude, but also on the orientation of the drive
$F$, symmetric substrate perturbations are expected to induce a net
current in either direction.

Broadly speaking, ratchet devices fall into three categories
depending on how the applied perturbation couples to the substrate
asymmetry.

\subsubsection{Rocked ratchets}\label{rockedratchet}

Let us consider first, for simplicity, the Langevin equation
(\ref{oLE}) with the sinusoidal drive of Eq. (\ref{forceac}). When
applied to the reflection-symmetric sine potential of Eq.
(\ref{cosine}), $F(t)$ breaks instantaneously the symmetry of the
potential by tilting it to the right for $F(t) > 0$, and to the left
for  $F(t) < 0$. However, due to the spatio-temporal symmetry of the
process, $V(x)=V(-x)$ and $F(t)=F(-t)$, over one drive cycle,
$T_1=2\pi/\Omega_1$, the drift to the right and to the left
compensate one another: the net current is null. When placed in the
double-sine potential of Eq. (\ref{doublesine}), instead, an
overdamped particle is apparently more movable to the right than to
the left. Indeed, depinning occurs at $F = F_{3R}=3/4$ and
$F=-F_{3L}$ for $F(t)$ oriented respectively to the right and to the
left; moreover from the asymmetry condition $l_+>l_-$ folows
immediately the inequality $F_{3R}<F_{3L}$. The depinning thresholds
of the potential of Eq. (\ref{doublesine}) in the rescaled units of
Eq. (\ref{yLE}) are $F_{3R}=3/4$ and $F_{3L}=3/2$.

\begin{figure}[ht]
\includegraphics[width=6.0cm]{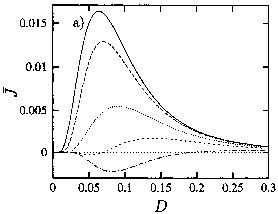}
\vglue 0.5truecm
\includegraphics[width=6.0cm]{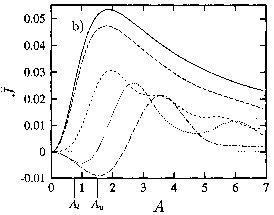}
\vglue 0.5truecm
\includegraphics[width=6.0cm]{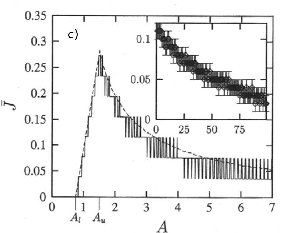}
\vglue 0.5truecm \caption{Rocked ratchet. Rectification current $j$
versus the noise intensity $D$, see Eq. (\ref{noiseoLE}), panel (a),
and versus ac modulation strength $A \equiv A_1$, panels (b) and (c)
for the ratchet potential in Eq. (\ref{doublesine}) with $L=1$. The
relevant simulation parameters are: (a) $A_1=0.5$ and
$\Omega_1=0.01, 1, 2.5, 4$ and $7$ (top to bottom); (b) $D=0.1$ and
$\Omega_1=0.01, 1, 4$ and $7$ (top to bottom). In both panels (a)
and (b) the solid line for $\Omega_1=0.01$ coincides with the
adiabatic limit in Eq. (\ref{rrcurrent});  (c) deterministic regime
with $D=0$, and $\Omega_1=0.01$ (dashed line) and $0.25$ (solid
line). After \textcite{bartussek1994EPL28}.}\label{Frratchet2}
\end{figure}

If the forcing frequency $\Omega_1$ is taken much smaller than all
the intrawell relaxation constants, $\omega_0^2 T_1 \gg 1$ and $T_1
\gg \langle t(L,0) \rangle$, than the net transport current can be
computed in the adiabatic approximation by averaging the
instantaneous velocity $\dot x(t)=\mu[F(t)]F(t)$, see Eq.
(\ref{mobility}), over one forcing cycle, that is
\cite{bartussek1994EPL28,borromeo2002PRE65}
\begin{equation} \label{rrcurrent}
j\equiv \frac{\langle \dot x
\rangle}{L}=\frac{1}{T_1}\int_0^{T_1}\frac{\hbox{sgn}[F(s)]}{\langle
t[L,F(s)]\rangle}ds.
\end{equation}
The adiabatic current in Eq. (\ref{rrcurrent}) is positive definite,
as the natural ratchet direction is defined by the choice
$F_{3R}<F_{3L}$. However, its range of validity is restricted to
extremely low temperatures $T$ and drive frequencies $\Omega_1$. On
raising $\Omega_1$, the rectification current develops a more
complicated dependence on the noise intensity $D$ and the drive
amplitude $A_1$, as proven by the numerical simulations reported in
panels (a)-(c) of Fig. \ref{Frratchet2}. This is due to the fact
that for forcing periods $T_1$ shorter than the mean first-passage
time $\langle t(L,0)\rangle$, the driven particle oscillations span
over fewer substrate cells. A few properties of
the rocked ratchet current in Fig. \ref{Frratchet2} are remarkable:\\
(a) At finite temperature, the maximum rectification effect occurs
right in the adiabatic regime, $\Omega_1\to 0$, and with natural
orientation, panels (a) and (b); \\
(b) For $\Omega_1>\omega_0$, the drive and the interwell
oscillations may combine to reverse the current orientation. Such
{\it current inversions} are restricted to select $D$ and $A_1$
ranges, but never in the absence of noise, panel (a) and (b);
\\
(c) In the noiseless regime, $T\equiv 0$, the incommensuration
between forced oscillation amplitude and substrate periodicity
causes the quantized locking structure of panel (c). Such a
structure disappears either in the limit $\Omega_1 \to 0$, as the
forced oscillation amplitude grows infinitely large, or in the
presence of noise, when, due to randomness, the finer steps for
$A_1>F_{3L}$ merge into the broad oscillations of panel (b);\\
(d) As the temperature vanishes, the adiabatic current in Eq.
(\ref{rrcurrent}) gets suppressed for $A_1<F_{3R}$ and enhanced
elsewhere [see Fig. \ref{Fasym}(a)]. For $F_{3R}<A_1<F_{3L}$ the
particle can only move to the right, so that $j$ grows almost
linearly with $A_1$, while for $A_1>F_{3L}$ the particle can drift
in both directions, thus making $j$ decrease.

Rectifiers, like a rocked ratchet, can work against an external dc
drive. On adding a dc component $A_0$ to the sinusoidal force,
namely, on applying the drive (\ref{forcedcac}) instead of
(\ref{forceac}), we can determine the value of $A_0$, termed {\it
stopping force}, such that for a certain choice of the other
perturbation parameters, $A_1$, $\Omega_1$ and $D$, the net current
vanishes.

Rocked ratchets with massive particles exhibit strong inertial
effects also capable of reversing their current. A noiseless
inertial rocked ratchet is naturally subject to developing chaotic
dynamics; indeed, its current turns out to be extremely sensitive to
the initial conditions
\cite{jung1996PRL76,mateos2000PRL84,mateos2003PhysicaA325,borromeo2002PRE65,machura2007PRL98}.
However, an appropriate noise level suffices to stabilize the
rectification properties of an inertial ratchet and makes it a
useful device with potential applications in science and technology
\cite{machura2004PRE70,marchesoni2006PRE73}.

In the notation of Sec. \ref{dcdrive}, slightly generalized to
account for the asymmetry of the potential in Eq.
(\ref{doublesine}), we define two additional pairs of thresholds,
besides the depinning thresholds $F_{3R,L}$: the repinning
thresholds $F_{1R,L}$ and the underdamped transition thresholds
$F_{2R,L}$, with $F_{iR}<F_{iL}$ for $i=1,2,3$ [the suffix $R,L$
denoting the orientation of the drive]. Under the adiabatic
condition $\Omega_1\ll \gamma, \omega_0$, the following
approximations have been obtained in
\textcite{borromeo2002PRE65,marchesoni2006PRE73}:\\
(a) in the zero-temperature limit, $T\to 0$:
\begin{eqnarray}\label{rrcurrentgamma0}
j(A_1)&=&j_R(A_1),
~~~~~~~~~~~~~F_{2R}<A_1<F_{2L}\nonumber\\
&=&j_R(A_1)-j_L(A_1)
~~~~~~~~~~A_1>F_{2L}
\end{eqnarray}
and $j(A_1)=0$ for $A_1<F_{2R}$. The right/left current components
are
$j_{R,L}(A_1)=2\sqrt{A_1^2-F_{2R,L}^2}/(\pi \gamma L)$.\\
(b) in the noiseless regime, $T\equiv 0$:
\begin{eqnarray}\label{rrcurrentT0}
j(A_1)&=&j_R(A_1), ~~~~~~~~~~~~~~~
F_{3R}<A_1<F_{3L}\nonumber\\&=&j_R(A_1)-j_L(A_1),
~~~~~~~~~~~ A_1>F_{3L}
\end{eqnarray}
and $j(A_1)=0$ for $A_1<F_{3R}$. Here,
$j_{R,L}(A_1)=[\sqrt{A_1^2-F_{1R,L}^2}+\sqrt{A_1^2-F_{3R,L}^2}]/(\pi
\gamma L)$. Note that $j(A_1)$ in Eq. (\ref{rrcurrentT0}) is
discontinuous at the depinning thresholds $F_{3R,L}$ as expected for
the response curve of any underdamped rocked ratchet operated in the
deterministic regime (see Sec. \ref{dcdrive}).

\subsubsection{Pulsated ratchets}\label{pulsatedratchet}

Let us consider now the overdamped process of Eq. (\ref{G1}) with
$F_1(t)=0$ and $F_2(t)=A_2 \hbox{sgn}[\cos(\Omega_2t+\phi_2)]$. As
discussed in Sec. \ref{gating}, modulating the amplitude of a
symmetric potential $V(x)$ in the presence of an uncorrelated,
time-symmetric perturbation $\xi(t)$ does not suffices to induce
rectification. This state of affair changes when $F_2(t)$ couples,
instead, to an asymmetric substrate, like our reference ratchet
potential (\ref{doublesine}). For the sake of simplicity, we set
$A_2=1$, so that the effective substrate potential $[1+F_2(t)]V(x)$
appears to switch on and off at every half period
$T_2/2=\pi/\Omega_2$. The interplay between time modulation and
spatial asymmetry generates a non-zero drift current, which, at low
(but not too low!) frequencies, is oriented in the positive
direction. Devices operated under similar conditions are termed {\it
pulsated ratchets}.

\begin{figure}[ht]
\includegraphics[width=7.0cm]{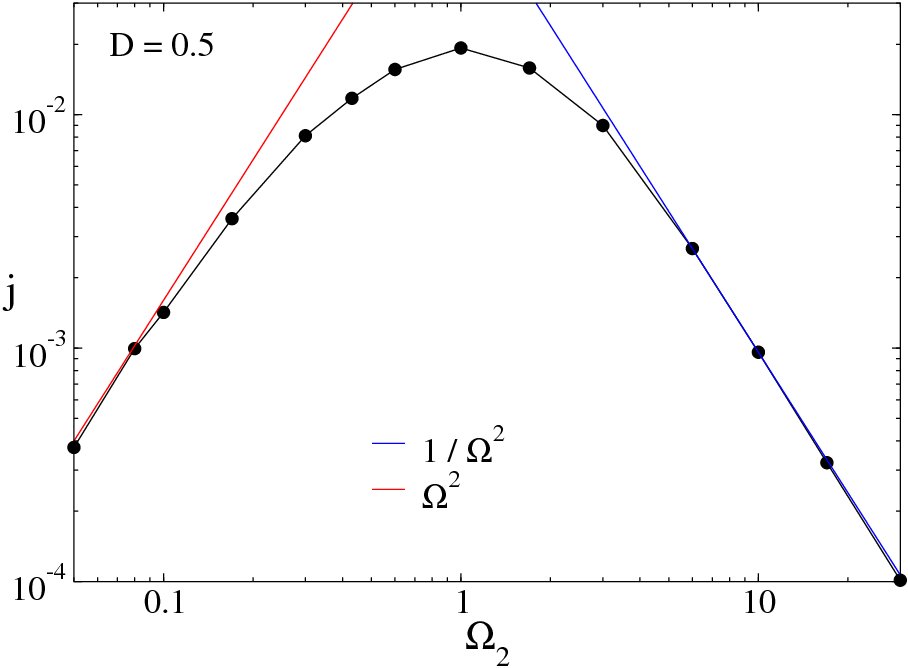}
\caption{(Color online) Pulsated ratchet. Rectification current
$j=\langle \dot x \rangle$ versus $\Omega_2$ (dimensionless units),
for the ratchet potential of Eq. (\ref{doublesine}) with $L=2\pi$
and  $V_0=1$. Other simulation parameters are: $A_2=0.8$ and $D=0.5$
The low frequency $\Omega_2^2$, and high frequency, $\Omega_2^{-2}$
asymptotic limits are denoted by straight lines. Courtesy of
Marcello Borromeo.}\label{Fpratchet}
\end{figure}

Rectification by a pulsated ratchet can be explained qualitatively
by looking at the cartoon of Fig. \ref{Fpratchet}. The probability
density of a particle initially placed at the bottom of a potential
well (top), is mostly confined to that well, as long as $T_2/2$ is
not exceedingly longer than the escape time $\langle t(L,0)\rangle$,
which is the case at low temperatures. When, during the following
half cycle, the substrate is switched off (middle), the particle,
still subject to noise, diffuses freely with Einstein's constant $D$
and, for $T_2\gg L^2/D$, its probability density approaches a
Gaussian spanning over many a unit cell. By switching on again, the
substrate cuts such a probability density into smaller peaks of
different size, one for each well surrounding the initial one. For
potential barriers skewed like in figure, $l_+>l_-$, wells on the
right are expected to be more populated than wells on the left,
hence a positive net current. This is also the case of the potential
in Eq. (\ref{doublesine}), which has positive natural orientation
when pulsated, and negative orientation when rocked. The mechanism
illustrated in Fig. \ref{Fpratchet} can work even in the presence of
dc force, $F_1(t)=A_0$, pointing in the opposite direction, $A_0<0$.
Such an upward directed motion is clearly powered by the thermal
fluctuations $\xi(t)$.

The directed particle current is clearly bound to vanish in the
adiabatic limit, $\Omega_2 \to 0$, when thermal equilibrium is
approached. A similar conclusion holds true for very fast substrate
modulations, $\Omega_2 \gg \omega_0^2$ (in rescaled units). In
perturbation analysis \cite{savelev2004PRE70a}, one finds the
noteworthy result that the current decays to zero in both asymptotic
regimes remarkably fast, namely like $\Omega_2^{2}$ in the slow
modulation limit, and  $\Omega_2^{2}$, in the fast modulation limit
respectively. Moreover, for intermediate modulation frequencies, the
current in a pulsated ratchet is generally not oriented in its
natural direction; current inversions are possible when the forcing
frequency matches some intrinsic relaxation rate of the process.

Finally, pulsated ratchets can be operated under very general
amplitude modulations $F_2(t)$, as well. This ratchet effect proved
robust with respect to (a) modifications of the potential shape
\cite{reimann2002PhysRep361}; (b) changes in the switching sequence.
A generic discrete modulation signal $F_2(t)$ is characterized by
different residence times, namely, $t_{\rm on}$ in the ``on" state,
$+A_2$, and $t_{\rm off}$ in the ``off" state, $-A_2$, respectively.
For periodic modulations, $t_{\rm on}+t_{\rm off}=T_2$ defines the
so-called duty cycle of the device
\cite{bug1987PRL59,ajdari1992ComptRend315}; for random modulations,
instead, $t_{\rm on}$ and $t_{\rm off}$ must be interpreted as the
average residence times in the respective ``on", ``off" state
\cite{astumian1994PRL72,faucheux1995PRL74}. In both cases the
modulation is asymmetric for $t_{\rm on} \neq t_{\rm off}$; (c)
replacement of pulsated with flashing substrates. The substrate is
made switch, either periodically or randomly in time, among two or
more discrete configurations, which do not result from the amplitude
modulation of a unique substrate profile
\cite{gorman1996PRL76,chen1997PRL79,borromeo1998PLA249,lee2005PRL94}.
The rectification process is controlled by the spatial asymmetry of
the single substrate configurations or by the temporal asymmetry of
the switching sequence, or by a combination of both; (d) modulation
of the temperature \cite{reimann1996PLA215,bao1999PhysicaA273}.
Modulating the intensity of the ambient noise $\xi(t)$ or its
coupling to the device, corresponds to introducing the time
dependent temperature $T(t)=[1+F_2(t)]T$, for an appropriate control
signal $F_2(t)$ with amplitude smaller than one. Pulsating the
potential amplitude $V_0$ or the temperature $T$ yield very similar
rectification effects, being the net current of a pulsated ratchet
mostly controlled by the barrier-to-noise ratio; (e) the addition of
constant or spatially modulated damping terms
(\textcite{luchsinger2000PRE62,suzuki2003PRE68}; see also
\textcite{reimann2002PhysRep361}). Inertial effects make the
rectification mechanism sensitive to the particle mass, so that
selective transport and segregation of mixed species becomes
possible \cite{derenyi1995PRL75,derenyi1995PRE54}.

\subsubsection{Correlation ratchets}\label{thermalratchet}

To better understand the role of asymmetry in the rectification of a
nonequilibrium process, we consider now an overdamped Brownian
particle  with Langevin equation (\ref{oLE}) diffusing in the
ratchet potential of Eq. (\ref{doublesine}) subject to a zero-mean,
colored Gaussian noise $F\equiv \eta(t)$, with
\begin{equation}
\label{tratchet1} \langle \eta(t) \eta(0)\rangle = \frac{Q}{\tau}
\exp\left ( -\frac{|t|}{\tau}\right ).
\end{equation}
The substrate asymmetry is capable {\it per se} of rectifying the
correlated fluctuation $\eta(t)$, even in the absence of external
modulations or thermal noise. The early prediction of this transport
effect, based on simple perturbation arguments
\cite{magnasco1993PRL71,doering1994PRL72,bartussek1994EPL28,millonas1994PLA185,luczka1995EPL31},
kindled a widespread interest in the thermodynamics of molecular
motors. A more sophisticated path-integral analysis
\cite{bartussek1996PRL76} led later to the following low-noise
estimates of the correlation ratchet current:\\
(a) in the weak color limit, $\omega_0^2\tau \ll 1$,
\begin{eqnarray}
\label{tratchet2} j(\tau)&=& j_+(\tau)-j_-(\tau),
\end{eqnarray}
where $j_{\pm}(\tau)=r_K\exp[-\tau^2Qc_{\pm}/(D+Q)^2]$ and $r_K$ is
the equilibrium Kramers rate for the particle to exit a potential
well through one side, $r_K=\overline{\omega}_0\omega_{\pm}\exp[-
\Delta V/(D+Q)]/(2\pi)$.
Here, $x_0$ and $x_{\pm}$ are the extremal points of $V(x)$ defined
as in Fig. \ref{Frratchet1}; $\overline{\omega}_0^2=V''(x_0)$,
$\omega_{\pm}^2=|V''(x_{\pm})|$,
and $c_{\pm}=\int_{x_0}^{x_{\pm}}[V''(x)]V'(x)dx$;\\
(b) in the strong color limit, $\tau \to \infty$,
\begin{eqnarray}
\label{tratchet3} j(\tau)&=&j_0~\left[1-\exp\left
({-\frac{QL}{2D^2\tau}}(l_+-l_-)\right )\right],
\end{eqnarray}
where $j_0$ is a positive definite constant. Equation
(\ref{tratchet2}) can be regarded as the difference of two exit
currents $j_{\pm}(\tau)$, respectively through $x_+$ (forward) and
$x_-$ (backward), in the presence of colored noise
\cite{millonas1994PLA185}.

\begin{figure}[ht]
\includegraphics[width=8.5cm]{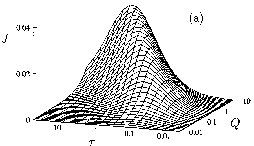}
\caption{Correlation ratchet. Rectification current $j$ versus
$\tau$ and $Q$ for $D=0.1$. The parameters of the ratchet potential,
Eq. (\ref{doublesine}), are as in Fig. \ref{Frratchet1}. More data
can be found in \textcite{bartussek1996PRL76}.}\label{Ftratchet}
\end{figure}

The orientation of the current of a correlation ratchet is extremely
sensitive to the substrate geometry. In the strong color limit, the
condition $(l_{+} - l_{-}) > 0 $ between the two ratchet lengths
guarantees that the current is positive, no matter what the choice
of $V(x)$; all correlation ratchets work in the natural direction of
the corresponding rocked ratchet.
\begin{figure*}[ht] 
\includegraphics[width=12cm]{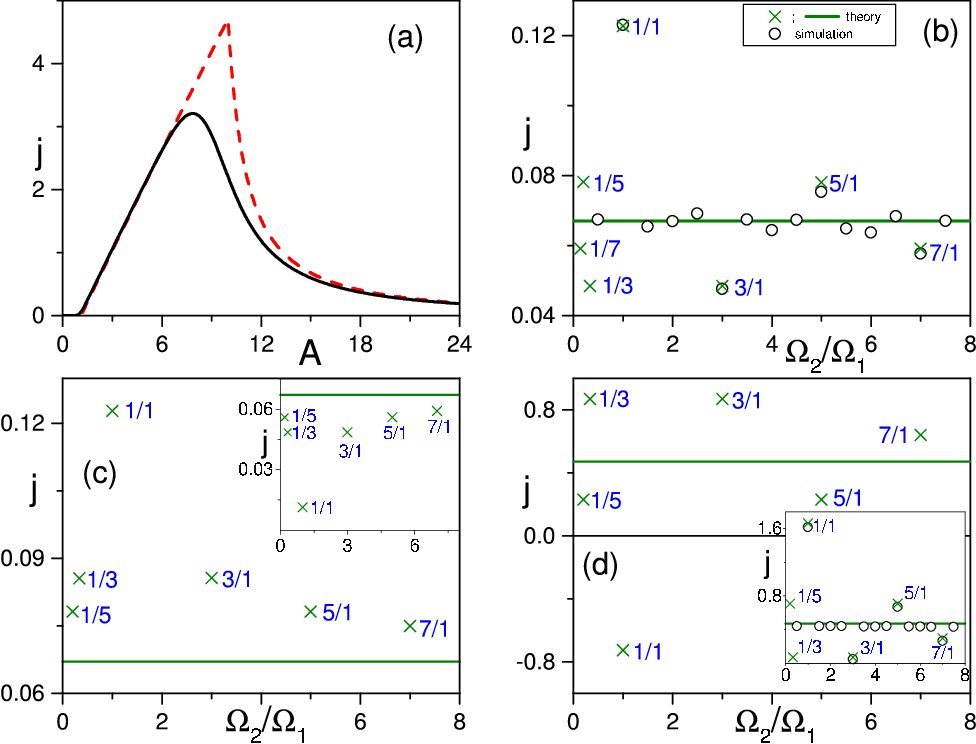}
\caption{(Color online) Net current in a piecewise linear potential
with amplitude $\Delta V$, driven by two rectangular waveforms, Eq.
(\ref{HM5}): (a) one-frequency rocked ratchet; (b), (c) harmonic
mixing of Sec. \ref{HM}; (d) gating mechanism, Eq. (\ref{G1}). The
ratchet potential is: $V(x)=x\Delta V/L_2$ for $-L_2 <x <0$ and
$V(x)=x\Delta V/L_1$ for $0<x <L_1$, with $L_1=0.9$ and $L=1$;
$\Delta V=1$ in (a)-(c) and $\Delta V=2$ in (d). Panel (a): Response
curve $j(A)\equiv \langle \dot x \rangle/L$ to a low-frequency
rectangular force with amplitude $A$ at zero temperature $D = 0$
(dashed curve), and low temperature $D/\Delta V= 0.05$ (solid
curve). Panel (b): Numerical simulations for a doubly rocked ratchet
with $\phi_1=\phi_2=\pi$ and $\Omega_1=0.01$ (open circles) versus
adiabatic approximation, Eqs. (\ref{gate.mob}) and (\ref{asym.HM})
(green line and green crosses). The off-set $j_{\rm avg}$, Eq.
(\ref{asym.HM}), is indicated by the green line; the spikes of Eq.
(\ref{gate.mob}) at some selected integer-valued odd harmonics are
marked with green crosses ($\times$); Panel (c): Adiabatic
approximation for $\phi_1=\phi_2=3\pi/2$ (main panel) and
$\phi_1=3\pi/2$, $\phi_2=\pi/2$ (inset). In both cases $A_1=3$,
$A_2=2$, $D=0.6$; Panel (d): Numerical simulations versus adiabatic
approximation, Eqs. (\ref{gate.mob}) and (\ref{asym.gating}), for a
rocked-pulsated ratchet with $A_1=4$, $A_2=0.5$ and $\Omega_1=0.01$;
noise level: $D=0.4$. Main panel: $\phi_1=\phi_2=\pi$ (adiabatic
approximation); inset: simulation (open circles) versus the fully
adiabatic approximation ($\times$) for $\phi_1=\pi$ and $\phi_2=0$.
After \textcite{borromeo2005CHAOS15}.}\label{Fasym}
\end{figure*}

In the weak color limit, the current in Eq. (\ref{tratchet2}) has
the same sign as the difference
$c_--c_+=\int_{0}^{L}[V''(x)]V'(x)dx$, which, in turn, depends on
the detailed profile of $V(x)$, regardless of $\hbox{sgn}[l_+-l_-]$.
Numerical simulation of correlation ratchets with potential given in
Eq. (\ref{doublesine}), confirmed that, consistently with the
inequality $c_--c_+>0$, $j(\tau)$ is positive definite, as
illustrated in Fig. \ref{Ftratchet}; as $j(\tau\to 0)=j(\tau \to
\infty)=0$, optimal rectification is achieved for an intermediate
noise correlation time, $\omega_0^2\tau \sim 1$. Modifying the
potential profile, e.g., by adding an appropriate higher order
Fourier component without changing $l_+-l_-$, suffices to reverse
the sign of $c_--c_+$ and, thus, introduce at least one current
inversion \cite{bartussek1996PRL76}.

Correlation ratchets have the potential for technological
applications to ``noise harvesting", rather than to nano-particle
transport. The low current output and its extreme sensitivity to the
substrate geometry and the particle mass
\cite{marchesoni1998PLA237,lindner1999PRE59} makes the design and
operation of correlation ratchets as mass rectifiers questionable.
However, the asymmetry induced rectification of nonequilibrium
fluctuations, no matter how efficient and hard to control, can be
exploited by a small device to extract from its environment and
store the power it needs to operate.

\subsubsection{Further asymmetry effects} \label{asymmetry}

The rectification mechanisms introduced in Sec.
\ref{nonlinearmechanism} apply to any periodic substrate,
independently of its spatial symmetry. In the presence of spatial
asymmetry, the relevant drift currents get modified as follows:

(i) {\it Current off-sets.} At variance with Secs. \ref{HM} and
\ref{gating}, two ac drives applied to an asymmetric device are
expected to induce a net current  $j_{\rm avg}=\langle \dot x
\rangle_{\rm avg}$ also for incommensurate frequencies $\Omega_1$,
$\Omega_2$. This is a mere ratchet effect that in most experiments
is handled as a simple current off-set. This conclusion is apparent,
for instance, in the case of low-frequency rectangular input waves,
where the device output current can be expressed as a linear
combination of two known ratchet currents
\cite{savelev2004EPL67,savelev2004PRE70b},
namely: \\
(a) Harmonic mixing current (Sec. \ref{HM}). When the drive is a
double rectangular wave, Eq. (\ref{HM5}), $j_{\rm avg}$ can be
regarded as the incoherent superposition of currents from two rocked
ratchets driven by rectangular waves with amplitudes $A_1+A_2$ and
$|A_1-A_2|$, respectively, i.e.
\begin{equation} \label{asym.HM}
j_{\rm avg}=\frac{1}{2}[j_0(|A_1-A_2|)+ j_0(A_1+A_2)];
\end{equation}
(b) Gating current (Sec. \ref{gating}). The rocked-pulsated device
of Eq. (\ref{G1}) can be regarded as the incoherent superposition of
two ratchets with potentials $V_{\pm}(x)=(1\pm A_2)V(x)$,
respectively, both rocked with amplitude $A_1$; hence
\begin{equation} \label{asym.gating}
j_{\rm avg}=\frac{1}{2}[j_+(A_1)+j_-(A_1)].
\end{equation}
In Eqs. (\ref{asym.HM}) and (\ref{asym.gating}) $j_{0,\pm}(A)$ are
the net currents of Eq. (\ref{rrcurrent}) for a rocked ratchet
driven by a rectangular wave with amplitude $A$ and vanishing
frequency, the suffixes $0,\pm$ referring to the regular, high/low
amplitude potential configurations, $V(x)$ and $V_{\pm}(x)$.

(ii) {\it Asymmetry induced mixing.} As anticipated in Sec.
\ref{HM}, a double rectangular wave is not capable of rectifying a
Brownian particle in a symmetric potential, that is, $\langle \dot x
\rangle=0$ also under harmonic mixing conditions,
$\Omega_2/\Omega_1=m/n$ with $m,n$ coprime integers and $m+n$ odd.
This is no more the case if $V(x)$ is asymmetric. The non-zero odd
moments of the process $x(t)$, determined by the substrate
asymmetry, generate additional harmonic couplings of the ``odd''
harmonics of the drive, i.e., for $\Omega_2/\Omega_1=(2m-1)/(2n-1)$
with $2m-1$ and $2n-1$ coprime, as shown in Fig. \ref{Fasym}. The
total rectification current, including both the incommensurate term
of Eq. (\ref{asym.HM}) and this new harmonic mixing spikes, was
calculated analytically in
\textcite{savelev2004EPL67,savelev2004PRE70b}:
\begin{equation}
\label{asym.HMtotal} j =j_{\rm avg}
-\frac{{(-1)^{m+n}p(\Delta_{n,m})}}{{(2m-1)(2n-1)}}~j_{\rm odd},
\end{equation}
where $p(\Delta_{n,m})$ is defined after Eq. (\ref{gate.mob}) and
$j_{\rm odd}=\frac{1}{2}[j_0(|A_1-A_2|)-j_0(A_1+A_2)]$.

The competition between current off-set, item (i), and asymmetry
induced spikes, item (ii), may cause surprising current reversals
for special values of the driving frequencies, as observed in
several experimental setups (cf. Secs. \ref{nanopore},
\ref{coldatom} and \ref{fluxon2Darrays}). Moreover, replacing the
rectangular waveform, Eq. (\ref{HM5}), with a more conventional
linear superposition of two sinusoids of the same frequency, Eq.
(\ref{2freq}), leads to an even more complicated interplay of
nonlinearity and asymmetry induced harmonic mixing
\cite{savelev2004PRE70b}. A special limit of bi-harmonically driven
rocked ratchet is discussed in Sec. \ref{vibratedratchet}.

\begin{figure}[ht]
\includegraphics[width=6.0cm]{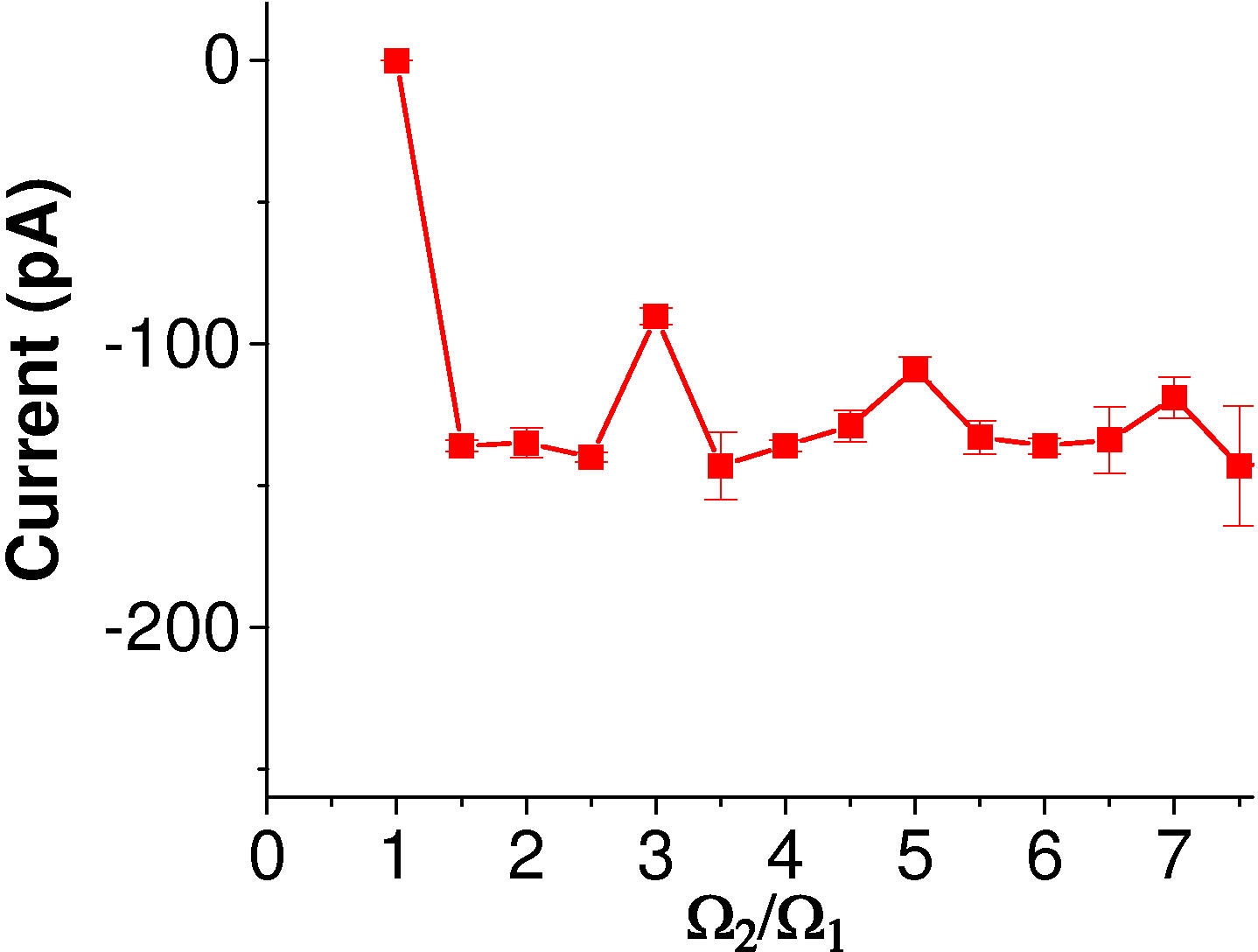}
\caption{(Color online) Net current through a single conical
nanopore for the sum of two voltage signals applied across the pore,
$0.5 \hbox{sgn} [\sin(\Omega_1 t)]+0.5 \hbox{sgn}[\sin(\Omega_2
t+\pi)]$, with different $\Omega_2/\Omega_1$. The measurements were
taken in a 0.1M KCl solution at pH 8.0. For further details see
\textcite{kalman2007EPL78}.} \label{SiwyEPL}
\end{figure}

A brilliant demonstration of the combination of harmonic mixing and
asymmetry effects in the context of particle transport at the
nanoscales, has been recently reported by
\textcite{kalman2007EPL78}, who managed to drive dilute ions through
conical nanopores by applying a biharmonic rectangular voltage.
Harmonic mixing and gating currents could be separated, as predicted
by the theory, and the resulting commensuration effects are
displayed in Fig. \ref{SiwyEPL}.

\subsection{Efficiency and control issues} \label{efficiency}

The analysis of Sec. \ref{molecularmotor} suggests that the output
of a real ratchet device is hard to control experimentally, let
alone to predict. As a matter of fact, if we use a ratchet to
rectify an assigned signal, the only tunable variables are the
temperature $T$, the damping constant $\gamma$ (or the mass $m$),
and the substrate profile $V(x)$. Under certain experimental
circumstances these quantities may prove not directly accessible to
this purpose or inconvenient to change. This is why control of
particle transport in small devices sometimes requires introducing
auxiliary signals, auxiliary particle species (Sec.
\ref{binarymixture}), or even the use of a feedback control scheme
to optimize the Brownian motor current
\cite{craig2008EPL81,craig2008AnnPhysik17,feito2007EPJB59,feito2008PhysicaA387,son2008PRE77}.

\subsubsection{Optimization}\label{optimization}

The most common definition of efficiency for a loaded Brownian motor
is
\begin{equation} \label{R}
\eta_0=\langle \dot x \rangle A_0/\langle P \rangle_{\rm in},
\end{equation}
where $\langle \dot x \rangle A_0$ is the average mechanical work
done per unit of time against a working load $A_0$ and $\langle P
\rangle_{\rm in}$ is the average net power pumped into the system by
the  external drives, no matter how applied.
The quantity $\eta_0$ has been interpreted in terms of macroscopic
thermodynamics by \textcite{sekimoto1998ProgTheorPhys130} (for
further developments on issues of energetics and efficiency see also
\textcite{parrondo2002APA75}), and references therein. In their
scheme, a ratchet operates like a Carnot cycle, where the lower
temperature is determined by the thermal noise, Eqs. (\ref{noise}
and \ref{noiseoLE}), and the higher temperature is related to the
magnitude of the external modulation. Operating far from thermal
equilibrium, the efficiency is typically smaller than the upper
limit of a thermal reversibly operating Carnot cycle.

The ensuing question, whether a ratchet can be operated at the
maximal Carnot efficiency, spurred an intense debate (see for a
review, Sec. 6.9 in \textcite{reimann2002PhysRep361}). The issue has
been clarified by an analysis performed within the validity regime
of the framework of linear irreversible thermodynamics by
\textcite{vandenbroeck2007AdvChemPhys135}: The key towards obtaining
maximal Carnot efficiency for a Brownian motor is zero overall
entropy production. This can be achieved by use of architectural
constraints for which the (linear) Onsager matrix has a determinant
equal to zero, implying vanishing (linear) irreversible heat fluxes.
Maximizing efficiency subject to ``maximum power" also leads to the
same condition of a vanishing determinant of the Onsager matrix (i.e
perfect coupling) \cite{vandenbroeck2005PRl95}, yielding the
Curzon-Ahlborn limit \cite{curzon1975AmJPhys43}, which in turn at
small temperature difference just yields half the Carnot efficiency.
For the archetype Smoluchowski-Feynman Brownian motor device
\cite{smoluchowski1912ZP13,feynman1963} the efficiency at maximum
power has recently been evaluated for different coupling schemes by
\textcite{tuJPhysA2008}: Interestingly enough, the typical upper
bound set by \textcite{curzon1975AmJPhys43} can then even be
surpassed.

In spite of the combined efforts of theorists and experimenters, as
of today the question remains unanswered. At present, ratchet
devices operating under controllable experimental conditions hardly
achieve an efficiency $\eta_0$ larger than a few percent.

On the other hand, definition (\ref{R}) of efficiency is not always
adequate to determine the optimal performance of a Brownian motor.
Firstly, $\eta_0$ assumes that work is being done against a load
$A_0$, which is not always the case. For example, one clearly finds
a vanishing efficiency whenever no load $A_0)$ is present. Secondly,
the only transport quantifier used in Eq. (\ref{R}) is the drift
velocity $\langle v \rangle$, whereas the fluctuations of
$\dot{x}(t)=v(t)$, i.e., the variance $ \sigma _{v}^2 = \langle v^2
\rangle - \langle v \rangle^2$, are also of practical importance.
Note that in this Section we use $v(t)$ to denote the particle
velocity in Eq. (\ref{yLE}) as a proper stochastic process. If
$\sigma_v > \langle v \rangle$, and even more so if $\sigma_v \gg
\langle v \rangle$, the Brownian motor can possibly move for some
time against its drift direction $\langle v\rangle$.

A load-independent rectification efficiency, $\eta_r$, has been
introduced by \textcite{suzuki2003PRE68} and then generalized by
\textcite{machura2004PRE70}. They computed $\eta_r$ as the ratio of
the dissipated power associated with the directed motion of the
motor against both the friction and the load, and the input power
from the time-periodic forcing. The result assumes the explicit form
\begin{eqnarray}
\label{optimization1} \eta_r=\frac{\langle v \rangle(A_0+\langle v
\rangle)} {|\langle v^2 \rangle + \langle v \rangle A_0 - D_0|},
\end{eqnarray}
where $D_0$ is the free diffusion coefficient of Sec.
\ref{diffusionpeak}. This definition holds evidently also for
$A_0=0$, while numerical evidence indicates that $\langle v^2
\rangle \geq D_0$, consistently with the inequality $\eta_r \leq 1$.
Note that the definition (\ref{optimization1}) assumes that $x(t)$
is a damped process with finite $\gamma$ (no matter how large). In
this way $\eta_r$ accounts explicitly also for the power dissipated
as velocity fluctuations during the rectification process; in
particular, it increases upon decreasing $\sigma_v$.
\textcite{machura2004PRE70} noticed that, for a rocked ratchet,
$\sigma_v$ exhibits pronounced peaks in correspondence with the
rectification thresholds $F_{3R,L}$ (large damping) and $F_{2R,L}$
(low damping), where the diffusion coefficient also has a maximum
(Sec. \ref{diffusionpeak}). That lead to the general conclusion that
transport by a Brownian motor can be optimized by operating away
from activation thresholds, in regimes of large net currents, where
the velocity fluctuations are intrinsically small.

\begin{figure}[htb]
\centering
\includegraphics[width=7.0cm,angle=-90]{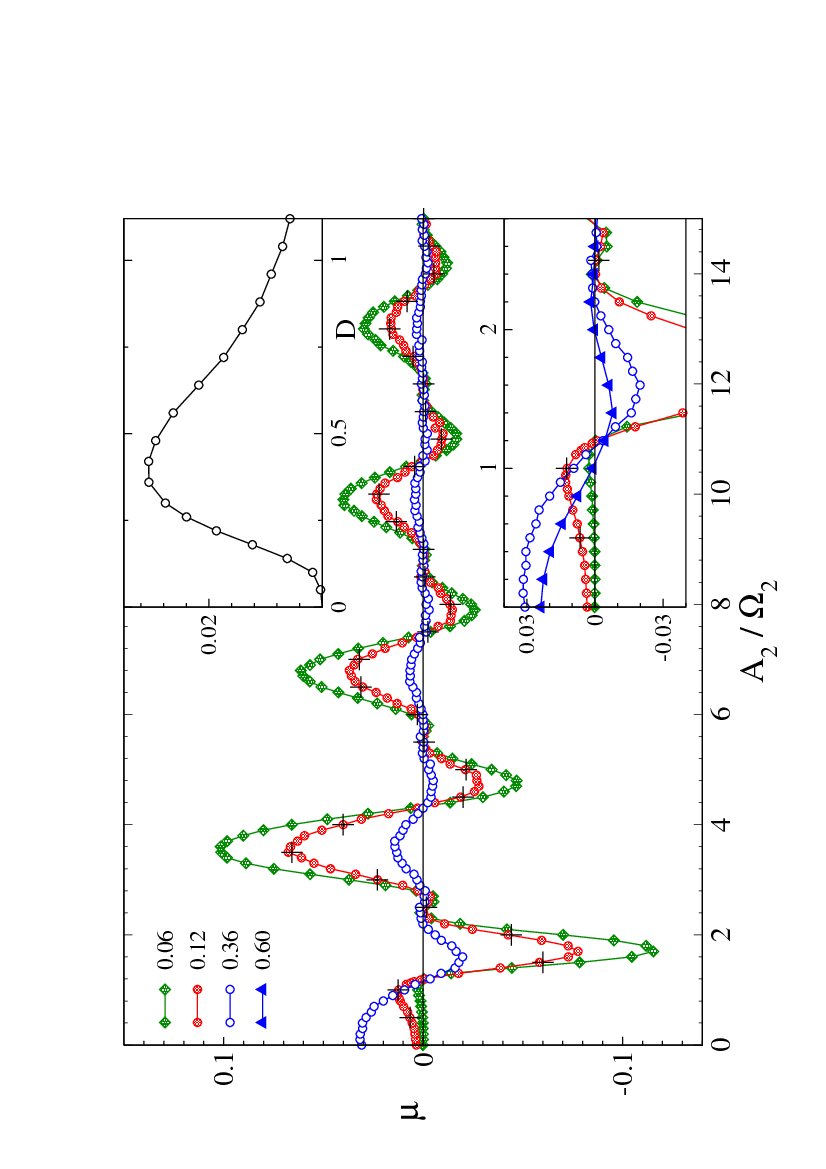}
\caption{ (Color online) Mobility versus $A_2/\Omega_2$ for a
vibrated ratchet with $A_1=0.5$, $\Omega_1=0.01$, $\phi_1=\phi_2=0$,
and different values of the noise intensity $D$ (see legend, main
panel). All simulation data have been obtained for $\Omega_2=10$,
but the black crosses where we set $D=0.12$ and $\Omega_2=20$.
Bottom inset: simulation data for $\mu$ as in the main panel with an
additional curve at $D=0.6$. Top inset: $\mu$ versus $D$ for
$A_2=0$, $A_1=0.5$, and $\Omega_1=0.01$; circles: simulation data;
solid curve: adiabatic formula (11.44) in \cite{risken1984}. After
\textcite{borromeo2005EPL72}.} \label{Fvratchet}
\end{figure}

\subsubsection{Vibrated ratchets}\label{vibratedratchet}

Besides the remarkable exceptions presented in Sec. \ref{coldatom},
transport control in a rocked ratchet cannot be obtained
experimentally by deforming the substrate potential ``on demand". As
shown by \textcite{borromeo2005EPL72}, this goal can more easily be
achieved by means of an external tunable signal. Let $V(x)$ be
assigned a fixed profile, say, the standard double-sine in Eq.
(\ref{doublesine}). The ratchet current can be still varied by
injecting an additional control signal with frequency $\Omega_2$,
that is, by replacing the harmonic drive in Eq. (\ref{forceac}) with
the bi-harmonic drive in Eq. (\ref{2freq}); in the adiabatic regime
$\Omega_2 \gg \Omega_1$ the system response is very different than
in Sec. \ref{asymmetry}. A rocked ratchet operated under such
conditions is termed {\it vibrational} ratchet.

Following the perturbation approach of
\textcite{bleckman2000,landa2000JPA33,baltanas2003PRE67}, the
variable $x(t)$ in Eq. (\ref{oLE}) can be separated as $x(t)
\longrightarrow x(t) + \psi(t)$: in the remainder of this Section
$x(t)$ will represent a slowly time-modulated stochastic process and
$\psi(t)$ the particle free spatial oscillation $\psi(t) = \psi_0
\sin (\Omega_2 t + \phi_2)$, with $\psi_0=A_2/\Omega_2$. On
averaging out $\psi(t)$ over time, the Langevin equation for the
slow reduced spatial variable $x(t)$ can be written as
\begin{equation}
\label{vibrated1} \dot x= -{\overline V}'(x) + A_1\cos(\Omega_1 t +
\phi_1) + \xi(t),
\end{equation}
where
\begin{equation}
\label{vibrated2} {\overline V}(x) = -V_0\left [J_0(\psi_0) \sin x +
\frac{1}{4} J_0(2\psi_0) \sin 2x\right ]
\end{equation}
and $J_0(x)$ is the Bessel function of zero-th order. The symmetry
of the effective vibrated potential (\ref{vibrated1}) is restored if
one of its Fourier components vanishes, namely for $A_2/\Omega_2 =
\frac{1}{2} j_1, j_1, \frac{1}{2} j_2, \frac{1}{2} j_3, j_2,
\frac{1}{2} j_4, \frac{1}{2} j_5, j_3, \dots$, where $j_1=2.405$,
$j_2=5.520$, $j_3=8.654$, $j_4=11.79$, $j_5=14.93 \dots$, are the
ordered zeros of the function $J_0(x)$; correspondingly, the ratchet
is expected to vanish, as confirmed by the simulation data displayed
in Fig. \ref{Fvratchet}.

Not all zeros of this sequence mark an inversion of the ratchet
current. For instance, for $A_2/\Omega_2 < \frac{1}{2} j_1$ the
current is oriented in the natural direction of the effective
potential, Eq. (\ref{vibrated2}); for $\frac{1}{2} j_1< A_2/\Omega_2
<j_1$, the coefficient of $\sin 2x$ changes sign and so does the
ratchet natural orientation, or polarity; on further increasing
$A_2/\Omega_2$ larger than $j_1$, the sign of both Fourier
coefficients in Eq. (\ref{vibrated2}) become reversed: this is
equivalent to turning $V(x)$ upside-down (beside slightly
re-modulating its profile), so that the polarity of ${\overline
V}(x)$ stays negative. Following this line of reasoning one predicts
{\it double} zeros (i.e. no current inversions) at $\psi_0 = j_1,
j_2, j_3, j_4, \dots$.

This control technique can be easily applied to the transport of
massive Brownian particles both in symmetric
\cite{borromeo2007PRL99} -- see inset of Fig. \ref{Fvratchet} -- and
asymmetric devices \cite{borromeo2006PRE73}.
Likewise, the use of a delay in a feedback control signal
\cite{craig2008EPL81,craig2008AnnPhysik17,feito2007EPJB59,feito2008PhysicaA387,son2008PRE77}
can, via synchronization mechanisms, efficiently improve the
performance of Brownian motor currents; a scheme that  as well can
be implemented readily in experimental flashing ratchets
\cite{craig2008AnnPhysik17}.


\section{TRANSPORT IN NANOPORES} \label{nanopore}

Membranes in biology encase cells and their organelles and allow the
compartmentalization of cellular processes, thereby  operating far
from thermal equilibrium, a condition that is essential to life.
Membranes separate two phases by creating an active or passive
barrier to the transport of matter between them. As a first
classification, membranes can be divided into biological and
artificial membranes. The latter term is applied to all membranes
made by man with natural, possibly modified, materials and with
synthetic materials (synthetic membranes). Synthetic membranes can
be further divided into organic (made with polymers) and inorganic
membranes (made with alumina, metals, etc.).

Transport across a membrane occurs through channels, or pores
\cite{kolomeisky2007PRL98, berezhkovskii2003JCP119}.
In several cases, the membrane transport properties must be regarded
as a collective effect, where the function of an individual channel
is influenced by the presence of (possibly diverse) neighboring
channels. This is the case, for instance, of ion pumps
\cite{lauger1991Sinauer,im2002JMolBiol322} in cellular membranes
(Sec. \ref{ionpump}) or for coupled ion channels which experience a
common transmembrane voltage. Single molecule techniques, however,
allow for the study and characterization of rectification properties
of individual entities, such as single ion channels.
Nowadays, synthetic membranes with assigned pore density and
patterns are made commercially available. Moreover, by means of
increasing sophisticated growth methods, irradiation
\cite{fleisher1975}, and nanofabrication techniques
\cite{li2001Nature412,storm2003NatMat2,dekker2007NatNano2,healy2007nanomedicine2b},
the cross section of membrane pores can be modulated along their
axis. As a result, transport in artificial ion channels fabricated
from asymmetric single pores of the most diverse geometries has
become accessible to experiments
\cite{siwy2004AmJPhys72,healy2007nanomedicine2b}, see also in Sec.
\ref{siwynanopore}.

In this Section we focus on devices where the size of the
transported particles is comparable to the pore cross section.
Transport in these devices can be analyzed in terms of the
one-particle 1D mechanisms illustrated in Sec. \ref{singleparticle}.
Larger pores, commonly used to channel fluids (liquid or gaseous) or
colloidal particles suspended in a fluid, are considered in Sec.
\ref{microfluidic} in the context of microfluidic devices.

\subsection{Ion pumps } \label{ionpump}

As remarked already before, biology teaches us useful lessons that
can guide us in the design of artificial Brownian motors. Biological
membranes knowingly form lipophilic barriers and have embedded a
diverse range of units which facilitate the selective movement of
various ionic and polar segments, or the pumping of protons and
electrons  across channel like membrane openings. These biological
nanodevices make typically use of electrochemical gradients that
enable them to pump a species against its concentration gradient at
the expense of yet another gradient. Such devices thereby convert
nondirectional chemical energy, from the resource of the hydrolysis
of adenosintriphosphate (ATP), into directed transport of charged
species against electrochemical gradients.

Although many details of the underlying mechanism are far from being
understood, these machines apparently make use of mechanisms that
characterize the physics of Brownian motors
\cite{astumian2007Physchemchemphys9}. A characteristic feature is
that changes  in the binding affinity at selective sites in the
transmembrane region must be coupled  to conformational changes
which, in turn, control motion into the desired direction. Thus, in
contrast to rocked or pulsated ratchets, where the
particle-potential interaction acts globally along the whole
periodic substrate landscape, the transport mechanism at work here
can be better described as an ``information'' ratchet
\cite{astumian1998EurBiophysJ27,parrondo2002APA75}. Indeed, the
effective potential bottlenecks to transduction of Brownian motion
get modified locally according to the actual location of the
transported unit; as a result, information gets transferred from the
unit to the potential landscape. For example, this scheme can be
used to model the pumping of Ca$^{++}$ ions in Ca$^{++}$-ATPase
\cite{xu2002JMolBiol316}. Related schemes have been invoked also in
the theoretical and experimental demonstration of pumping of Na$^+$-
and K$^+$-ions via pulsed electric field fluctuations in Na- and
K-ATPase \cite{xie1994BiophysJ67, tsong2002JBiolPhys28,
freund1999PRE60}, or for the operation of a catalytic wheel with
help of a ratchet-like, electro-conformational coupling model
\cite{rozenbaum2004JPhysChemB108}.

Yet another mechanism can be utilized to pump electrons in
biomolecules. It involves nonadiabatic electron tunneling in
combination with asymmetric, but non-biased, nonequilibrium
fluctuations, as proposed by \textcite{goychuk2006MolSim32}. The
nonequilibrium fluctuations originate from either random binding of
negatively charged ATP or externally applied asymmetric, but
non-biased, electric stochastic fields. Likewise, unbiased
nonequilibrum two-state fluctuations (telegraphic noise) can induce
directional motion across an asymmetric biological nanopore, as
numerically investigated for an aquaglyceroporin channel, where
water and glycerol get transported by means of a rocked ratchet
mechanism \cite{kosztin2004PRL93}.

Finally, we stress that in such realistic complex biological
channels the physical implications of an externally applied control
are often difficult to predict. This is due to the multi-facetted
consequences any control action may have in terms of chemical
variations, conformational changes, polarization effects, and the
alike. Moreover, at variance with single-molecule type experiments,
single channel recordings are not easily accessible when dealing
with biological molecules. Nanopores of lesser complexity are thus
synthetic nanopores which can be fabricated by use of bottom-down
nanoscience techniques -- the theme we review next.

\begin{figure}[btp]
\centering
\includegraphics[width=7.5cm,angle=0]{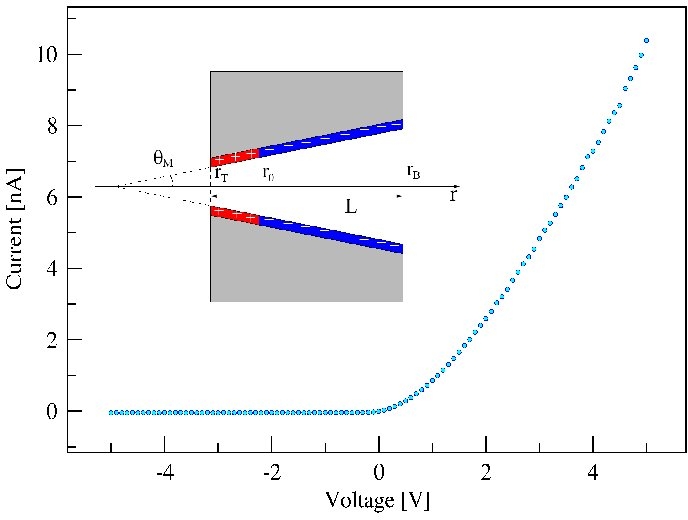}
\caption{(Color online) Ion current recorded experimentally by
\textcite{constantin2007PRE76} at $0.1M$ KCl, $p$H$=5.5$ through a
single conical nanopore with diameters were 5 nm and 1000 nm,
respectively. The pore rectification power is $\eta=217$ at $5V$.
Inset: geometry of a conical nanopore with schematic representation
of surface charge distribution creating a bipolar nanofluidic diode.
Figure provided by Zuzanna Siwy.} \label{Fnanopore2}
\end{figure}

\subsection{Artificial nanopores} \label{siwynanopore}

With the recent advances of track-etching \cite{fleisher1975} and
silicon technologies
\cite{li2001Nature412,storm2003NatMat2,dekker2007NatNano2}, charge
transport in a single nanopore became experimentally accessible.
This is a substantial leap forward with respect to ion pump and
zeolite transport experiments, where experimental data are taken
over a relatively high channel density. Fabricated nanopores in
polymer films and silicon materials are being investigated in view
of their potential applications as biomimetic systems, that is, for
modeling biological channels, and as biosensors.

Siwy and coworkers
\cite{siwy2002PRL89,siwy2005PRL94,vlassiouk2007NanoLett7,constantin2007PRE76}
have recently prepared a nanofluidic diode, which had been predicted
to rectify ion current in a similar way as a bipolar semiconductor
diode rectifies electron current \cite{daiguji2005NanoLett5}. This
diode is based on a {\it single} conically shaped nanopore
track-etched in a polymer film with openings of several nm and 1
$\mu$m, respectively (see sketch in Fig. \ref{Fnanopore2}). The
surface charge of the pore is patterned so that two regions of the
pore with positive and negative surface charges create a sharp
barrier called the transition zone. This nanofluidic diode is
bipolar in character since both positively and negatively charged
ions contribute to the measured current. Majumdar and coworkers
\cite{karnik2007NanoLett7} fabricated a similar nanofluidic diode
with a sharp barrier between a positively charged and a neutral side
of the pore. The presence of only one type of surface charge causes
the latter device to be unipolar.

Ion rectification was achieved by applying a longitudinal ac voltage
(Fig. \ref{Fnanopore2}); the system thus operate as a one-cell 1D
rocked ratchet (Sec. \ref{rockedratchet}), where the spatial
asymmetry is determined by the interaction of a single ion with the
inhomogeneous charge distribution on the pore walls. The
rectification power of the pore is defined as the ratio of the ionic
currents recorded for positive and negative driving voltages, i.e.,
$\eta(V)=|I(V)|/|I(-V)|$. Of course, due to their asymmetric
geometry, conical pores can rectify diffusing ions also for uniform
wall-charge distributions \cite{siwy2005PRL94}; however, the
corresponding $\eta$ factor would be at least one order of magnitude
smaller than reported here.

As a serious limitation of the present design, it is not possible to
control the rectification power of a given conical nanopore without
introducing changes to its built-in electro-chemical potential. An
alternate approach has been proposed by \textcite{kalman2007EPL78},
where two superposed rectangular voltage signals of zero mean were
used to control the net ion current through a nanopore of
preassigned geometry. By changing the amplitude, frequency and
relative phase of these signals these authors used the gating effect
of Sec. \ref{gating} to gain control over the orientation and the
magnitude of ion flow through the pore. Their experimental data were
found in excellent agreement with the theoretical prediction of Eq.
(\ref{asym.HMtotal}). The magnitude of ion current variation
achieved by asymmetric signal mixing was comparable with the
incommensurate current off-set Eq. (\ref{asym.HM}), which means that
the nanopore diode could be operated disregarding the details of its
intrinsic rectification power.

We conclude this Section on artificial nanopores anticipating that
asymmetric micropores etched in silicon membranes also work as
microfluidic ratchet pumps for suspended colloidal particles
\cite{kettner2000PRE61,matthias2003Nature424}. For instance,
entropic effects on the rectification efficiency of the conical
nanopores of Siwy and coworkers have been analyzed by
\textcite{kosinska2008PRE77}. However, in-pore diffusion in a liquid
suspension requires a fully 3D analysis of the pumping mechanism,
which sets the basis for the fabrication of more complicated ionic
devices \cite{stein2004PRL93,vanheyden2006PRL96}. This category of
devices is reviewed in Sec. \ref{microfluidic}.

\begin{figure}[h]
\centering
\includegraphics[width=6.0cm,angle=0]{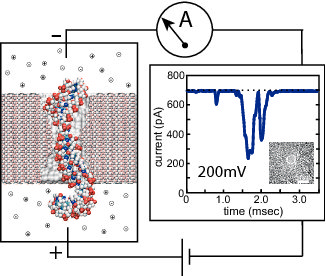}
\centering
\includegraphics[width=7.0cm,angle=0]{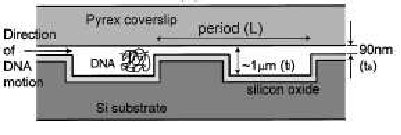}
\caption{(a) (Color online) Electric detection of individual DNA
molecules with a nanopore. A constant voltage bias induces a
steady-state ionic current through a single nanopore (left panel,
from simulation); adding DNA to the negatively biased compartment
causes transient reductions of the ionic current (right panel, from
experiment). This reduced conductance is associated with the
translocation of DNA through the pore, which partially blocks the
ionic current. After \textcite{aksimentiev2004BiophysJ87}. (b)
Schematic diagram of the entropic trap in \textcite{han1999PRL83}.}
\label{Fnanopore3}
\end{figure}

\subsection{Chain translocation}

The opposite limit of composite objects passing through a much
narrower opening is often called ``translocation". Using the
experimental set-up sketched in Fig. \ref{Fnanopore3}(a),
\textcite{kasianowicz1996PNAS93} measured, for the first time, the
blockage currents of single-stranded RNA and DNA electrophoretically
driven through the a transmembrane pore. Recent developments of this
technique demonstrated single nucleotide resolution for DNA hairpins
\cite{vercoutere2003NuclAcidRes31,ashkenasy2005AngewChem44,gerland2004PhysicalBiology1},
thus raising the prospect of creating nanopore sensors capable of
reading the nucleotide sequence directly from a DNA or RNA strand,
see for
recent comprehensive reviews
\cite{zwolak2008RMP80,healy2007nanomedicine2,movileanu2008SoftMatter4}.
Nowadays, translocation mechanisms are investigated in both protein
channels (mainly the bacterial $\alpha$-hemolysin pore
\cite{kasianowicz1996PNAS93}) and synthetic pores. Both these
approaches have advantages and disadvantages
\cite{healy2007nanomedicine2,dekker2007NatNano2}. Protein channels
can be engineered with almost angstrom precision, but the lipid
membrane in which they may be incorporated is very mechanically
unstable. Synthetic pores, on the other hand, offer robustness to
the system, which allows to better characterize the physical aspects
of translocation phenomena.

As a main difference with ion transport, the translocation of a long
polymer molecule in a 1D device involves entropic effects, which
become important when the opening cross section grows comparable
with the radius of gyration, $R_0$, of the polymer. These effects
were predicted in \cite{arvanitidou1991PRL67} to account for the
conformation changes a chain undergoes to move past a conduit
constriction and, finally, observed by
\textcite{han1999PRL83,han2000Science288} in an artificial channel.
As a model pore-constriction system, these authors fabricated a
channel consisting in a periodic sequence of regions of two
different depths, as shown in Fig. \ref{Fnanopore3}(b). The thick
regions were $1$mm deep, i.e., comparable with the $R_0$ of the
double stranded DNA molecules they used in their experiment, whereas
the depth of the thin region, $90$nm, was much smaller than $R_0$.

The thick regions act like ``entropic traps", as the DNA molecules
are entropically prevented from entering the thin regions. For the
same reason, a chain caught in between two traps, tends to fall back
into the trap that contains the most of it.  In the presence of an
external electric field, escape from a trap is initiated by the
introduction of a small portion of DNA into an adjacent thin region,
just enough to overcome the escaping activation barrier. This
initiation process is local in nature, and the energy barrier does
not depend on the total length of the trapped DNA molecule. Once a
DNA molecule is in the transition state (once a proper length of
``beachhead'' is formed), it readily escapes the entropic trap,
regardless of the length of the remaining molecule in the trap.
Quite counter intuitively, Han and coworkers found that escape of
DNA in longer  entropic traps occurs faster than in shorter ones.

A theoretical interpretation of these results was given by
\textcite{park1999JCP111}, who treated the dynamics of a flexible
polymer surmounting a 1D potential barrier as a multidimensional
Kramers activation process \cite{hanggi1990RMP62}. To determine the
activation free energy, Park and Sung computed the free energy of
the polymer at the transition state. For a small-curvature barrier,
the polymer keeps its random coil conformation during the whole
translocation process, giving rise to essentially the same dynamics
as that of a Brownian particle. For a large-curvature barrier, on
the other hand, a conformational transition (coil–stretch
transition) occurs at the onset of the barrier crossing, which
significantly lowers the activation free energy and so enhances the
barrier crossing rate. As the chain length varies, the rate shows a
minimum at a certain chain length due to the competition between the
potential barrier and the free energy decrease by chain stretching.
Synthetic nanopores can thus be used as Coulter counter devices to
selectively detect single DNA molecules, resolving their length and
diameter.

Spatial asymmetry can rectify translocation through synthetic and
biological pores, alike. This is the case, for instance, of a
hydrophobic polymer translocating across a curved bilayer membrane.
Extensive simulation \cite{baumgartner1995PRL74} showed that the
polymer crosses spontaneously and almost irreversibly from the side
of lower curvature to the side of higher curvature, so as to
maximize its conformational entropy (``entropy" ratchets, Sec.
\ref{boundaryeffect}). Moreover, at variance with artificial
channels, in protein translocation through a biomembrane, the
chemical structure of the pores can come into play by helping
rectify the thermal fluctuations of the stretched molecule: When
specific predetermined segments of the protein cross the membrane,
chemicals acting as chaperons bind on the segments to prevent their
backward diffusion
\cite{hartl1996Nature381,nigg1997Nature386,julicher1997RMP69}. The
ensuing chemical asymmetry then competes with the entropic asymmetry
to determine the translocation current.

\subsection{Toward a next generation of mass rectifiers}
\label{nextgeneration}

\subsubsection{Zeolites} \label{zeolite}

Zeolites are three dimensional, nanoporous, crystalline solids
(either natural or synthetic) with well-defined structures that
contain aluminum, silicon, and oxygen in their regular framework.
The silicon and aluminum atoms are tetrahedrally connected to each
other through shared oxygen atoms; this defines a regular framework
of voids and channels of discrete size, which is accessible through
nanopores of well-defined molecular dimensions. The negative
electric charge of the zeolite framework is compensated by
(inorganic or even organic) cations or by protons (in the acidic
form of the zeolites). The ions are not a part of the zeolite
framework, and they stand in the channels. The combination of many
properties -- such as the uniform cross section of their pores, the
ion exchange properties, the ability to develop internal acidity,
high thermal stability, high internal surface area -- makes zeolites
unique among inorganic oxides and also leads to unique activity and
selectivity. As a result, zeolites can separate molecules based on
size, shape, polarity, and degree of unsaturation among others
\cite{karger1992,karger2008MolSieves}.

Nuclear magnetic resonance (NMR) provides direct access to the
density and the mobility of molecules in zeolites. By implementing a
time-dependent NMR technique, termed pulsed field gradient NMR,
K\"arger and coworkers (see for a review
\textcite{karger2008bMolSieves}) investigated the problem of
particle diffusion in a narrow zeolite nanopore. For instance, they
observed that, owing to the constrained pore geometry (0.73nm
across), the mean square displacement of CF$_4$ molecules (0.47nm in
diameter) diffusing in zeolite AlPO$_4$-5, increases linearly with
the square root of the observation time rather than with the
observation time itself.
That was an early experimental demonstration of the single-file
diffusion mechanism introduced in Sec. \ref{singlefile}. In the
meantime, however, real zeolite crystals have been repeatedly found
to deviate notably from their ideal textbook structure
\cite{schemmert1999EPL46}, so that, their pore cross-sections ought
to be regarded as longitudinally corrugated. To what extent this may
affect particle diffusion is still matter of ongoing research
\cite{karger2005NJP7,taloni2006PRL96}.

Most zeolite structure types exhibit 3D pore networks.  The most
studied example are a synthetic zeolites of type MFI. Their pore
network is formed by mutually intersecting straight (in
crystallographic $y$-direction) and sinusoidal (in crystallographic
$x$-direction) channels. Although there is no corresponding third
channel system, molecular propagation has been observed also in
$z$-direction. Experimental data are consistent with a simple law of
{\it correlated} diffusion anisotropy \cite{fenzke1993ZPD25},
$a_z^2/D_z =a_x^2/D_x +a_y^2/D_y$, where $a_i$ and $D_i$, with
$i=x,y,z$, denote respectively the lattice and the diffusion
constants in the $i$-direction. This means that the molecular
``memory" is shorter than the mean traveling time between two
adjacent network intersections.

With the advent of synthetic zeolites and zeolitic membranes,
researchers are fascinated by the option that the existence of
different channel types within one and the same material may be used
for an enhancement of the performance of catalytic chemical
reactions. As the diffusion streams of the reactant and product
molecules tend to interfere with each other, rerouting them through
different channels may notably speed up a catalytic reactor; hence,
the idea of reactivity enhancement by ``molecular traffic control"
\cite{derouane1980JCatal65,neugebauer2000JCatal194}.

\subsubsection{Nanotubes} \label{nanotube}

Another interesting category of artificial nanopore are carbon
nanotubes \cite{dresselhaus1996}. Single-walled carbon nanotubes are
cylindrical molecules of about $1$ nm in diameter and $1$-$100$
$\mu$m in length. They consist of carbon atoms only, and can
essentially be thought of as a layer of graphite rolled-up into a
cylinder. Multiple layers of graphite rolled in on themselves are
said to form a multi-wall carbon nanotube. The electronic properties
of nanotubes depend strongly on the tube diameter as well as on the
helicity of the hexagonal carbon lattice along the tube (chirality).
For example, a slight change in the pitch of the helicity can
transform the tube from a metal into a large-gap semiconductor,
hence they potential use as quantum wires in nanocircuits
\cite{dekker1999PhysToday52,collins2000SciAm283}. An even wider
range of geometries and applications became available recently with
the synthesis of various oxide nanotubes
\cite{remskar2004AdvMater16}. With their hollow cores and large
aspect ratios, nanotubes are excellent conduits for nanoscale
amounts of material. Depending on the filling material,
experimenters have thus realized nanoscale magnets, hydrogen
accumulators, thermometers and switches.

Nanotubes also provide an artificial substrate for controllable,
reversible atomic scale mass transport.
\textcite{regan2004Nature428} attached indium nanocrystals to a
multi-wall carbon nanotube and placed it between electrodes set up
in the sample chamber of an electron microscope. Applying a voltage
to the tubes, they observed that the metallic particles at one end
of the tube gradually disappeared, while those at the other end
grew. While the details of the underlying driving mechanisms
(thermo- versus electro-migration) remain unclear, they concluded
that the voltage dictates the directionality of that nanoscale mass
conveyor. Experimenters even succeeded to synthesize carbon
nanotubes encapsulating metallic atoms and characterize the
electro-mechanical properties of such nanochannels (see e.g.,
\textcite{gao2002Nature415}). In the next future, carbon nanotubes
will also be combined to form molecular ``gears", whose feasibility
has been proven so far only by simulation \cite{drexler1992}. A
conceptual example is provided by a double-walled carbon nanotube
consisting of two coaxial single-walled nanotubes with different
chirality, immersed in an isothermal bath. In the presence of a
varying axial electrical voltage, this system would exhibit a
unidirectional ratchet-like rotation as a function of the chirality
difference between it constituents \cite{marchesoni1996PRL77}.

Nanotubes have also been used to realize prototypes of thermal
diodes for phonon transport \cite{chang2006science314}. With this
concept in mind, one can further devise a phonon Brownian motor
aimed to ratchet a net heat flux from ``cold to hot", as numerically
demonstrated in a recent work by \textcite{nianbei2008arxiv}. This
class of devices has the potential to allow an efficient control of
heat fluxes at the nanoscales.

In spite of the recent advances in nanotechnology, (inner or outer)
transport along nanotubes is still controlled by external gradients.
Nevertheless, nanotube based ratchets, though not immediately
available, are like to be one of the next frontiers in artificial
Brownian motor research. Moreover, synthetic nanotubes are potential
building blocks also for nanofluidic devices
\cite{holt2006Science312}, as discussed in Sec. \ref{microfluidic}.


\section{COLD ATOMS IN OPTICAL LATTICES} \label{coldatom}

Optical lattices are periodic potential for atoms created by the
interference of two or more laser fields
\cite{jessen1996AdvAtMol37,grynberg2001PhysRep355}. In near resonant
optical lattices the laser fields produce simultaneously a periodic
potential for the atoms and a cooling mechanism. The optical
potential for an atom in an optical lattice is given by the light
shift, or ac Stark shift, of the atomic energy level that acts as
the ground state of an assigned optical transition. The simplest
case is represented by the $J_g = 1/2 \to J_e = 3/2$ atomic
transition in a 1D configuration.

Let us consider two counter-propagating laser fields, detuned below
the atomic resonance, with orthogonal linear polarizations and same
intensity and wavelength $\lambda$. Their interference results into
a spatial gradient with polarization ellipticity of period
$\lambda/2$. This in turn produces a periodic potential for the
atom. For instance, the atomic hyperfine ground states $|g,
\pm\rangle = |J_g = 1/2,M = \pm 1/2\rangle$ experience periodic
potentials in phase opposition along the direction $x$ of light
propagation, namely
\begin{equation}
\label{cold1} V_{\pm}(x) = \frac{V_0}{2}(-2 \pm \cos kx),
\end{equation}
where $k=2\pi/\lambda$ and the depth of the potential wells $V_0$
scales as $I_L/\Delta$, with $I_L$ the total laser intensity and
$\Delta$ the detuning from atomic resonance. As the laser fields are
near to resonance with the atomic transition $J_g = 1/2 \to J_e =
3/2$, the interaction with the light fields also leads to stochastic
transitions between the Zeeman sublevels $|g, \pm\rangle$ of the
ground state. The rate of this transitions can be quantified by the
photon-atom scattering rate $\Gamma'$ which scales as
$I_L/\Delta^2$. It is therefore possible to vary independently the
optical lattice depth $V_0$ and $\Gamma'$ by changing simultaneously
$I_L$ and $\Delta$.

The stochastic transitions between ground states also lead to
damping and fluctuations. The damping mechanism, named Sisyphus
cooling, originates from the combined action of light shifts and
optical pumping which transfers, through cycles of
absorption/spontaneous emission involving the excited state $J_e$,
atoms from one ground state sublevel to the other one. Moreover, the
stochastic transitions between the two potentials $V_{\pm}(x)$ also
generate fluctuations of the instantaneous force experienced by the
atom. In conclusion, the equilibrium between cooling and heating
mechanisms determines a stationary diffusive dynamics of the atoms,
a process confined to the symmetric ground-state optical lattice
$V_{\pm}(x)$.

Random amplitude pulsations of one symmetric potential, Eq.
(\ref{cold1}), do not suffice to produce rectification. The only
missing element to reproduce the rocked set-ups of Secs.
\ref{nonlinearmechanism} and \ref{molecularmotor} is the additive ac
force $F(t)$. In order to generate a time-dependent homogeneous
force, one of the lattice beams is phase modulated, and we will
indicate by $\phi(t)$ its time-dependent phase. In the laboratory
reference frame the phase modulation of one of the lattice beams
results into the generation of a moving optical lattice $V_{\pm}(x -
\phi(t)/2k)$, reminiscent of the periodic sieves in
\cite{borromeo2007PRL99}. In the accelerated reference frame $x \to
x - \phi(t)/2k$, the optical potential would be stationary; however,
an atom of mass $m$ experiences also an inertial force in the
$z$-direction proportional to the acceleration of the moving frame,
namely $F(t)= (m/2k) \ddot\phi(t)$. This is the homogeneous ac drive
needed to rock a cold atom ratchet.

The above described harmonic mixing/rocking ratchet set-up was
recently used  by Renzoni and coworkers
\cite{schiavoni2003PRL90,gommers2005PRL95,gommers2006PRL96,gommers2007PRA75,gommers2008PRL100}
to investigate experimentally the relationship between symmetry and
transport in 1D atom traps. We discuss here separately the
experimental results for two different cases: $(1)$ a bi-harmonic
driving including two harmonics at frequencies $\Omega_0$ and
$2\Omega_0$, and $(2)$ a multi-frequency driving obtained by
combining signals at three different frequencies.

\subsection{Bi-harmonic driving}\label{biharmonic}

These authors \cite{schiavoni2003PRL90,gommers2005PRL95} generated a
bi-harmonic drive, Eq. (\ref{2freq}), with $\Omega_2=2\Omega_1$ and
$\phi_1=0$, and searched for a harmonic mixing current proportional
to $\sin (\phi_2-\phi_0)$. At variance with Eq. \ref{HM1}, a phase
lag $\phi_0$ was introduced to account for the finite atom
dissipation. Indeed, by generalizing our argument of Sec. \ref{HM},
one can easily prove that increasing the damping makes $\phi_0$ drop
from $0$ down to $-\pi/2$. In the experiment, the atom dissipation
could be tuned continuously, without changing the optical potential
constant $V_0$, by varying simultaneously $I_L$ and $\Delta$ at a
constant $I_L/\Delta$ ratio; thus, harmonic mixing was
experimentally accessible in both limiting regimes of zero and
infinite damping.

\begin{figure*}[btp]
\centering
\hspace*{-1.0cm}\includegraphics[width=6.6cm]{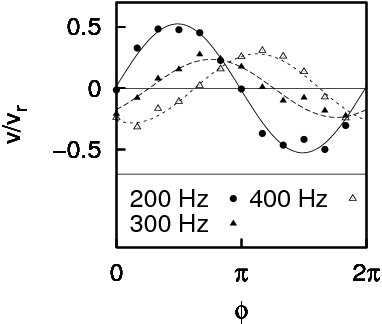}
\hspace*{1.0cm}\includegraphics[width=7.0cm]{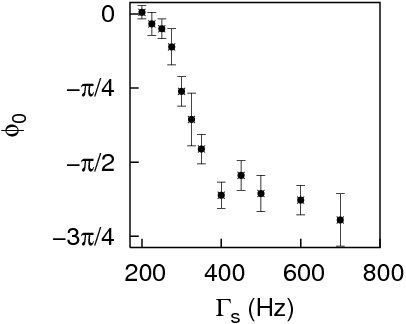}
\caption{Cold atoms in an optical lattice. (a) Average atomic
velocity as a function of the phase $\phi_2$ for different values of
$\Gamma_s = [\omega_v/(2\pi)]^2/\Delta$ (a quantity proportional to
the scattering rate $\Gamma'$). The data are labeled by the lattice
detuning $\Delta$, as the vibrational frequency at the bottom of the
wells was kept constant, $\omega_v/2\pi=170kHz$; the forcing
frequency is $\Omega_1/2\pi= 100kHz$. The lines are the best fit of
the data with the function $v/v_r = A_{\Delta}\sin(\phi-\phi_0)$,
where $v_r=\hbar k/m$ is the so-called atom recoil velocity. (b)
Experimental results for the phase shift $\phi_0$ as a function of
$\Gamma_s$. After \textcite{gommers2005PRL95}.} \label{Fcold1}
\end{figure*}

The experimental results of Fig. \ref{Fcold1} for cesium atoms
cooled in the $\mu$K range, clearly demonstrate the mechanism of
harmonic mixing at work. In fact, the net atom velocity plotted in
panel (a) is well fitted by $v/v_{r} = A_\Delta\sin(\phi_2 -
\phi_0)$, where the dependence of $\phi_0$ on $\Gamma'$ is as in
panel (b), and  $A_\Delta$ is a characteristic function of the
system \cite{gommers2005PRL95}. For the smallest scattering rate
examined in the experiment, no current was generated at
$\phi_2=l\pi$, with $l$ integer, as expected from Eq. (\ref{HM1}).
On the other hand, the magnitude of the phase shift $\phi_0$
increases with increasing the scattering rate, thus causing current
generation also for $\phi_2=l\pi$. Besides confirming the notion of
harmonic mixing, the experimental results in
\textcite{schiavoni2003PRL90,gommers2005PRL95} also suggest that
symmetry breaking can be controlled by dissipation.
Closely related results have been reported by
\textcite{ustinov2004PRL93} in their attempt to control fluxon
ratcheting in Josephson junctions by means of harmonic mixing.

\subsection{Multi-frequency driving} \label{multifrequency}

Recent experiments with multi-frequency driving
\cite{gommers2006PRL96,gommers2007PRA75} aimed to investigate the
transition from periodic to quasi-periodic driving, and to examine
how the analysis of Sec. \ref{nonlinearmechanism} is modified in
this transition. The multi-frequency driving was obtained by adding
a sinusoidal component to the ac drive employed in the previous
Section, i.e.,
\begin{equation}\label{cold2}
F(t)=A_1\cos(\Omega_1 t)+A_2\cos(2\Omega_1 +\phi_2)+A_3\cos(\Omega_3
t +\phi_3)
\end{equation}

For $\Omega_3/\Omega_1$ irrational the driving is quasi-periodic.
Clearly, in a real experiment $\Omega_3/\Omega_1$ is always a
rational number, which can be written as $p/q$, with $p, q$ two
coprime positive integers. However, as the duration of the
experiment is finite, by choosing $p$ and $q$ sufficiently large it
is possible to obtain a driving which is effectively quasi-periodic
on the time scale of the experiment.

Let us consider first the case of periodic driving, with
$\Omega_3/\Omega_1$ rational. Adding a third harmonic with phase
constant $\phi_3\neq 0$ in Eq. (\ref{cold2}) breaks the time
symmetry of $F(t)$ so that directed transport is allowed also for
$\phi_0=0$ and $\phi_2=l\pi$. In other words, for $\phi_3\neq 0$ the
third driving component leads to an additional phase shift of the
current as a function of $\phi_2$. As a results, the current will
retain its sinusoidal phase dependence $\sin(\phi_2-\phi_0)$, as
above, with the difference that now $\phi_0$ accounts for the phase
shifts produced both by dissipation and by the added driving
component. Following the discussion sketched in Sec.
\ref{asymmetry}, one easily concludes that in the quasi-periodic
regime the third harmonic at frequency $\Omega_3$ is not relevant to
characterize the time symmetry of the forcing signal, which is
entirely determined by the bi-harmonic terms at frequency $\Omega_1$
and $2\Omega_2$.

In later experiments \cite{gommers2006PRL96,gommers2007PRA75}, the
transition to quasi-periodicity was investigated by studying the
atomic current as a function of $\phi_2$ for different $p/q$. By
increasing $p$ and $q$ the driving was made more and more
quasi-periodic on the finite duration of the experiment, with the
quantity $pq$ an easily accessible measure of the degree of
quasi-periodicity. The data for the average atomic current were
fitted with the function $v/v_{\mathrm{max}} = \sin(\phi_2 -
\phi_0)$ and the resulting value for $\phi_0$ were plotted as a
function of $pq$. For small values of $pq$, i.e., for periodic
driving, the added component at frequency $\Omega_3$ lead to a shift
which strongly depends on the actual value of $pq$. For larger
values of $pq$, i.e., approaching quasi-periodicity, the phase shift
$\phi_0$ tends to a constant value independent of $\phi_3$, which
coincides with the purely dissipative phase-shift measured in the
bi-harmonic driving case of Sec. \ref{biharmonic}. The experimental
results confirm that in the quasi-periodic limit, the only relevant
symmetries are those determined by the periodic bi-harmonic driving
and by dissipation.

\subsection{More cold atom devices}\label{morecoldatom}

Cold atoms in optical traps proved to be a playground for
rectification experiments.

{\it (i) Gating effect.} For instance \textcite{gommers2008PRL100}
modified the experimental set-up described above, to demonstrate
experimentally a gating ratchet with cold rubidium atoms in a driven
near-resonant optical lattice. As suggested in Sec. \ref{gating}, a
single-harmonic periodic modulation of the optical potential depth
with frequency $\Omega_2$ was applied, together with a
single-harmonic rocking force with frequency $\Omega_1$. The
modulation of the optical potential depth $V_0$ was obtained by
modulating the intensity $I_L$ of the laser beams. This also
resulted in an unavoidable modulation of the optical pumping rate,
which affected only the fitting phase shift $\phi_0$. Directed
motion was observed for rational values of $\Omega_2/\Omega_1$, a
result due to the breaking of the symmetries of the system.

{\it (ii) Pulsated ratchets.} Although a {\it bona fide} rocked
ratchet for cold atoms in an optical lattice has not been realized
(see e.g. \textcite{ritt2006PRA74}), yet, a clever variation of 1D
optical lattice described above allowed an early demonstration of
randomly pulsated ratchet \cite{mennerat1999PRL82}. These authors
set the polarizations of the laser beams at an angle $\theta \neq
\pi/2$ and applied a weak Zeeman magnetic field orthogonal to the
optical lattice. The effect of the magnetic field consisted in
removing the degeneracy of the ground states by adding a $\lambda/2$
wavelength components to both potentials in Eq. (\ref{cold1}), which
thus acquired different asymmetric profiles. As a consequence, the
random optical transitions between the modified potentials
$V_{\pm}(x)$ turned out to propel trapped cold rubidium atoms in the
vertical direction, with the sign that depended on $\theta$ and on
the orientation of the magnetic field.

In Sec. \ref{pulsatedratchet} we reported that rectification can
occur on symmetric 1D substrates that shift {\it instantaneously}
back and forth in space with a fixed amplitude, namely, on substrate
that are subjected to a time-discrete phase modulation (periodic or
random, alike). In this case, breaking the supersymmetry condition
(\ref{susy1}) requires appropriate asymmetric space-dependent
switching rates \cite{gorman1996PRL76}. Based on this rectification
scheme, \textcite{Sjolund2006PRL96} realized an simple flashing
ratchet for cold atoms. It consisted again of a $\mu$K cold gas of
cesium atoms switching between two symmetric ground state optical
lattices, coupled via optical pumping. In the presence of induced
friction, although small, and for appropriate laser detunings and
intensities of the laser beams, the degeneracy of the ground states
was removed by making one ground state, say $V_+$, long lived and
the other one, $V_-$, short lived. If we further consider that the
two optical lattices were shifted one relative to the other,
$V_-(x)=V_+(x-x_0)$, then we conclude that the switching rates
between potentials were state {\it and} position dependent with
$k_{-\to +}(x)\gg k_{+\to -}(x)$. In this setup, the atoms execute
stationary time-asymmetric sequences of random jumps between $V_+$
and $V_-$. During the time spent in the short-lived lattice $V_-$,
they experience a potential with an incline that depends on $x_0$.
Thus, their diffusion is strongly enhanced in one specific
direction, and correspondingly reduced in the opposite direction. In
the experiment, the spatial shift $x_0$ and the transition rates
between the two optical lattices could be adjusted at will. A
directed motion with constant velocity was observed in the absence
of additional forcing terms, i.e., for $F(t)=0$, except for specific
system parameters, where symmetry was restored. Moreover, these
authors showed that their device, when operated in 3D
\cite{ellmann2003PRL90}, can generate directed motion in any
direction.


\section{COLLECTIVE TRANSPORT} \label{collectivetransport}

The studies on rectification mechanisms reviewed in Sec.
\ref{singleparticle}, have been conducted mostly for a single
particle; however, often systems contain many identical particles,
and the collective interactions between them may significantly
influence the transport of particle aggregates
\cite{aghababaie1999PRE59}.
For example, interactions among coupled Brownian motors can give
rise to novel cooperative phenomena like anomalous hysteresis and
zero-bias absolute negative resistance \cite{reimann1999EPL45}. The
role of interactions among individual motors plays  a particular
important role for the occurrence of unidirectional transport in
biological systems, e.g. see Sec. 9 in the review by
\textcite{reimann2002PhysRep361}.

In this section we shall restrict ourselves to the properties of 1D
and 2D systems of point-like interacting particles, although an
extension to 3D systems is straightforward. Let the pair interaction
potential ${\cal W}(r)$ be a function of the pair relative
coordinates ${\bf r}=(x,y)$, characterized by an appropriate
parameter $g$, quantifying the strength of the repelling ($g>0$) or
attracting force ($g<0$), and by a constant $\lambda$, defining the
pair interaction length. An appropriate choice of ${\cal W}(r)$,
with $g>0$, can be used to model hardcore particles with diameter
$\lambda$ \cite{savelev2003PRL91,savelev2004PRL92}. We term the pair
interaction long-range, if $\lambda$ is larger than the average
interparticle distance, and short-range in the opposite regime. In
the absence of perturbations due to the substrate or the external
drives, long-range repelling particles rearrange themselves to form
a triangular \textcite{abrikosov1957JETP5} lattice. When placed on a
disordered substrate or forced through a constrained geometry, such
an ideal lattice breaks up into lattice fragments punctuated by
point-like defects and dislocations and separated by faulty
boundaries. In the presence of a drive, such boundaries act like 1D
easy-flow paths, or rivers, for the movable particles. This process,
often referred to as ``plastic" flow, is analyzed numerically in
\textcite{reichhardt1998PRB57}.

Rectification of interacting particles occurs as a combined effect
of the configuration of the device and the geometry of its
microscopic constituents. In this context the dimensionality of the
underlying dynamics is also important: $(1)$ Transport in some 2D
geometries can often be reduced to the 1D mechanisms illustrated in
Sec. \ref{singleparticle} (reducible 2D geometries); $(2)$ Under
certain circumstances, however, a {\it bona-fide} 2D rectification
may occur at a non-zero angle with the driving force (irreducible 2D
geometries). Finally, we remark that particle interaction can be
exploited to rectify particles of one species by acting on particles
of another species, alone, either by an appropriate drive or a
specially tailored substrate geometry.

\begin{figure}[btp]
\centering
\includegraphics[width=7.0cm]{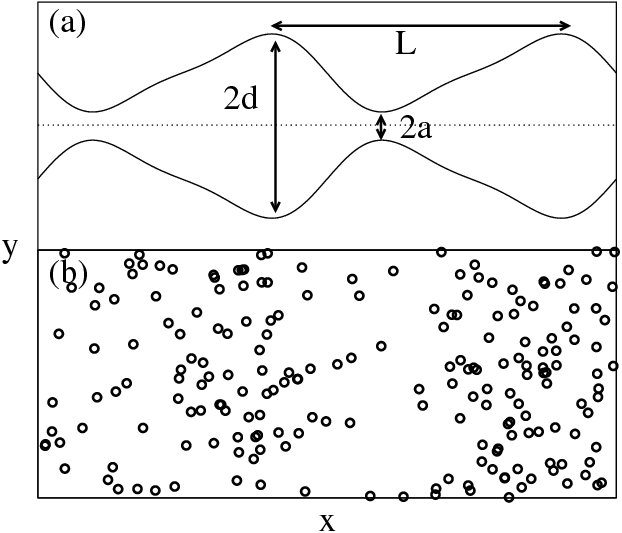}
\caption{1D reducible asymmetric geometries: (a) Channel with
asymmetric profile $y(x)=y_0-y_1[\sin (2\pi x/L)+\frac{1}{4}\sin
(4\pi x/L)]$; the parameters $y_1$ and $y_0$ control the width $2d$
and the bottleneck $2a$; (b) Disordered pattern of circular defects
generated with periodic distribution $y(x)$ along the $x$-axis}.
\label{Fcollective1}
\end{figure}

\subsection{Asymmetric 1D geometries} \label{1dgeometry}

We consider now two examples of 2D reducible geometries where the
transport of massless repelling particles, subjected to an external
ac drive, occurs as a collective effect, namely, under conditions
incompatible with the rectification of a single particle. In
particular, we show that collective effects make transport of
interacting particles even in the absence of an {\it ad hoc} ratchet
substrate.

\subsubsection{Boundary effects} \label{boundaryeffect}

Let us consider the asymmetric channel sketched in Fig.
\ref{Fcollective1}(a) filled with $n$ repelling particles per unit
cell. The corresponding particle density is $\rho=n/a_l$, where
$a_l$ is the area of the channel unit cell. The walls are rigid and
the particle-wall collisions are taken elastic and relatively
short-range (interaction length not much larger than $\lambda$). The
profiles $\pm y(x)$ of the upper ($+$) and lower ($-$) walls are
modeled by an appropriate double-sine function; $2d$ and $2a$
denote, respectively, the width of the channel and of its
bottlenecks. Due to the repulsive interactions, the particles are
pressed against the walls which corresponds to an effective
asymmetric spatial modulation. As a consequence, when driven by an
ac force, the particles are more likely to flow to the left than to
the right. The ensuing rectification mechanism is reminiscent of a
1D rocked ratchet with an inverted potential of Eq.
(\ref{doublesine}). Numerical simulations, besides supporting this
prediction, clearly indicate an optimal or "resonant", temperature
regime in which the particle drift is maximized. This observation
can be explained by noticing that at low temperatures, the ac drive
causes a moving particle to migrate to the center of the channel
where it no longer interacts with the boundaries; while at high
temperatures the driving force becomes irrelevant and thus the
particle is no longer pushed through the channel bottlenecks
periodically in time. The first simulation evidence of this
phenomenon was produced by \textcite{wambaugh1999PRL83}. Their
results obtained for the special case of magnetic vortex channeling
are discussed in Sec. \ref{fluxonchannel}.

At high enough particle densities, $\rho$, the net current in the
channel is expected to be suppressed, as the repelling particles end
up clogging the bottlenecks and thus hampering collective
longitudinal oscillations. In the opposite limit of, say, $n=1$, the
arguments above seem to rule out rectification, as a single
oscillating particle would be confined in the inner channel of
radius $a$. However, this conclusion holds only in the absence of
thermal fluctuations, $T=0$. At finite $T$, reducing the diffusion
in a 2D channel to a 1D process implies defining an effective
diffusion constant (\ref{diffusionpeak}),
$D(x)/D_0=[1+y'(x)^2]^{-1/3}$, equivalent to a periodic asymmetric
modulation of the temperature \cite{reguera2006PRL96} (a Seebeck
ratchet in the notation of \textcite{reimann2002PhysRep361}). A
particle crossing a channel bottleneck perceives a lower effective
temperature on the left, where the wall is steeper, than on the
right, so that it gets sucked forwards; the opposite happens in
correspondence to the largest channel cross-sections.  As long as
$\int dx/D(x)\neq 0$, the oscillating motion of a single particle
can indeed be rectified, with the sign that depends on the details
of the wall profile \cite{ai2006PRE74}.  No matter how weak, such a
mechanism, termed ``entropic" ratchet \cite{slater1997PRL78},
supports the conclusion of \textcite{wambaugh1999PRL83} that
rectification in a channel occurs only at finite $T$, as a certain
amount of noise is needed for the particle to explore the asymmetric
geometric of the device. Moreover, the argument above hints at the
occurrence of an optimal channel density $\rho$, as detected in real
superconducting devices (Sec. \ref{fluxonchannel}).

\subsubsection{Asymmetric patterns of symmetric traps} \label{asymmetricpattern}

\textcite{olson2001PRL87} proposed a new type of 2D ratchet system
which utilizes gradients of pointlike disorder, rather than a
uniformly varying substrate potential. Let us consider a 2D sample
containing a periodically graduated density of point defects, as
sketched in Fig. \ref{Fcollective1}(b). Each defect is depicted as a
circular micro-hole, which acts as a {\it symmetric} short-range
particle trap of finite depth. In real experiments, such defects can
actually be created by either controlled irradiation techniques or
direct-write electron-beam lithography \cite{kwok2002PhysicaC382}.
The defect density $\rho_l(x)$ was chosen to be uniform along the
vertical axis, and to follow a typical double-sine asymmetric
profile of Eq. (\ref{doublesine}) along the horizontal axis. Let us
now inject into the sample repelling massless particles with average
density $\rho$ and interaction length $\lambda$. The defect radius
controls particle pinning by defects and was taken much smaller than
$\lambda$. For a sufficiently high particle-to-defect density
ratios, the particles fill most of the pinning sites and create an
effective repulsive potential. If we further assume long-range
particle pair interactions ${\cal W}(r)$, such a 2D potential surely
gets insensitive to the details of the defect distribution; it only
retains the periodic horizontal modulation of $\rho_l(x)$, thus
resulting in a mean-field ratchet potential $V(x)$, like in Eq.
(\ref{doublesine}). The effective amplitude $V_0$ is a function of
at least three length scales: the interaction constant, the average
particle distance and the average defect spacing. A certain fraction
of the particles does not become pinned at individual defects but,
subjected to an applied ac drive, can move in the interstitial
regions between pinning sites. Although the moving interstitials do
not directly interact with the short-ranged defects, they feel the
long-range interaction of the particles trapped at the pinning sites
and described by the mean-field potential $V(x)$. As a consequence,
a horizontally applied ac drive can induce a longitudinal particle
transport, as proven in \textcite{olson2001PRL87} by means of
numerical simulation.

\begin{figure}[ht]
\centering
\includegraphics[width=5.0cm]{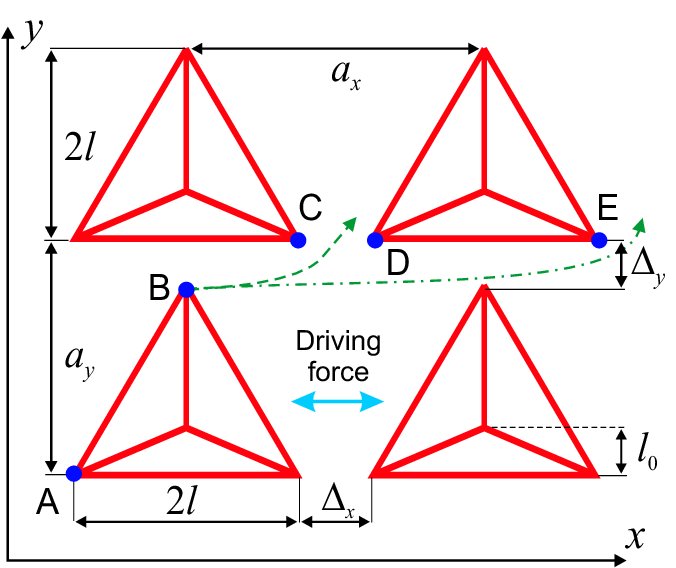}
\includegraphics[width=5.5cm]{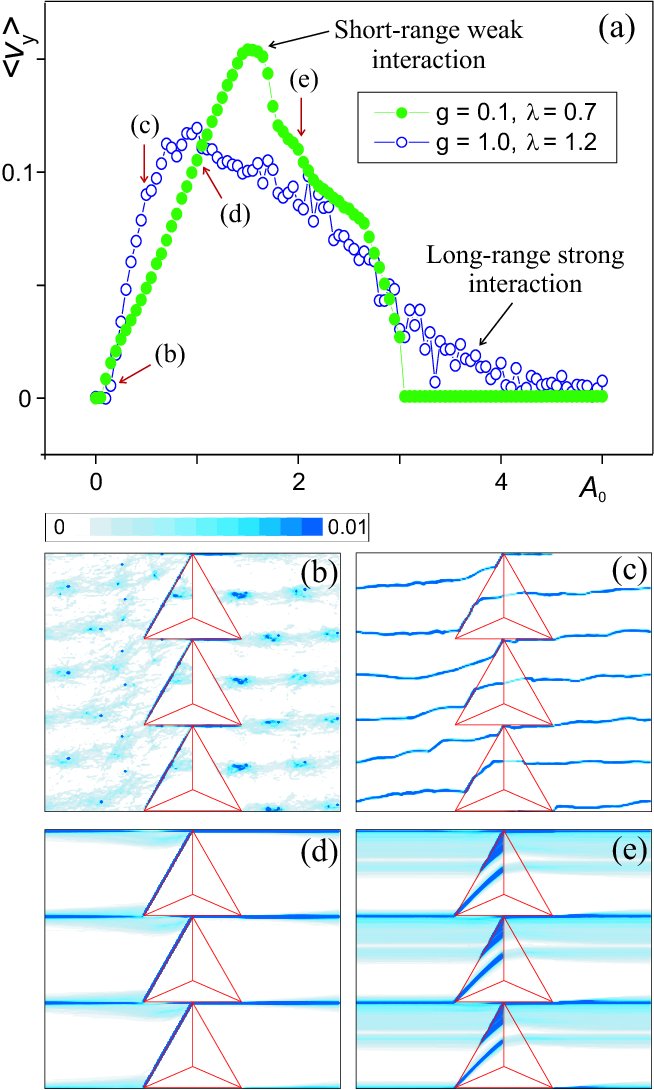}
\caption{(Color online) Top panel: asymmetric 2D arrays of potential
energy barrier/wells, top view. The parameters $l$, $l_0$, $a_x$,
$a_y$, $\Delta_x$, $\Delta_y$ define the geometry of the array. The
dashed and dash-dotted arrows represent trajectories perturbed by
thermal noise or particle-particle repulsion. Bottom panel: (a)
Transverse net velocity $\langle v_y\rangle$ versus $A_1$ for {\it
massless} interacting particles ac driven through a gapless
triangular chain: $a_y=2$, $a_x=6$ $l=1$, $\Delta_x=4$,
$\Delta_y=0$, and $T=0$. The repulsive potential ${\cal W}(r)$ is a
wedge function with $g=1$ and half-width $\lambda=0.05$. Green solid
circles are for weak short-range interacting particles, while blue
open circles are for strong long-range interacting particles.
Particle distributions for strong long-range (b,c) and weak
short-range (d,e) interactions for the $A_1$ values indicated by
arrows in (a). See for details
\textcite{savelev2005PRB71}.}\label{Fcollective2}
\end{figure}

Two conditions are instrumental to the onset of a rectification
current: $(1)$ a finite temperature, $T>0$, and $(2)$ a defect
filling fraction close to or larger than one, both conditions being
required for interstitial particles to get and stay unpinned. As in
Sec. \ref{boundaryeffect}, condition $(1)$ implies the existence of
an optimal rectification temperature. The dependence on the forcing
frequency and amplitude are as discussed in Sec.
\ref{rockedratchet}. These ideas have been implemented to control
transport of colloids and charge carriers in experimental setups
where point defect gradients could be engineered at will (cf. Secs.
\ref{colloid}, \ref{fluxon2Darrays} and \ref{QuantumdotQR}).

\subsection{2D lattices of asymmetric traps} \label{2dlattice}

Examples of substrates sustaining transverse rectification are
asymmetric potential barriers, height $q>0$, or wells, depth $q<0$,
either isolated or arranged into 1D chains and 2D lattices. Similar
lateral displacement devices, also known as bumper arrays, have been
proposed to separate particles by exploiting their mass and size
dispersion \cite{savelev2005PRB71,heller2008JMicromech18}. Consider
for instance the pyramidal potential barriers/wells with isosceles
triangular cross-section. This geometry, sketched in Fig.
\ref{Fcollective2}, is a generalization of the experimental setup by
\textcite{villegas2003Science302}.

The substrates discussed here combine two types of asymmetry: the
triangular shape of their building blocks and the asymmetry
associated with the pyramidal structure of each block. The latter
asymmetry affects the motion of the particles only if the drive is
strong enough to push them across the barriers/wells.
\textcite{savelev2005PRB71} have numerically simulated the dynamics
of a gas of {\it repelling} massive particles driven across 1D or 2D
lattices of barriers/wells for different parameters of the substrate
lattice (Fig. \ref{Fcollective2}, bottom panel), of the drive ${\bf
F}(t)$ and of the particle pair potential ${\cal W}(r)$. Although
\textcite{savelev2005PRB71} have shown that inertial effects often
enhance transverse rectification, we restrict our presentation to
the case of massless particles, certainly the most relevant for
technological applications. We consider next two distinct operating
regimes:

{\it (i) Deterministic} setups. An ac force applied along the
$y$-axis, i.e., parallel to the symmetry axis of the pyramid
cross-section, was known to induce a {\it longitudinal} particle
drift in the drive direction (see for more details in
\textcite{zhu2004PRL92} and Sec. \ref{superconductingdevice}).
Indeed, driving a (distorted) lattice of repelling particles along
the crystallographic axis of the substrate, oriented parallel to the
symmetry axis of the substrate blocks, makes the system reducible to
a mean-field 1D dynamics in that direction, along the line of Sec.
\ref{1dgeometry}. Due to the pyramidal shape of its building block,
the reduced 1D substrate is spatially asymmetric, which explains the
reported {\it longitudinal} particle flows.

{\it (ii) Diffusive} setups. Under appropriate conditions, instead,
dc or ac forces applied perpendicularly to the symmetry axis may
induce a {\it transverse} particle drift in the $y$-direction. For
all geometries considered, the net velocity of a gas of
non-interacting overdamped particles driven perpendicularly to the
symmetry axis vanishes for $T \to 0$. Indeed, sooner or later each
particle gets captured in a horizontal lane between two triangle
rows and then keeps oscillating back and forth in it forever. At
finite temperature, instead, fluctuations tend to push the particle
out of its lane, thus inducing the net transverse currents reported
in the earlier literature
\cite{duke1998PRL80,bier2000PRE61,huang2004Science304}. More
remarkably, \textcite{savelev2005PRB71} reported that the
interaction among particles not only contributes to transverse
rectification but actually plays a dominant role if the interaction
length $\lambda$, or the particle density are large enough. In
particular, particle-particle interaction was shown to control
transverse rectification for both weak short-range and strong
long-range inter-particle forces (see frame (a) of Fig.
\ref{Fcollective2}). Therefore, transverse rectification of
interacting colloidal particles (short range) and magnetic vortices
(long range) are expected to differ appreciably (see Secs.
\ref{colloid}, and \ref{superconductingdevice}).

To illustrate the key mechanism of transverse rectification, we
consider only rectification by pyramidal barriers, $q>0$, subjected
to the square ac force $F(t)=A_1 \mbox{sgn}[\cos(\Omega_1 t)]$. For
a fully detailed analysis the reader is referred to
\textcite{savelev2005PRB71}. Let us consider first the geometry in
frames (b)-(e) of Fig. \ref{Fcollective2} (bottom panel), where the
pyramids are stacked up in a close row ($\Delta_y= 0$). The particle
interactions have a strong impact on the equilibrium particle
distribution; a transition from an ordered lattice-like to a
disordered liquid-like phase is displayed in frames (b)-(e).
Nevertheless, the transverse currents $\langle v_y\rangle$ show a
qualitatively similar $A_1$ dependence for both weak short-range and
strong long-range interactions; only the decaying tail is longer for
the latter ones. The region of the linear growth of $\langle
v_y\rangle$ for the case of weak short-range interaction corresponds
to the regime when a constant fraction of particles (less than 1/2
because of the geometry of the system) is rectified into the
$y$-direction. This regime applies for increasing $A_1$ until when
particles start crossing over the triangles [frames (d) and (e)].
These results can be easily generalized to describe transverse
rectification in any 2D lattice of triangular shaped barriers/wells.

\textcite{savelev2005PRB71} also simulated the case of a 2D array of
pyramids, top panel of Fig. \ref{Fcollective2}, where a gap between
triangles along the $y$-axis, $\Delta_y> 0$, turned out to make the
net current sensitive to the particle-particle interaction length
$\lambda$. In the limit of low $T$, if $\lambda$ is smaller than a
certain threshold value $\lambda^c \sim \Delta_y$, then the net
current vanishes; if $\lambda$ exceeds $\lambda^c$ then the current
rapidly increases. This effect has been advocated to separate
particles according to their interaction length. More precisely,
particles having an interaction length smaller than $\lambda^c$
would pass through the array, or sieve of barriers separated by
$\Delta_y$. In contrast, particles with a longer interaction length
$\lambda>\lambda^c$ would be sifted sidewise. On connecting several
such sieves with different gap $\Delta_y$, one can construct a
device capable of separating the different fractions of a particle
mixture.

The overall conclusions of \textcite{savelev2005PRB71} do not change
on replacing barriers with wells of the same shape, nor do so for
lattices of asymmetric pins of different aspect-ratio and geometries
\cite{ertas1998PRL80,duke1998PRL80,bier2000PRE61,zhu2003PRB68,zhu2004PRL92,huang2004Science304,chepelianskii2005PRB71}.
This effect had been anticipated to some extent by
\textcite{lorke1998PhysicaB251}, who investigated the
magneto-transport properties of a square lattice of triangular
antidots of the type shown in Fig. \ref{Fcollective2}. Antidot
lattices are 2D electron gases with appropriately placed sub-micron
voids. These authors fabricated their lattices by electron beam
lithography on shallow high electron mobility transistor structures,
grown on semi-insulating gallium-arsenide substrates. In particular,
they reported evidence that under far-infrared irradiation, electron
sloshing in-between the antidot rows lead to a net electric current
along the symmetry axis of the antidots. Significant transverse
effects were observed in the presence of an orthogonal magnetic
field. Moreover, in most applications to the electrophoresis of
macromolecules, the particles moving through these sieve are
suspended in the fluid where they diffuse, thus involving additional
microfluidic effects (Sec. \ref{microfluidic}).

\subsection{Binary mixtures} \label{binarymixture}

We address now the problem as how to induce and control the net
transport of passive particles (target or $A$ particles), namely,
particles that are little sensitive to the applied drives and/or
substrates. \textcite{savelev2003PRL91,savelev2004PRL92} proposed to
employ auxiliary $B$ particles that $(1)$ interact with the target
species and $(2)$ are easy to drive by means of external forces. By
driving the auxiliary particles one can regulate the motion of
otherwise passive particles through experimentally accessible means.

Savel'ev and coworkers considered a mixture of two species of
pointlike overdamped Brownian particles $A$ and $B$ at temperature
$T$, diffusing on the 1D periodic substrates described by the
potentials $V_A(x)$ and $V_B(x)$, respectively. Particles of type
$A$ interact pair-wise with one another as well as with the $B$
particles via the potentials ${\cal W}_{AA}$ and ${\cal W}_{AB}$
($={\cal W}_{BA}$), while ${\cal W}_{BB}$ describes the interaction
of the $B$ pairs. The pair interaction is quantified by the tunable
strengths $g_{ij}$, with $i,j=A,B$, whereas the interaction
constants $\lambda_{ij}$ play no significant role as long as they
are conveniently small.

To illustrate the rectification mechanisms in a binary mixture, we
review in some detail the case of a pulsated device with oscillating
temperature, i.e. a temperature ratchet studied originally by
\textcite{reimann1996PLA215}, see also Sec. \ref{pulsatedratchet}).
The model can be made even simpler by assuming that (i) One
subsystem (say, the auxiliary $B$ particles) is subject to an
asymmetric ratchet potential $V_B(x)$, while the other one (the
target $A$ particles) is not, $V_A(x)=0$. This may happen, for
instance, in a mixture of neutral and charged particles with the
ratchet potential being produced by an electrical field; (ii) The
interaction among particles of the same type is repulsive, i.e.,
$g_{AA}> 0$, $g_{BB}> 0$.

\begin{figure}[btp]
\centering
\includegraphics[width=8.5cm]{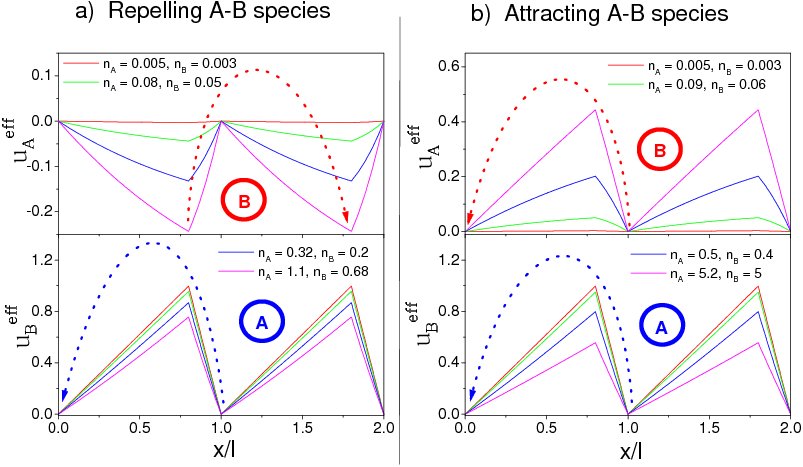}
\caption{(Color online) Spatial dependence of the effective
potentials $V^{\mathrm{eff}}_A(x)$, $V^{\mathrm{eff}}_B(x)$ at
different densities of the $A$ and $B$ particles. In both panels,
particles of the same type repel one another; the interaction
between particles of different species is repulsive in (a) and
attractive in (b). There is no substrate for the $A$ particles,
$V_A(x)=0$, whereas the ratchet potential $V_B(x)$ is piecewise
linear: $V_B(0<x<l_+)=x/l_+$, $V_B(l_+<x<l)=(1-x)/l_-$, with
$l_-=0.2$ and $l_+=0.8$. The other coupling parameters are
$g_{AA}=g_{BB}=|g_{AB}| = 1$ and $T=1$. [After
\textcite{savelev2005PRB71}].}\label{Fcollective3}
\end{figure}

Under these operating conditions, the $B$ particles condense
naturally at the minima of the asymmetric substrate potential [see
Figs. \ref{Fcollective3}(a) and (b), bottom panels]. In addition, if
the $B$ particles repel the $A$ particles, $g_{AB}>0$, then the
latter ones will accumulate in the regions where the density of the
$B$ particles is minimum, that is near the maxima of the substrate
potential $V_B(x)$. Vice versa, for attractive $AB$ interactions,
$g_{AB}<0$, the $B$ particles concentrate around the minima of
$V_B(x)$. Therefore, the target $A$ particles feel an effective
potential $V^{\rm eff}_A(x)$, which has opposite spatial asymmetry
with respect to $V^{\rm eff}_B(x)$ for $g_{AB}>0$, and has the same
asymmetry for $g_{AB} < 0$ [see Figs. \ref{Fcollective3}(a) and (b),
top panels]. Note that for low occupation numbers $n_A$ and $n_B$,
the bare potential $V_B(x)$ is only marginally affected by particle
interaction, i.e., $V^{\rm eff}_B(x) \simeq V_B(x)$. During the
lower $T$ half cycle, all particles are confined tighter around the
minima of the relevant effective potential, while, during the higher
$T$ half cycle, the particles of both species diffuse more easily
out of the $V^{\rm eff}_A(x)$, $V^{\rm eff}_B(x)$ potential wells.
As the asymmetry of the ratchet potentials $V^{\rm eff}_A(x)$ and
$V^{\rm eff}_B(x)$ for repelling $A$ and $B$ particles is opposite,
so is the orientation of their currents. On the contrary, two
attracting $A$ and $B$ species drift in the same direction. This
implies that the transport of both particle species can be
effectively and separately controlled by regulating their number
$n_A$ and $n_B$ per unit cell, without the need of tuning their
substrate.

As $g_{AA}, g_{BB}>0$, the potential wells of $V^{\rm{eff}}_A(x)$
($V^{\rm{eff}}_B(x)$) tend to flatten out when increasing the
corresponding density $n_B$ ($n_A$). This results in the decay of
the associated ratchet current $\langle v_B\rangle$ ($\langle
v_B\rangle$). In contrast, the ratchet asymmetry and current of one
species may be enhanced by increasing the density of the other
mixture component. For instance, in the case of $A$–$B$ attractive
forces, the $A$ particles tend to concentrate in the regions of
higher $B$ densities, that is around the minima of $V_B$, thus
attracting even more $B$ particles and making the
$V^{\mathrm{eff}}_B(x)$ potential wells deeper. In short, one can
enhance the transport of the ''target'' particles by adding a
certain amount of auxiliary particles. All these properties have
been reproduced analytically in the framework of the nonlinear
Fokker-Planck formalism \cite{savelev2003PRL91,savelev2004PRE70a}.

The influence of the interspecies interaction on the transport of a
binary mixture has been investigated further by
\textcite{savelev2004PRL92}, where dc and ac external forces were
applied at constant temperature, to the particles either of one
species, only, or of both, simultaneously. Their main results can be
summarized as follows: $(1)$ with increasing the intensity of an
applied dc driving force, there is a dynamical phase transition from
a ``clustered" motion of $A$ and $B$ particles to a regime of weakly
coupled motion \cite{marchesoni2006EPL73}; $(2)$ by applying a
time-asymmetric zero-mean drive to the $B$ species only, one can
obtain a net current for both the $A$ and $B$ species, even in the
absence of a substrate; $(3)$ when two symmetric ac signals act
independently on the $A$ and $B$ particles, and only the particles
of one species feel an asymmetric substrate, then, the two species
can be delivered in the same or opposite direction by tuning the
relative signal phase (for both attractive and repulsive $AB$
interactions). Superconducting devices based on these two-species
transport mechanisms have been realized experimentally in recent
years (Sec. \ref{anisotropic}).

\section{MICROFLUIDICS } \label{microfluidic}

Microfluidics \cite{squires2005RMP77} is playing an ever growing
role in controlling transport of particles, or even whole extended
phases, on the micro- and nanoscales. The ability to manipulate the
dynamics of liquids is crucial in various applications  such as for
a lab-on-a-chip technology. The Brownian motor concept has recently
been invoked in different contexts to face this challenge. In this
regard we remark that the inertial forces of small suspended
particles are typically quite small \cite{purcell1977AmJPhys45}.
Pointlike particles can then be considered to be advectively
transported by the fluid velocity field at the particle's actual
position. Then, for an incompressible liquid the particle dynamics
is volume conserving and consequently no dynamical attractors emerge
\cite{kostur2006PRL96}. However, for extended particles the local
velocity of a surface point need not coincide with the fluid
velocity which would act at this point in the absence of the
particle. Notably, for extended objects with internal degrees of
freedom the volume of the state space is no longer conserved by the
dynamics, thus giving rise to attractors for stationary flow fields
\cite{kostur2006PRL96}. Of course, Brownian diffusion provides an
additional transport mechanism that must also be taken into account.

We remark here that in spite of the intrinsic hydrodynamical
effects, microfluidic devices do not fall into the category of
collective ratchets as defined in Sec. \ref{collectivetransport},
because here {\it transported} objects are not required to interact
with one another. The suspension fluid still plays a central role:
(a) it powers particle transport, and (b) it determines how the
particle dynamics is coupled to the asymmetric geometry of the
substrate. Genuine collective ratchets are discussed in Sec.
\ref{superconductingdevice}.

\subsection{Transporting colloids} \label{colloid}

Many physical examples and technological applications involve
particles or molecules in solution that undergo a directed net
motion in response to the action of a ratchet. There, the ratchet
does not induce a mean flow of the solvent itself. For instance,
colloidal particles or macromolecules, suspended in solution, move
when exposed to a sawtooth electric potential that is successively
turned on and off \cite{rousselet1994Nature370}. Electrolytic
effects can be avoided by shuttling microsized Brownian polystyrene
particles by optically trapping them with repeatedly applied on-off
cycles in an optical tweezer, thus mimicking a flashing Brownian
motor
\cite{faucheux1995PRL74,faucheux1995FaradayTrans91,marquet2002PRL88}.
As a function of the cycle frequency one can even detect flux
reversal of diffusing colloidal spheres in an optical three-state
thermal Brownian motor \cite{lee2005PRL94}. This modus operandi of a
Brownian motor can therefore be put to work to pump or separate
charged species such as fragments of DNA. A micromachined
silicon-chip device that transports rhodamine-labeled fragments of
DNA in water has been demonstrated with  a flashing on-off Brownian
motor scheme by \textcite{bader1999PNAS96}. Yet other devices are
based on ideas and experimental realizations of entropic ratchets
\cite{slater1997PRL78,duke1998PRL80,ertas1998PRL80,tessier2002ApplPhysA75,chou1999PNAS96,oudenaarden1999Science285}
which make use of asymmetry within geometric sieve devices to
transport and separate polyelectrolytes.

\begin{figure}
\includegraphics[width=6.0truecm]{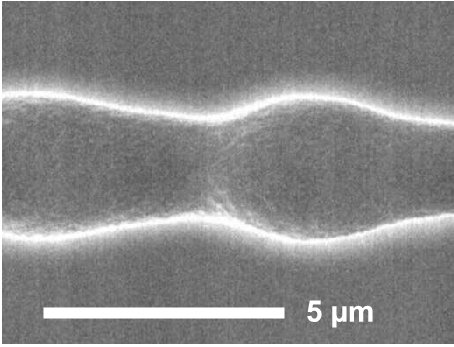}
\hfil
\includegraphics[width=6.0truecm,clip]{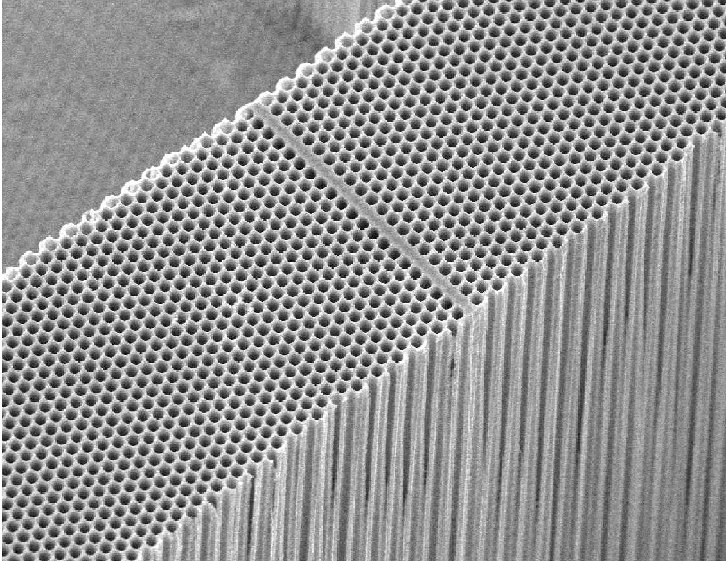}
\caption{Top Panel: Scanning-electron-microscope picture of a single
pore with ratchet longitudinal profile. The length of one period is
$8.4\mu  \text{m}$. The maximum pore diameter is $4.8 \mu  \text{m}$
and the minimum pore diameter is $ 2.5 \mu  \text{m}$. Bottom panel:
Scanning-electron-microscope picture of a silicon wafer pierced by
practically identical pores about $1.5  \text{mm}$ apart and 1mm in
diameter. This picture illustrates the enormous potential for
particle separation of a parallel 3D ratchet architecture. Figure
provided by Frank M\"uller, Max-Planck-Institute of Microstructure
Physics, Halle, Germany.} \label{Fmicro_pore}
\end{figure}

\begin{figure}
\includegraphics[width=6.0truecm]{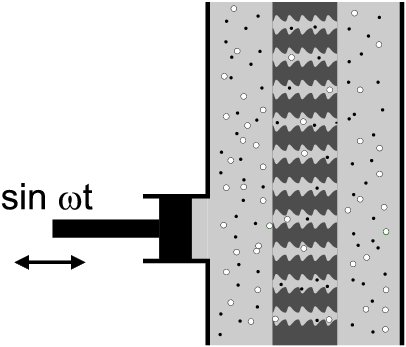}
\includegraphics[width=7.0truecm]{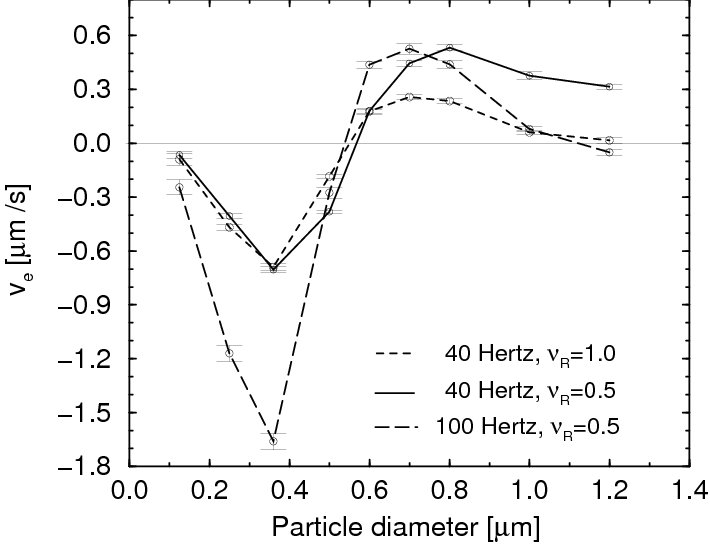}
\caption{Microfluidic drift ratchet of \textcite{kettner2000PRE61}.
Top panel: A macro-porous silicon wafer is connected at both ends to
basins. The pores with their ratchet-shaped profile are sketched in
dark grey. The basins and the pores are filled with liquid;
microparticles of two different species are represented. The fluid
is pumped back and forth by the piston on the left. Figure provided
by Christiane Kettner. Bottom panel: theoretically predicted net
particle current, $v_e$, versus the particle diameter for different
driving frequencies and viscosities (relative to water) $\nu_R$.
Note in particular the very sharp velocity reversal around $0.5
\,\mu \text{m}$. The results are for a pore of infinite length
consisting of periodic units with a pore length of $\text{L} = 6 \mu
\text{m}$, a  maximum pore diameter of $ 4 \mu$ \text{m} and a
minimum pore diameter of $1.6 \mu \text{m}$, see Fig.
\ref{Fmicro_pore} (a). For further details see \textcite
{kettner2000PRE61}.} \label{Fmicro_pump}
\end{figure}

A similar concept is employed to selectively filter mesoscopic
particles through a micro-fabricated macro-porous silicon membrane
(Fig. \ref{Fmicro_pore}), containing a parallel array of etched
 asymmetrical bottleneck-like pores
\cite{matthias2003Nature424,muller2000physstatsolidiA182,kettner2000PRE61}.
The working principle and the predicted particle flow for this
microfluidic Browinan motor device are shown in Fig.
\ref{Fmicro_pump}.
A fluid such as water containing immersed, suspended polystyrene
particles is pumped back and forth with no net bias through the 3D
array of asymmetric pores of Fig. \ref{Fmicro_pore}. Such an
artificial Brownian motor is thus kept far away from equilibrium by
the periodically varying pressure across the membrane. Due to the
asymmetry of the pores, the fluid develops asymmetric flow patterns
\cite{kettner2000PRE61}, thus providing the ratchet-like 3D force
profile in which a Brownian particle of finite size can both: (i)
undergo Brownian diffusion into liquid layers of differing speed,
and/or (ii) become reflected asymmetrically from the pore walls.
Both mechanisms will then result in a driven non-equilibrium net
flow of particles. Note that the direction of the net flow cannot be
easily guessed {\it a priori}. Indeed, the direction of the Brownian
motor current is determined by the interplay of the Navier-Stokes
flow in this engineered geometry and hydrodynamic thermal
fluctuations.

\begin{figure*}
\includegraphics[width=15.0truecm]{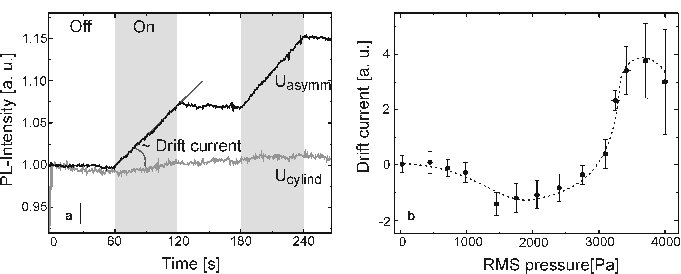}
\caption{(Color online) Parallel Brownian motors in a macroporous
silicon membrane containing ca. 1.7 million asymmetric pores with 17
etched modulations each (see Fig. \ref{Fmicro_pore}, top panel).
Left panel: When the pressure oscillations of the water are switched
on, the measured photoluminescence signal and thus the number of
particles in the basin located to the right, see top panel of Fig.
\ref{Fmicro_pump}, increases linearly in time, corresponding to a
finite transport velocity. In contrast, for symmetric
cylindrical-shaped pores no net drift is observed. Right panel: The
experimentally observed net transport behavior strongly depends on
the amplitude of the applied pressure and qualitatively shows the
theoretically predicted current inversion \cite{kettner2000PRE61}.
The pressure oscillations are toggled on and off each $60\,
\text{s}$. Other experimental parameters are as follows: the
suspended luminescent polystyrene spheres in water are $0.32 \mu
\text{m}$ across, the frequency of the pressure oscillations is 40Hz
and the root mean square (RMS) of the applied pressure during the
`on' phase is 2000Pa. A control experiment using  straight
cylindrical pores with a diameter of $ 2.4  \mu  \text{m}$ showed no
directed Brownian motor transport.   [Image: Max-Planck-Institute of
Microstructure Physics, Halle, Germany.]} \label{Fmicro_halle}
\end{figure*}

A detailed quantitative interpretation of the experimental data is
plagued by several complications such as the influence of
hydrodynamic interactions and, possibly, electric response effects
due to residual charge accumulation near the boundaries. A most
striking feature of this setup, however, is the distinct dependence
of current reversals on particle size. The sharply peaked
current-size characteristics curves of this directed flow, that is,
the theoretical flow {\it vs.} size prediction in Fig.
\ref{Fmicro_pump} and the experimental current-pressure
characteristics in Fig. \ref{Fmicro_halle}, suggest a highly
selective particle separation efficiency \cite{kettner2000PRE61}.
This microfluidic artificial Brownian device has recently been
implemented in experiments with suspended polystyrene spheres of
well defined diameters ($100$ nm, $320$ nm, $530$ nm and $1000$ nm)
by \textcite{matthias2003Nature424}: Their experimental findings are
in good qualitative agrement with the theory as shown in Fig.
\ref{Fmicro_halle}.

%
%
Remarkably, this device has advantageous 3D scaling properties: A
massively parallel architecture composed of about $1.7$ million
pores illustrated in Fig. \ref{Fmicro_pore}, is capable to direct
and separate suspended microparticles very efficiently. For this
reason, such type of devices have clear potential for bio-medical
separation applications and therapy use.

The separation and sorting of cellular or colloidal particles is
definitely a topic attracting wide interest in different areas of
biology, physical chemistry and soft matter physics. The powerful
toolbox of optical manipulation \cite{grier2003Nature424,
dholakia2007lasermanip82} uses the optical forces exerted on
colloids by focussed laser beams to move and control objects ranging
in size from tens of nanometers to micrometers. If the optical
forces are sufficiently strong to rule transport, the stochastic
Brownian forces play only a minor role, so that transport occurs as
a deterministic process with an  efficiency close to unity, a
circumstance known as (optical) peristalsis
\cite{koss2003APL82,bleil2007PRE75}. Such a scheme can then be
implemented also to experimentally realize ratchet cellular automata
capable of performing logical operations with interacting objects.
All together, in combination with this new optical technology,
colloids provides ideal model systems to investigate statistical
problems near and far away from thermal equilibrium
\cite{babic2005pRL94,babic2005CHAOS15}.

\subsection{Transporting condensed phases} \label{continuous phase}

Most present applications use ratchet concepts to transport or
filter discrete objects, like colloidal particles or macromolecules.
However, ratchets may also serve to induce macroscopic transport of
a continuous phase using local gradients, only.
One  such realization, strongly related to the above cases of
particle transport in a ``resting'' liquid phase, is the Brownian
motion of magnetic particles in ferrofluids subjected to an
oscillating magnetic field \cite{engel2004PRE70}. In contrast to the
cases reported above, here the motion is also transmitted to the
solvent by viscous coupling.

Another class of systems does not require colloidal particles to
drive the motion of the liquid phase.  In a first example, a
secondary large scale mean flow is triggered in
{M}arangoni-{B\'e}nard convection over a solid substrate with
asymmetric grooves \cite{stroock2003Langmuir19}. The direction and
strength of the mean flow can be controlled by changing the
thickness of the liquid layer and the temperature gradient across
the layer.

\begin{figure}[bpt]
\centering
\includegraphics[width=7.0truecm]{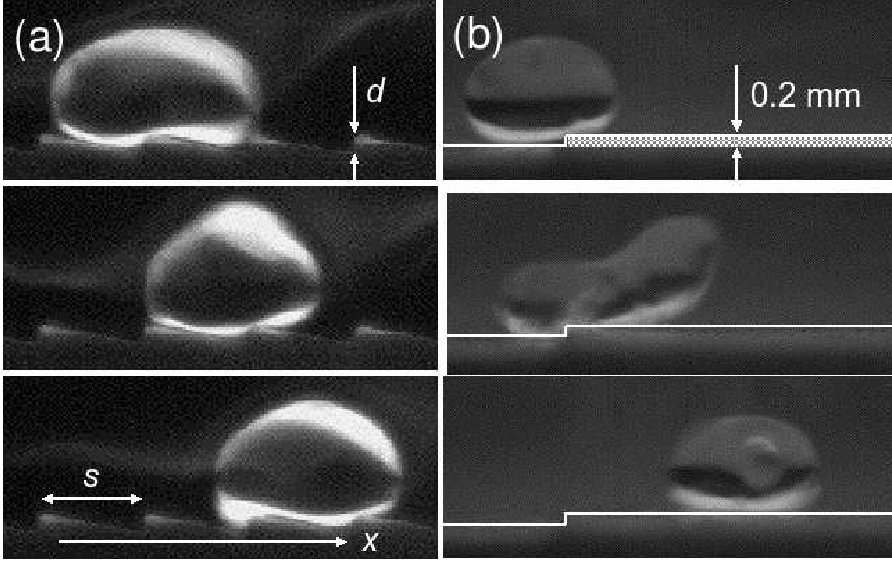}
\caption{(a) Video sequence (time interval 32 ms) demonstrating
droplet motion perpendicular to a thermal gradient. A droplet of the
refrigerant tetrafluorethane (R134a; boiling point $T_b= 26.1
^\circ$C) was placed on a room-temperature, horizontally leveled
brass surface with a ratchet-like grooved profile ($d= 0.3$mm, $s=
1.5$mm). See also http//www.aip.org/pubservs/epaps.html (b) Droplet
of liquid nitrogen on a flat surface moving with a small initial
velocity towards a piece of tape (shaded). Figure provided by Heiner
Linke.} \label{Fmicro_linke1}
\end{figure}

Yet another intriguing situation involves self-propelled Leidenfrost
drops placed on a hot surface also with a ratchet-like grooved
profile (Fig. \ref{Fmicro_linke1}). The locally asymmetric geometry
of the support induces a directed drop motion. This effect has been
observed for many liquids and in wide temperature ranges within the
film boiling (Leidenfrost) regime
\cite{quere2006NatMat5,linke2006PRL96}.

Moreover, micro-drops confined to the gap between asymmetrically
structured plates, move when drop shape or wetting properties are
changed periodically, for instance by vibrating the substrate,
applying an on/off electric field across the gap or a low-frequency
zero-mean electric field along the gap
\cite{gorre1996EPL33,buguin2002ApplPhysA75}. In a related experiment
the motion of drops on a global wettability gradient resulted to be
strongly enhanced when a periodic force was applied
\cite{daniel2004Langmuir20}. In that work the ratchet aspect was
determined by the intrinsically asymmetric shape of the drops and
the hysteresis of their contact angle on the gradient substrate.

The theory for particle transport operated by Brownian motors is
presently well developed
\cite{hanggi1996LNP476,astumian1997Science276,julicher1997RMP69,
reimann2002ApplPhysA75,astumian2002PhysToday55,hanggi2005AnPhys14}.
In clear contrast, models for ratchet-driven transport of a
continuous phase are a rather scarce commodity, although the concept
is thought to be important for emerging microfluidic applications
\cite{squires2005RMP77}.

\begin{figure}
\begin{center}
\includegraphics[width=8.0truecm]{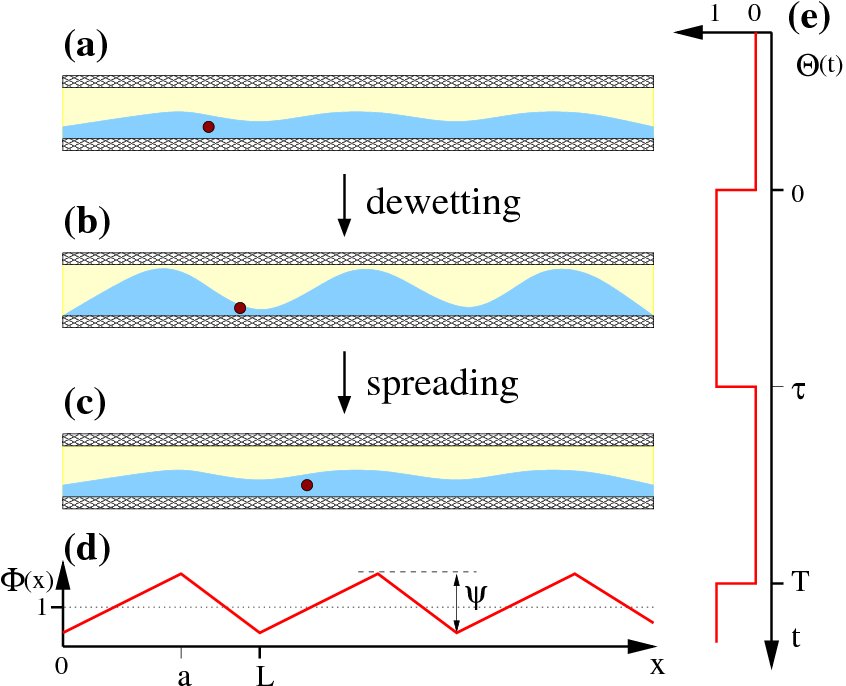}
\caption{(Color online) Panels (a) to (c) illustrate the working
principle of a fluidic ratchet based on dewetting-spreading cycles.
(d) illustrates the spatial asymmetric periodic interaction profile
$\Phi(x)$ responsible for the wettability pattern and (e) indicates
the time dependence $\Theta(t)$ of the switching in relation to the
dewetting and spreading phases (a)-(c). The filled circle indicates
a fluid element that gets transported during one ratchet cycle
although the evolution of the interface profile is exactly
time-periodic. For further details and resulting directed flow
pattern see \textcite{john2008SoftMatter}.} \label{Fmicro_thiele}
\end{center}
\end{figure}

In recent work, a model for the flow of two immiscible layered
liquids confined between two walls and driven by a flashing
``wettability ratchet'' has been put forward by
\textcite{john2007APL90} and \textcite{john2008SoftMatter}. In doing
so, one employs the general concept of wettability comprising all
effective interactions between the liquid-liquid free interface and
the solid walls. Notably, any interaction that is applicable in a
time-periodic, spatially periodic (but locally asymmetric) fashion,
is capable to drive a flow, i.e. will result in directed transport.
There exist several possibilities for experimental realizations of
spatially inhomogeneous and time switching interactions. A practical
setup consists of thin films of dielectric liquids in a capacitor
with a periodic, locally asymmetric voltage profile, that can be
periodically switched on and off.
A spatially homogeneous electric field would destabilize the
interface between two dielectric liquids. Therefore, an on-off
ratchet with a constant lateral force can result in the
dewetting-spreading cycle sketched in Fig. \ref{Fmicro_thiele}. The
process underlying this motor scheme may be termed
``electro-dewetting''.

\subsection{Granular flows} \label{granular}

The concept of ratchet physics had found an early application in the
field of granular matter \cite{jaeger1996RMP68}. Indeed, to explain
the vertical grain size segregation under vibration, the so-called
{\it Brazil-nut} effect \cite{rosato1987PRL58},
\textcite{bug1987PRL59} had initially investigated directed
diffusion in sawtooth-like inclined grooves. In the meantime, this
problem spurred several computer simulations aimed at exploring the
wonderful world of granular ratchets. Another unusual phenomenon,
observed both experimentally and in computer simulations, is
horizontal size segregation.
It involves the occurrence of  horizontal flows of  granular layers
that are vibrated vertically on plates whose surface profile
consists of sawtooth-like grooves
\cite{derenyi1995PRL75,rapaport2002CompPhys147}.

The spontaneous anelastic clustering of a vibrofluidized granular
gas has been employed, for instance, to generate phenomena such as
{\it granular fountains} (convection rolls) and {\it ratchet
transport} in compartmentalized containers, where symmetry breaking
flow patterns emerge perpendicular to the direction of the energy
input
\cite{eggers1999PRL83,brey2001PRE65,farkas2002PRE65,vandermeer2004PRL92}.
Granular systems exhibit indeed a rich scenario of intriguing
collective transport phenomena, which have been tested and validated
both computationally and experimentally
\cite{vandermeer2007JSM2007}.

\begin{figure}
\begin{center}
\includegraphics[width=8.0truecm]{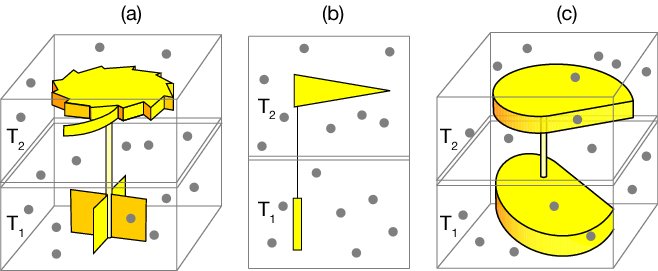}
\caption{(Color online) Taming thermal motion. Panel (a) shoes a
schematic representation of the Smoluchowski-Feynman rachet device,
which is capable to rectify thermal Brownian motion of gases held at
two different temperatures \cite{feynman1963}. An idealized version
suitable for computer studies is presented in panel (b) with the
ratchet replaced by a triangular unit and the pawl replaced by a
rigidly joined blade residing in the lower gas compartment. Panel
(c) illustrates a 3D rotational construction. Figure provided by
\textcite{vandenbroeck2008PRL100}.} \label{Fmicro_motorina}
\end{center}
\end{figure}

All these systems are kept away from thermal equilibrium by vertical
vibrations, while the presence of random perturbations play a role
as a source of complexity, due to the intrinsic chaotic nature of
granular dynamics. Closer in spirit of the present review are,
however, those devised granular Brownian motors which are truly
capable to tame thermal Brownian motion
\cite{cleuren2007EPL77,costantini2007PRE75,costantini2008EPL82}. In
those setups, an asymmetric object of mass $m$ (Fig.
\ref{Fmicro_motorina}), is free to slide, without rotation, in a
given direction. Its motion with velocity $v$ is induced by {\it
dissipative} collisions with a dilute gas of surrounding particles
of average temperature $T$. On each collision, only a certain
fraction (called restitution coefficient) of the total kinetic
energy of the object-particle system is conserved. At variance with
the case of elastic collisions, here isotropy, implying $ \langle v
\rangle = 0$, and energy equipartition, implying $\langle v^2
\rangle = kT/m$, do not apply: Dissipation is responsible for
breaking the time reversal symmetry. As a result one finds that the
asymmetric object undergoes (approximately) a directed random walk
with nonzero average velocity, which can be modeled in terms of an
effective biased Langevin dynamics for the velocity variable $v$.
Such a Langevin dynamics assumes the form of an Ornstein-Uhlenbeck
process \cite{hanggi1982PhysRep88,risken1984} governed by an
effective Stokesian friction and an effective external force, both
depending on the restitution coefficient
\cite{cleuren2007EPL77,costantini2007PRE75}. The directed motion of
the object is accompanied by a continuous flow of heat from the gas
to the object, in order to balance the dissipation of the
collisional processes; as a consequence the average temperature of
the asymmetric object is lower than the gas temperature.

Ever since the formulation of the second law of thermodynamics there
has been haunting debates about its validity at the small scales.
In this context the ratchet gadget invented by
\textcite{smoluchowski1912ZP13} and later beautifully popularized by
\textcite{feynman1963} points at the heart of a controversy, which
concerns the very working principle of Brownian motors: To be
consistent with this law, no rectification of thermal fluctuations
can take place at thermal equilibrium for which all dynamics is
intrinsically governed by the principle of detailed balance. A
steadily working Brownian motors necessarily requires a combination
of asymmetry and non-equilibrium, such as a temperature gradient.
This point has been convincingly elucidated in recent molecular
dynamic studies by Van den Broeck and his collaborators
\cite{vandenbroeck2005NJP7,vandenbroeck2004PRL93,vandenbroeck2006PRL96},
who variously stylized a Maxwell daemon operating far away from
equilibrium (Fig. \ref{Fmicro_motorina}). Several computer versions
of Smoluchowski-Feynman ratchets operating in a dilute gas of
colliding granular particles, have thus been confirmed to generate
directed transport in the presence of a finite temperature
difference. With the possibility of today's observation and
manipulation of physical, chemical and biological objects at the
nano- and meso-scale, such devices no longer represent a theorist's
thought experiment, but are rather becoming a physical reality. It
is right on these length scales that thermal fluctuations cannot be
ignored, as they combine with non-equilibrium forces to yield
constructive transport.

\section{SUPERCONDUCTING DEVICES} \label{superconductingdevice}

Magnetic vortices in type-II superconducting devices provide a
flexible and experimentally accessible playground for testing our
stochastic transport models. In recent years an impressive effort
has been directed toward the design and operation of a new
generation of vortex devices with potential applications to
terahertz (THz) technology
\cite{hechtfischer1997PRL79,zitzmann2002PRB66}, fluxon circuitry
\cite{shalom2005PRL94}, and quantum computation
\cite{you2005PhysToday58}, to mention but a few. We anticipate that
magnetic vortices are inherently quantum objects that, under most
experimental conditions, behave like (massless, point-like) {\it
classical} particles. Genuine quantum effects in the vortex
transport are briefly discussed in Sec. \ref{quantumdevice}.

A magnetic field applied to a type-II superconducting film, with
intensity $H$ comprised between the critical values $H_{c1}$ and
$H_{c2}$, penetrates into the sample producing supercurrent
vortices, termed Abrikosov vortices
\cite{abrikosov1957JETP5,blatter1994RMP66}. The supercurrent
circulates around the normal (i.e. non-superconducting) core of the
vortex; the core has a size of the order the superconducting
coherence length $\xi$ (parameter of a Ginzburg-Landau theory); the
supercurrents decay on a distance $\lambda$ (London penetration
depth) from the core. In the following we assume that $\lambda \gg
\xi$. The circulating supercurrents produce magnetic vortices, each
characterized by a total flux equal to a single flux quantum
$\Phi_0=hc/2e$ (hence the name fluxon for magnetic vortex) and a
radially decaying magnetic field
\begin{equation}\label{abrikosov1}
B(r) = B_0~K_0(r/\lambda),
\end{equation}
where $K_0(x)$ is a modified Bessel function of zero-th order and
$B_0=\Phi_0/(2\pi\lambda^2)$. Far from the vortex core, $r\gg
\lambda$, the field decays exponentially,
$B(r)/B_0=\sqrt{{\lambda\pi}/{2r}}~e^{-r/\lambda}$, whereas,
approaching the vortex center, it diverges logarithmically for $\xi
< r \ll \lambda$, $B(r)/B_0=\ln(\lambda/r)$, and then saturates at
$B(0)=B_0\ln(\lambda/\xi)$ inside the core $r < \xi$. As a
consequence, the coupling between vortex supercurrents and magnetic
fields determines a long range repulsive vortex-vortex interaction.

Due to mutual repulsion, fluxons form a lattice (also called
Abrikosov vortex lattice, usually triangular, possibly with defects
and dislocations) with average vortex density $\rho$ approximately
equal to $H/\Phi_0$. A local density $\bf I$ of either a transport
current or a magnetization current or both, exerts on each fluxon a
Lorentz force ${\bf F}_L=\Phi_0{\bf I}\times {\bf H}/(cH)$.

The distribution and microscopic properties of pinning centers can
qualitatively influence the thermodynamic and vortex transport
properties of the superconducting sample.
Pinning forces created by isolated defects in the material oppose
the motion of the fluxons, thus determining a critical (or
threshold) current, below which the vortex motion is suppressed.
Many kinds of artificial pinning centers have been proposed and
developed to increase the critical current, ranging from the
dispersal of small non-superconducting second phases to creation of
defects by 
irradiation \cite{altshuler2004RMP76}. A novel approach to the
problem came with advances in lithography, which allowed for regular
structuring and modulation of the sample properties over a large
surface area \cite{harada1996Science274}. Long-range correlation in
the position of the pinning centers resulted in the interplay
between the length scales characterizing the pin lattice and the
vortex lattice. Indeed, the magnitude of the commensuration effects
is readily controlled by tuning the intensity of the magnetic field
$H$: at the first matching field $H_1=\Phi_0/a_l$, pin and vortex
lattices are exactly commensurate, with one fluxon per pin lattice
cell of area $a_l$; at the $n$-th matching field, $H(n)=nH_1$, each
pin lattice cell is occupied by $n=\rho a_l$ fluxons; for irrational
$H/H_1$ pin and vortex lattices are incommensurate. Commensuration
effects are responsible for a variety of dynamical superconducting
states of great relevance to the problem of vortex rectification
\cite{reichhardt1998PRB57}.

\begin{figure}[htb]
\centering
\includegraphics[width=8.5cm]{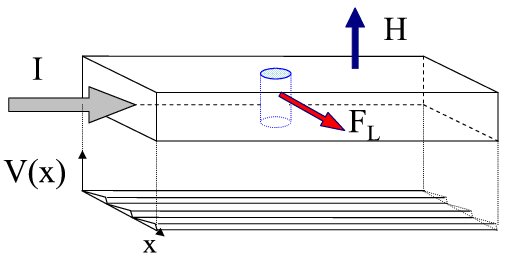}
\caption{(Color online) Diagram of a superconductor in the presence
of an external magnetic field $H$. A direct current with density $I$
flowing along the $y$ direction (indicated by the large arrow)
induces a Lorentz force ${\bf F}_L$ that moves the vortex in the $x$
direction (upper panel). The superconductor is patterned with a
pinning potential (lower panel). The potential is periodic and
asymmetric along the $x$ direction, and is invariant under
translation along $y$; the potential $V(x)$ is an effective ratchet
potential (see Sec. \ref{molecularmotor}). } \label{Fvortex1}
\end{figure}

As a first attempt to design a vortex ratchet
\textcite{lee1999Nature400} showed that an ac electric current
applied to a superconductor patterned with the asymmetric pinning
potential shown in Fig. \ref{Fvortex1}, can induce vortex transport
whose direction is determined only by the asymmetry of the pattern.
The fluxons were treated as zero-mass point-like particles moving on
a saw-tooth potential along the $x$-axis, according to the scheme of
Sec. \ref{rockedratchet}. These authors demonstrated theoretically
that, for an appropriate choice of the pinning potential, such a
rocked ratchet can be used to manipulate single vortices in
superconducting samples under realistic conditions. The
rectification mechanisms of an extended overdamped string, like a
fluxon, on a ratchet potential had been anticipated by Marchesoni
and coworkers \cite{marchesoni1996PRL77,costantini2002PRE65}.
Brownian motor ratcheting of single oscillating fluxons have been
observed experimentally in asymmetrically etched Nb strips
\cite{plourde2000PhysicaC341}, on planar patterns of columnar defect
\cite{kwok2002PhysicaC382}, and in asymmetric linear arrays of
underdamped Josephson junctions \cite{lee2003APL83}, in an annular
Josephson junction \cite{ustinov2004PRL93}, to mention just a few
recent experiments.

\subsection{Fluxon channels} \label{fluxonchannel}

\begin{figure}[htb]
\vglue 0.5truecm
\includegraphics[width=8.5cm]{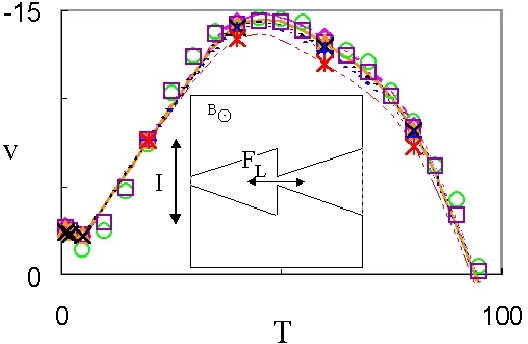}
\caption{(Color online) Rectified average fluxon velocity $\langle v
\rangle $ (in units of $\lambda/\tau$) versus $T$ for the sawtooth
channel geometry in the inset (width: $7\lambda$, bottleneck:
$\lambda$). The magnetic field ${\bf H}$ is directed out of the
figure. A vertically applied alternating current square wave drives
the fluxons back and forth horizontally along the channel with
period $\tau=100$ MD steps. The number of fluxons per unit cell is:
1 (circles), 25 (diamonds), 50 (squares), 75 (triangles), 100
($\times$), 150 ($+$), 250 (*). [After
\textcite{wambaugh1999PRL83}]. }\label{Fvortex2}
\end{figure}

A more viable approach to achieve controllable stochastic transport
of fluxons in superconducting devices consists in exploiting
collective boundary effects as suggested Sec. \ref{boundaryeffect}.
Fluxon rectification in asymmetric channels has been investigated by
\textcite{wambaugh1999PRL83} by means of a molecular dynamics (MD)
code, originally developed to study magnetic systems with random and
correlated pinning \cite{reichhardt1998PRB57}. They simulated a thin
slice of a zero-field cooled, type-II superconducting slab, taken
orthogonally to the magnetic vortex lines generated by a tunable
external magnetic field of intensity H. On regarding them as fairly
rigid field structures, magnetic vortices formed a confined 2D gas
of zero-mass repelling particles with density $\rho=H/\Phi_0$. The
simulated sample had very strong, effectively infinite pinning
except the ``zero pinning" central sawtooth-shaped channel, sketched
in the inset of Fig. \ref{Fvortex2}. In the channel fluxons moved
subject to fluxon-fluxon repulsions, an externally applied ac
driving Lorentz force, forces due to thermal fluctuations, and
strong damping. The fluxons outside the channel could not move, thus
impeding the movable fluxons in the middle to cross the channel
walls. As a consequence, the interaction length of the fluxon-wall
collisions was the same as of the fluxon-fluxon collisions. The net
fluxon velocity $\langle v \rangle$ versus temperature shown in Fig.
\ref{Fvortex2} is negative and exhibits an apparent resonant
behavior, both versus $T$ and $H$ (i.e., the fluxon density $\rho$),
as anticipated in Sec. \ref{boundaryeffect}.

These conclusions have been recently corroborated experimentally by
\textcite{togawa2005PRL95} and then by \textcite{yu2007PRB76}. The
authors of the more recent work fabricated triangular channels from
bilayer films of amorphous niobium-germanium, an extremely
weak-pinning superconductor, and niobium nitride (NbN), with
relatively strong pinning. A reactive ion etching process removed
NbN from regions as narrow as 100 nm, defined with electron-beam
lithography, to produce weak-pinning channels for vortices to move
through easily. In contrast, vortices trapped in the NbN banks
outside of the channels remain strongly pinned. The vortex motion
through such asymmetric channels exhibited interesting asymmetries
in both the static depinning and the dynamic flux flow. The vortex
ratchet effect thus revealed a even richer dependence on magnetic
field and driving force amplitude than anticipated by the simplified
model simulated by \textcite{wambaugh1999PRL83}.

Devices like that simulated in Fig. \ref{Fvortex2} serve as
excellent fluxon rectifiers. By coupling two or more such devices,
one can design fluxon lenses, to regulate the fluxons concentration
in chosen regions of a sample, and eventually channel networks or
circuits, for fluxons in superconducting films.

\begin{figure*}[htb]
\hspace*{-1.0cm} \includegraphics[angle=-90,width=5.5cm]{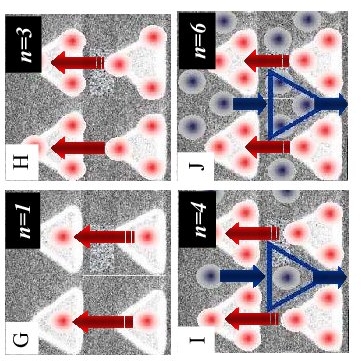}
\hspace*{1.0cm} \includegraphics[angle=-90,width=6.0cm]{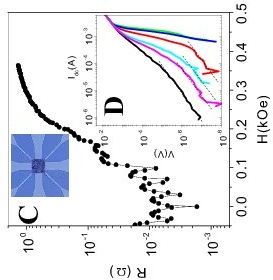}
\caption{(Color online) Top panel: Sketch of the positions of the
vortices for several matching fields $H(n)$, i.e., for $n$ vortices
per unit cell. Vortices pinned on the triangles (no more than 3 per
unit cell) are shown in red and interstitial vortices in blue. The
pinning substrate is a square lattice with constant $750nm$; the
triangular pinning sites have been grown on a Si support, are $40nm$
thick with side about $600nm$ long. Bottom panel: dc
magneto-resistance $R(H)$ versus $H$. The temperature is $T
=0.98T_c$ and the injected dc current density is $I_{dc}=12.5
kA\cdot cm^{-2}$. The upper inset shows a micrograph of the
measuring bridge, which is $40\mu m$ wide. The darker central area
of the inset is the $90\times90\mu m^2$ array of magnetic triangles.
The lower inset shows the characteristics curves $V_{dc}$-$I_{dc}$
at the matching magnetic fields $H(n)=nH_1$ with $H_1=32Oe$ and
$n=1$ (red), 3 (cyan), 4 (blue), 6 (green), 8 (magenta), and 10
(black).
The $V(I_{dc})$ curves change abruptly at magnetic fields from $n
=3$ to $n =4$, because an ohmic regime appears at low currents. This
is a clear signature of the presence of interstitial vortices.
[After \textcite{villegas2003Science302}]. }\label{Fvortex3}
\end{figure*}

Fluxon channeling was observed experimentally for the first time in
an effective 1D vortex rectifier by
\textcite{villegas2003Science302}. Their device has the pedagogical
merit of showing all collective transport effects summarized in Sec.
\ref{collectivetransport} at work. A $100\mu m$ thick Niobiun film
was grown on an array of nanoscale triangular pinning potentials
oriented as in Fig. \ref{Fvortex3} (top panel), with bases lined up
along the $x$-axis and vertices pointing in the $y$-direction.
Magneto-resistance $R(H)$ experiments were done with a magnetic
field $H$ applied perpendicular to the substrate in a liquid Helium
system. The dc magneto-resistance curves in the bottom panel of Fig.
\ref{Fvortex3} exhibit commensurability effects in which dissipation
minima develop as a consequence of the geometrical matching between
the vortex lattice and the underlying periodic structure. At these
matching fields, the vortex lattice motion slows down, and $R(H)$
minima appear at the equally spaced values $H(n)=nH_{1}$, with
$H_{1}=32$~Oe. The $R(H)$ minima are sharp at the $n=1,2,3$ matching
fields, but shallow and not well-defined for higher $n$. This effect
is a consequence of the appearance of interstitial vortices at
increasing $H$ beyond $H_{3}$, three being the maximum number of
vortices contained in each triangle.

In order to interpret the experimental results, one separates all
vortices in two groups: (i) pinned vortices, which move from one
triangular-shaped pinning trap to another one and, thus, are
directly affected by an effective 1D rocked ratchet substrate with
positive polarity in the $y$-direction; and (ii) interstitial
vortices, which are channeled in-between triangles and do not
directly interact with the pinning traps
\cite{savelev2003PRL91,zhu2003PRB68}. However, as apparent in the
top panel of Fig. \ref{Fvortex3} and discussed in Sec.
\ref{binarymixture}, pinned vortices, being strongly coupled to the
substrate potential, determine a weaker mean-field potential that
acts on the interstitials with opposite asymmetry
\cite{savelev2003PRL91}. As a consequence, when all fluxons are
subjected to the same ac drive force, the fluxon-fluxon repulsion
pushes the interstitials in the direction opposite to the pinned
vortices (for an animation see http:$//$dml.riken.go.jp/vortex).

\begin{figure}[htb]
\includegraphics[width=7.0cm]{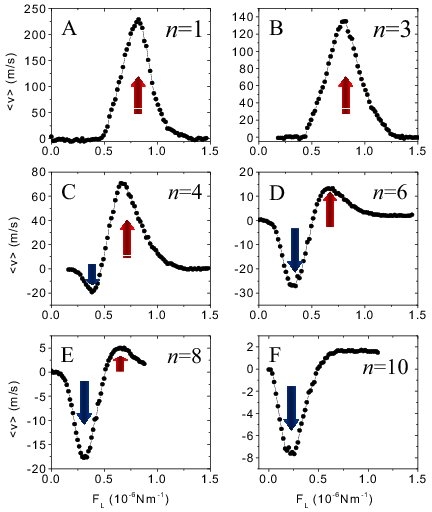}
\caption{(Color online) (a)-(f) Net velocity $\langle v \rangle$ of
vortices versus the amplitude $F_L$ of the ac Lorentz force for
$\Omega=10kHz$ and different matching magnetic fields $H(n)$; other
parameter values are as in Fig. \ref{Fvortex3}. Red and blue arrows
show the direction of the net flow, dominated respectively, by the
pinned and the interstitial vortices (see panel (a) of Fig.
\ref{Fvortex3}). [After \textcite{villegas2003Science302}. }
\label{Fvortex4}
\end{figure}

The effects of this mechanism are illustrated in Fig.
\ref{Fvortex4}. At constant temperature (close to $T_c$ to avoid
random pinning) and $H$ multiple of $H_{1}$, an ac current density
$I(t)=I_{ac}{\rm sin}(\Omega t)$ was injected along the $x$-axis.
This yields a sinusoidal Lorentz force with amplitude $F_L$ acting
on the vortices along the $y$-axis. Despite the zero time average of
such Lorentz force, a non-zero dc voltage drop $V_{dc}$ in the
$x$-direction was measured, thus proving that the vortices actually
drift in the $y$-direction. Note that, due to the peculiar substrate
symmetry, an ac current $I(t)$ oscillating along the $y$-axis, can
rectify the fluxon motion only parallel to it (transverse
rectification, Sec. \ref{fluxon2Darrays}). The amplitude of the dc
voltage signal decreases with increasing $H$ because the effective
pinning is suppressed by the inter-vortex repulsion. Moreover, when
$n>3$ (corresponding to more than three vortices per unit cell), a
reversed $V_{dc}$ signal begins to develop with a maximum (marked by
blue arrows) occurring at a lower Lorentz force $F_L$ than the
positive dc maxima (red arrows): The interstitials, which feel a
weak inverted ratchet potential with respect to the pinned vortices,
dominate the rectification process. This current reversal effect is
enhanced when further increasing the magnetic field [panels (c) and
(d)] and finally, at very high magnetic fields, close to the normal
state, the voltage reversal, although suppressed in magnitude, spans
over the entire $F_L$ range [panel (f)].

Moshchalkov and coworkers
\cite{silva2006Nature440,vandevondel2007EPL80} also demonstrated
longitudinal fluxon rectification. \textcite{silva2006Nature440}
observed multiple current reversals in regular square lattices of
asymmetric double-well traps periodically driven along an ``easy"
direction, as in the simulations by
\textcite{zhu2003PhysE18,zhu2003PRB68}. More interestingly,
\textcite{vandevondel2007EPL80} have detected inverted fluxon
currents in large triangular arrays, similar to those reviewed here,
but at magnetic fields (or fluxon densities) so high that a
collective ratchet mechanism would not be plausible. They attributed
their finding to a new intra-antidot rectification mechanism
controlled by the magnetic flux quantization condition synchronously
satisfied at the edge of each asymmetric antidot.

\subsection{Fluxon rectification in 2D arrays}
\label{fluxon2Darrays}

The rectification of a fluxon lattice, no matter how distorted or
disordered, is an inherently 2D process. Following the theoretical
and numerical investigations reviewed in Sec. \ref{2dlattice}, a
number of experimental groups
(\textcite{vanlook2002PRB66,crisan2005PRB71,menghini2007PRB76} to
mention a few) prototyped devices aimed at controlling vortex
ratcheting both in the direction parallel (longitudinal rectifiers)
and perpendicular to an applied ac Lorentz force (transverse
rectifiers).

Longitudinal transport in a 2D array of asymmetric pinning traps has
been experimentally obtained, for instance, by
\textcite{vandevondel2005PRL94}, who reported that, under
appropriate operating conditions, fluxon rectification can be
enhanced for $F_L$ amplitudes comprised between two critical pinning
forces of the underlying asymmetric substrate. The shape of the net
dc voltage drop $V_{dc}$ as a function of the drive amplitude
indicated that their vortex ratchet behaved in a way very different
from standard overdamped models. Rather, as the repinning force
necessary to stop vortex motion, is considerably smaller than the
depinning force, their device resembles the behavior of the inertial
ratchets of Sec. \ref{rockedratchet}. More recently,
\textcite{shalom2005PRL94} reported longitudinal fluxon
rectification in square arrays of Josephson junctions, where the
coupling energies had been periodically modulated along one symmetry
axis. Similarly, longitudinal fluxon rectification has been observed
also in square arrays of magnetic antidots with size-graded period
\cite{gillijns2007PRL99}. Both are nice experimental implementations
of the asymmetric patterns of symmetric traps introduced in Sec.
\ref{asymmetricpattern}. Finally, \textcite{ooi2007PRL99} trapped
symmetric fluxon lattices in a triangular dot lattice, they had
photo-lithographed on a thin Bi$_2$Sr$_2$CaCu$_2$O$_{8+\delta}$
(Bi2212) single-crystal film, by setting the applied magnetic field
at the lowest multiples of $H_1$. By subjecting the fluxon lattice
to a bi-harmonic Lorentz force oriented along the crystallographic
axes of the dot lattice, they obtained a neat demonstration of
harmonic mixing on a symmetric substrate. However, also the results
of these experiments can be easily explained in terms of simple 1D
models.

Transverse transport, instead, requires genuinely irreducible 2D
geometries. When designing and operating a device capable of
transverse rectification, experimenters can vary the orientation of
the pinning lattice axes, the symmetry axes of the individual traps
(if any) and the direction of the injected current (i.e., of the
Lorentz force). Many authors explored by numerical simulation the
geometries best suitable for the experimental implementation of this
concept \cite{zhu2001PRB64,olson2005PhysicaC432}. The most recent
realizations of transverse fluxon rectifiers fall into two main
categories:

{\it (i) Symmetric arrays of asymmetric traps.}
\textcite{gonzales2007APL91} modified the experimental set-up of
\textcite{villegas2003Science302} to investigate the rectification
mechanisms presented in Secs. \ref{1dgeometry} and \ref{2dlattice}.
They confirmed the numerical simulations by
\textcite{savelev2005PRB71}, who had predicted that the same device
can exhibit either longitudinal or transverse output current
depending on orientation of the ac drive with respect to the pinning
lattice axes. In the longitudinal ratchet configuration a sinusoidal
driving current $I(t)$ was applied perpendicular to the triangle
reflection symmetry axis ($x$-axis) and the output voltage signal
$V_{dc}$ was recorded in the same direction. We recall that the
Lorentz force induced vortex motion, parallel to the triangle
reflection symmetry axis ($y$-axis), corresponds to a voltage drop
in the direction of the injected current.  To observe an optimal
transverse rectification effect, the axes of the current injection
and the voltage drop were inverted. The asymmetry of the traps with
respect to the direction of the Lorentz force is responsible for the
observed fluxon drift in the $y$ direction.

The number $n$ of vortices per lattice cell was controlled by
varying the intensity of the magnetic field orthogonal to the
device, $H=nH_1$. On increasing $n$, the voltage associated to the
longitudinal ratchet current changes sign and gets amplified, as to
be expected due to the presence of the $n-3$ interstitials per unit
cell (Sec. \ref{fluxonchannel}). The $H$ dependence the transverse
fluxon rectification is very different: (a) the transverse net
current shows no inversions; (b) increasing the number of the
interstitials transverse rectification is suppressed.

\begin{figure}[htb]
\centering
\includegraphics[width=8.5cm]{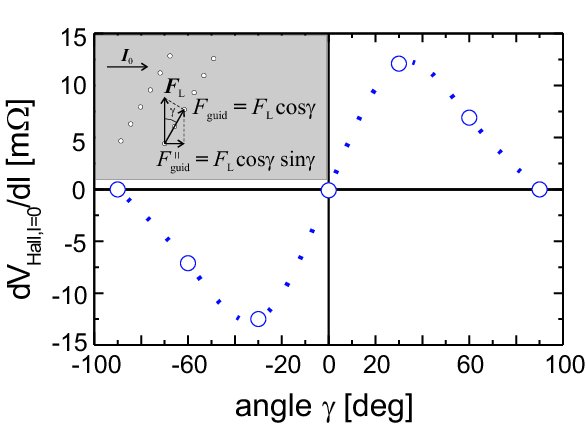}
\caption{Angular dependence of $R_H$ measured on a circular shaped
$90nm$ thick sample at $T=30K$, $H=143mT$ and with dc current
$I_{dc}=10mA$. Inset:  portion of a $20\mu m \times 10\mu m$
rectangular lattice of circular antidots with radius of $1 \mu m$;
$I_{dc}$ is horizontally oriented from left to right corresponding
to a Lorentz force ${\bf F}_{L}$ pointing upwards. The angular
dependence is fitted by the sinusoidal law $\sin 2\gamma$, as
sketched in the inset. [After \textcite{wordenweber2004PRB69}].
}\label{Fvortex5}
\end{figure}

{\it (ii) Asymmetric arrays of symmetric traps}. Controlled
transport of vortices through rows of antidots was measured by
\textcite{wordenweber2004PRB69} via standard 4-probe Hall-type
experiments \cite{altshuler2004RMP76}. $100$-$150nm$ thin
YBa$_2$Cu$_3$O$_2$ (YBCO) films were deposited on CeO$_{2}$ buffered
sapphire and then covered with a $50nm$ thick Au layer. Asymmetric
lattices of symmetric antidots in the shope of circular micro-holes,
were patterned via optical lithography and ion beam etching. The
inset of Fig. \ref{Fvortex5} shows a typical antidot lattice. A
tunable dc current $I_{dc}$ was then injected so that the
corresponding Lorentz force ${\bf F}_{L}$ was at an angle $\gamma$
with respect to the orientation of the antidot rows.

Reference measurements on samples without antidots as well as
temperature-dependent measurements of the Hall resistance $R_H$ for
different angles $\gamma$ clearly indicate that for $T<83K$ and low
dc drive current, the Hall resistance is dominated by the directed
vortex motion along the antidot rows.  Under this circumstances,
$R_H$ quantifies the fluxon current in the direction of the dc
current $I_{dc}$ (with negative Hall voltage), or equivalently, {\it
transverse} to the drive force ${\bf F}_L$. In view of the symmetry
of the traps, for currents perpendicular to the antidot rows, no
transverse fluxon current was observed; as shown in Fig.
\ref{Fvortex5}, changing the sign of $\gamma$ led to a inversion of
the transverse current. Most notably, the detected drift was not
restricted by a current threshold, that is, fluxons seemed not to
jump from antidot to antidot, like individual vortices in a
defective superconductor sample (Sec. \ref{asymmetricpattern}), but
rather obey an Ohmic behavior. The absence of an activation
threshold was explained in \textcite{wordenweber2004PRB69} by
invoking a combination of (i) finite size effects, as the area where
vortices can be nucleated is severely restricted by the device
geometry; (ii) screening effects, as trapped fluxons induce
spatially non-uniform current distributions around neighboring
antidots. Due to current screening, the antidot rows serve as
easy-flow (i.e., threshold-less) paths for the driven vortices,
which thus acquire a transverse velocity proportional to $\sin
2\gamma$ (see Fig. \ref{Fvortex5}, inset and main panel).

At higher temperatures and stronger drives, this effect becomes
negligible and no transverse rectification was detected
\cite{wordenweber2004PRB69}. Moreover, on replacing $I_{dc}$ with a
symmetric ac drive $I_{ac}(t)$, no net current is expected, neither
longitudinal nor transversal.

\begin{figure}[htb]
\centering
\includegraphics[width=7.5cm]{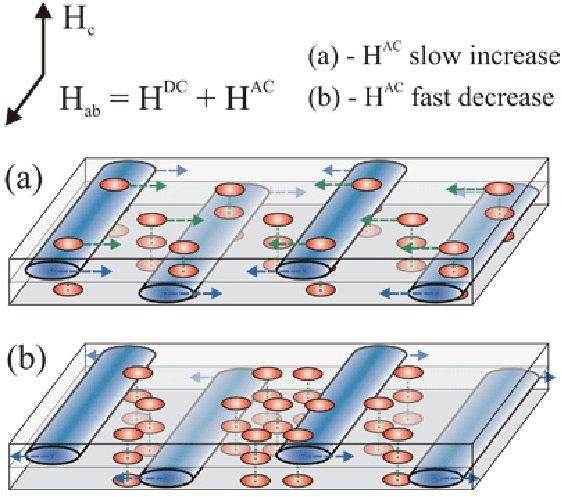}
\caption{(Color online) Vortex lens. Subjected to a superposition of
dc and time asymmetric ac in-plane magnetic fields, $H_{||}(t) =
H_{||}^{dc}+ H_{||}^{ac}(t)$, the JV (horizontal cylinders) are
asymmetrically pushed in and out of the sample. If the ac component
of $H_{||}(t)$ increases slowly, the PV stacks (pancakes) remain
trapped on the JV, and both move together towards the center of the
lens [panel (a)]. If, on the way back, $H_{||}^{ac}(t)$ decreases
rapidly, the JV leave the PV behind them [(panel (b)]. Courtesy of
Sergey Savel'ev.}\label{Fvortex6}
\end{figure}

\subsection{Anisotropic fluxon rectifiers} \label{anisotropic}

Transport control in a binary mixture can be achieved in samples
with no ratchet substrate. As reported in Sec. \ref{binarymixture},
when one of the components of the binary mixture is driven, the
moving particles drag along the other non-driven component,
interacting with it. A time asymmetric ac drive produces
rectification of both components of the binary mixture, which can be
tuned by means of the ac drive parameters. The device acts like a
ratchet, but it has no ratchet substrate.

\textcite{savelev2002NatMat1} proposed to implement this new ratchet
concept in a layered superconductor. In this class of materials, a
magnetic field, tilted away from the high-symmetry crystalline
$c$-axis, penetrates the sample as two perpendicular vortex arrays
\cite{ooi1999PRL82,matsuda2001Science294,grigorenko2001Nature414},
known as ``crossing" vortex lattices \cite{koshelev1999PRL83}. One
vortex sublattice consists of stacks of pancake vortices (PV)
aligned along the $c$-axis, whereas the other is formed by Josephson
vortices (JV) confined between CuO$_2$ layers (Fig. \ref{Fvortex6}).
A weak attractive PV-JV interaction has been experimentally observed
\cite{grigorenko2001Nature414}. The JV are usually very weakly
pinned and can be easily driven by changing either the in-plane
magnetic field, $H_{||}$, or applying an electrical current, $I_z$,
flowing along the $c$-axis.

Time-asymmetrically driven JV can drag along the PV, resulting in a
net motion of the PV, as in the PV lens device illustrated
schematically in Fig. \ref{Fvortex6}. The simplest possible
operating mode consists of slowly increasing the in-plane field,
$H_{||}$, from $0$ to $H^{\rm max}_{||}$ for a fixed value of the
out-of-plane magnetic field, $H_z$. The increasing in-plane field
slowly drives the JV from the edges to the middle of the sample. In
turn, the JV drag the PV along with them towards the sample center.
As a result, asymmetrically cycling causes either pumping (focusing)
or antipumping (defocusing) of the PV at the center of the lens.

\begin{figure}[ht]
\centering
\includegraphics[width=8.0cm]{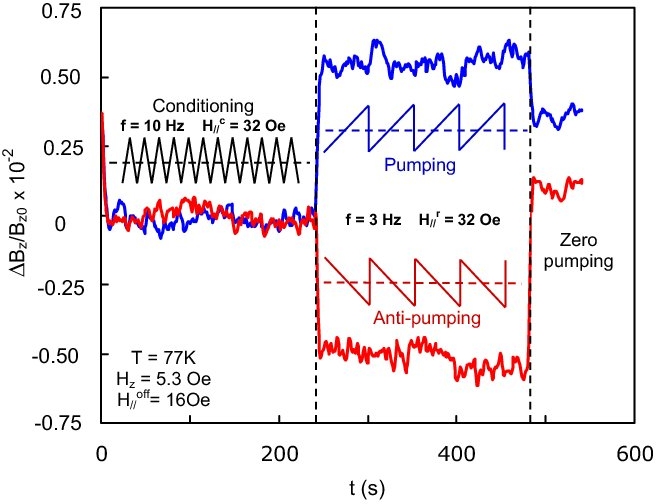}
\caption{(Color online) Vortex lens operation. The curves show the
measured percentage change of the magnetic induction (PV density) at
the sample's center, when applying an initial 'conditioning' signal
followed by pumping/antipumping time-asymmetric drives. The
conditioned PV density increases (decreases) during several cycles
of the pumping (antipumping) drive and then saturates. As soon as
the drive is switched off, the PV density starts to relax from the
non-equilibrium pumping (antipumping) state towards an equilibrium
state. Courtesy of Sergey Savel'ev.}\label{Fvortex7}
\end{figure}

\textcite{cole2006NatMat5} actually performed a vortex lensing
experiment on an as-grown single Bi2212 crystal. The changes in
magnetic induction, arising from PV lensing/anti-lensing, were
detected using one centrally placed element of a micro-Hall probe
array \cite{altshuler2004RMP76}. In order to realize the asymmetric
ac driven mode for the vortex lens, the following steps sketched in
Fig. \ref{Fvortex7} were carried out: $(1)$ The sample was cooled in
fixed $H_z$ at $H_{||}=0$; $(2)$ A ``conditioning" triangular wave
was run for 4 min to equilibrate the PV system; $(3)$ A
time-asymmetric, zero-mean ac drive, was switched on for 4 min (for
an animation see http:$\\$dml.riken.go.jp). The magnetic induction,
$B_z$, was then monitored in real time starting from step (2) at a
centrally located Hall element and then related to the local PV
density. The measured efficiency of this vortex lens, displayed in
Fig. \ref{Fvortex7}, is quite low. However, the same experiment,
when performed on a symmetric substrate, yields a much higher
efficiency, as recently proven by \textcite{cole2006NatMat5} for a
film of Bi2212 patterned with a square array of circular antidots.

The two main advantages of this class of vortex devices over the
earlier proposals reviewed in Secs. \ref{fluxonchannel} and
\ref{fluxon2Darrays} are: $(1)$ the possibility to guide particles
with no tailored spatial asymmetry; $(2)$ the tuning of the vortex
motion by simply changing the parameters of the externally applied
drive. The first feature allows to avoid expensive and cumbersome
nanofabrication processing; the second property becomes very
important if the transport properties of a device must be frequently
varied, something which is generally hard to achieve in standard
ratchet devices.

\section{QUANTUM DEVICES} \label{quantumdevice}

In the foregoing sections, we focused on directed transport within
the realm of classical dynamical laws and classical statistical
fluctuations. Totally new scenarios open up when we try to
consistently incorporate quantum mechanical laws in the operation of
an artificial Brownian motor.  The quantum nature of fluctuations
and the quantum evolution laws, as governed by quantum statistical
mechanics, allow for unexpected transport mechanisms. In particular,
quantum mechanics provides the doorway to new features such as
under-barrier(-threshold) tunneling, above barrier-reflection and
the interplay of coherent (i.e. with oscillatory relaxation) and
incoherent (i.e. with overdamped relaxation) quantum transport. Last
but not least, quantum mechanics generates the possibility for
non-classical correlations, including entanglement among quantum
states in presence of coupling.

\subsection{Quantum dissipative Brownian transport} \label{quantumdissipation}

To set the stage and in order to elucidate the complexity  involving
directed quantum Brownian transport, we consider a 1D quantum
particle of coordinate $q$ and mass $m$, moving in a typically
time-dependent rachet-like potential landscape $V_{R}(q,t)$. The
particle  is bi-linearly coupled with strength $c_i$ to a set of $N$
harmonic oscillators $x_i$, with $i=1,\dots,N$. The latter set of
oscillators models the heat bath, with the oscillators being
prepared in a canonical state with density matrix
$\hat\rho_\text{bath}$, corresponding to an isolated, bare bath
Hamiltonian and  a temperature $kT$. Accordingly, the total dynamics
is governed by the Hamiltonian:
\begin{align}
\label{eq:hamiltonian}
H&=\frac{p^2}{2m}+V_R(q,t)\\
&\qquad+\sum_{i=1}^N\left[\frac{p_i^2}{2m_i}+
\frac{m_i}{2}\omega_i^2x_i^2-qc_ix_i+q^2\frac{c_i^2}{2m_i\omega_i^2}
 \right]\nonumber \;.
\end{align}

The last term, which depends only on the system coordinate,
represents a potential renormalization term which is needed to
ensure that the potential $V_R(q,t)$ coincides with the bare ratchet
potential also in presence of coupling to the bath degrees of
freedom. This Hamiltonian has been studied since the early Sixties
for systems which are weakly coupled to their environment. Only in
the Eighties it was realized by Caldeira and Leggett
\cite{caldeira1983AnnPhys149,caldeira1984AnnPhys153} that this model
is also applicable to strongly damped systems and may be employed to
describe, for example, dissipative tunneling in solid state and
chemical physics \cite{hanggi1990RMP62}.

One may convince oneself that the Hamiltonian in Eq.
(\ref{eq:hamiltonian}) models indeed dissipation. Making use of the
solution of the Heisenberg equations of motion for the external
degrees of freedom, one derives a reduced system operator-valued
equation of motion, the so-called  inertial \textit{generalized
quantum} Langevin equation, that is

\begin{align}
&m\ddot q (t) + m\int_{t_0}^{t} d s\gamma(t-s)\dot q(s) \nonumber \\
&\qquad\qquad + \frac{dV_R(q,t)}{dq} = \eta(t) -
m\gamma(t-t_0)q(t_0). \label{eq:QLE}
\end{align}
The friction kernel is given by
\begin{equation}
\gamma(t)= \gamma
(-t)=\frac{1}{m}\sum_{i=1}^N\frac{c_i^2}{m_i\omega_i^2}\cos(\omega_i
t), \label{eq:dampingkernel}
\end{equation}
and the quantum Brownian force operator reads
\begin{align}
\label{eq:xi} \eta(t)&= \sum_{i=1}^Nc_i\bigg(x_i(t_0)\cos(\omega_i [t-t_0])\\
&\qquad\qquad\qquad+\frac{p_i(t_0)}{m_i\omega_i}\sin(\omega_i[t-t_0])\bigg),
\nonumber
\end{align}

The random force $\eta(t)$ is a stationary Gaussian operator noise
with correlation functions
\begin{align}
\label{eq:noiseaverage}
\langle \eta(t) \rangle_{\hat\rho_\text{bath}}&= 0 \\
\label{eq:noisecorrelation} S_{\eta\eta}(t-s)&=
\frac{1}{2}\langle \eta(t)\eta(s)+\eta(s)\eta(t)\rangle_{\hat\rho_\text{bath}}\\
&=\frac{\hbar}{2}\sum_{i=1}^N\frac{c_i^2}{m_i\omega_i}
\cos\big(\omega_i(t-s)\big)\coth\left(\frac{\hbar\omega_i}{2kT}\right).
\nonumber
\end{align}
Moreover, being an operator-valued noise, the $\eta$ commutators do
not vanish, i.e.,
\begin{equation}
\label{noisecommutator} [\eta(t),\eta(s)] = -i\hbar
\sum_{i=1}^N\frac{c_i^2}{m_i\omega_i} \sin\big(\omega_i(t-s)\big)\,.
\end{equation}

This complex structure for driven quantum Brownian motion in a
potential landscape follows the fact that a consistent description
of quantum dissipation necessarily requires the study of the
dynamics in the {\em full} Hilbert space of the system plus
environment. This is in clear contrast to the classical models,
where the Langevin dynamics is directly formulated in the system
state space \cite{hanggiingold2005Chaos15}. Caution applies in
making approximations to this structure, even if done
semi-classically. The interplay of quantum noise with the commutator
structure is necessary to avoid inconsistencies with the
thermodynamic laws, such as a spurious finite directed current in an
equilibrium quantum ratchet, i.e., even in absence of rocking, i.e.,
for $V_R(q,t) = V_R(q)$
\cite{hanggiingold2005Chaos15,machura2004PRE70b}. Simplifications
are possible only under specific circumstances like in the case of
very strong friction, when quantum corrections can consistently be
accounted by a semiclassical quantum Smoluchowski equation operating
on the state space of the classical system, only
\cite{ankerhold2001PRL87, machura2004PRE70b,machura2007Acta38}. The
situation is presently less settled for weakly damped quantum
Brownian motors  \cite{carlo2005PRL94,denisov2008arxiv} and for
systems at high temperatures, where reduced descriptions are often
in conflict with the laws of thermodynamics; this is true in
particular for the second law, central to this review
\cite{hanggiingold2005Chaos15,zueco2005PhysicaE29}.

An alternative approach to the quantum Langevin description, which
in addition allows powerful computational methods, is based on the
real time path integral technique. Starting from the quantum
statistical representation of the density operator evolution of the
total dynamics of a system coupled to its environment(s), one traces
over the bath degrees of freedom to end up with a path integral
representation for the reduced density operator, the so-called
influence functional, which consistently incorporates quantum
dissipation (see, for instance,
\cite{hanggi1990RMP62,grifoni1998PR304,hanggiingold2005Chaos15,kohler2005PhysRep406}).

\subsection{Josephson Brownian motors} \label{JosephsonBM}
In the following we consider the role of quantum effects on
artificial Brownian motors made of coupled Josephson junctions.

\begin{figure}
\includegraphics[width=8.0truecm]{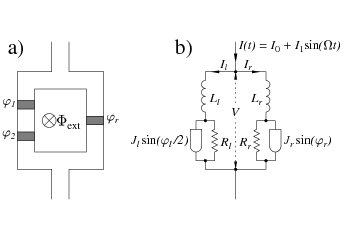}
\caption {(a) Schematic picture of an asymmetric SQUID with three
junctions threaded by an external magnetic flux. (b) Equivalent
electric circuit: the two junctions in series of the left branch
behave like a single junction with $\varphi$ replaced by
$\varphi/2$. For further details see \textcite{zapata1996PRL77}.}
\label{FSquidscheme}
\end{figure}

\subsubsection{Classical regime} 
\begin{figure}
\includegraphics[width=8.0truecm]{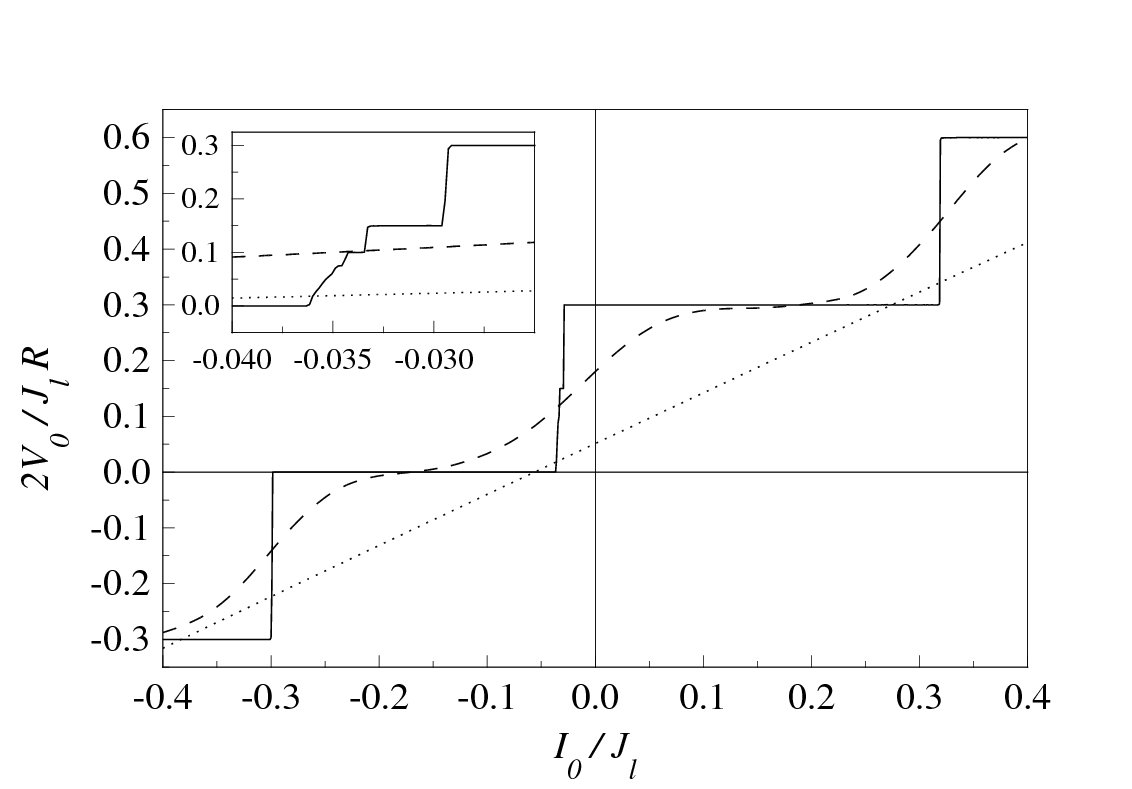}
\caption {Theoretically evaluated dc~current-voltage  (Josephson
relation: $V_0=(\hbar/2e) \left\langle\dot \varphi(t)\right\rangle$)
characteristics for the SQUID ratchet of Fig. \ref{FSquidscheme}.
Simulation parameters: (non-adiabatic) rocking frequency $\omega = 2
\hbar\Omega/eRJ_l = 0.3$, ac drive amplitude $A =I_0/J_l = 1.7$, and
different noise levels: $D = ek_B T/ \hbar J_l= 0$ (solid), $0.01$
(dashed), and $0.5$ (dotted). Inset: magnified plot showing
characteristics steps at fractional values of $\omega$ for $D=0$.
Here, $R$ and $J_l$ denote the resistance and the critical  current
amplitude  of the two identical Josephson junctions placed in the
left branch of the device, cf. in Fig. \ref{FSquidscheme}.}
\label{FSquid}
\end{figure}

A first case refers to the limit
when quantum coherence and tunneling events, such as photon-assisted
tunneling, can safely be neglected, which is often the case at
sufficiently high temperatures. In this regime the description of
Josephson quantum systems can well be approximated by an effective
Fokker-Planck dynamics in the context of the Stewart-McCumber model
\cite{barone1982}, which includes effective parameters containing
$\hbar$. An artificial Brownian motor then can be experimentally
realized for instance with the asymmetric superconducting quantum
interference device (SQUID) illustrated in Fig. \ref{FSquidscheme}
\cite{zapata1996PRL77}. In the overdamped regime, where capacitative
effects can be ignored, such a device maps precisely onto the rocked
ratchet of Sec. \ref{rockedratchet}. Under such operating
conditions, the phase of the device is a classical variable which
can be adequately described by the aforementioned ``resistively
shunted junction'' model. Thermal Brownian motion at temperature $T$
is included by adding Nyquist noise.  The behavior of the so induced
rachet voltage is displayed in Fig. \ref{FSquid} for a
non-adiabatically rocked SQUID-ratchet \cite{zapata1996PRL77}.

The results for this ratchet setup have been experimentally
validated by use of high temperature superconducting dc SQUID's by
the T\"ubingen group
\cite{weiss2000EPL51,sterck2002ApplPhysA75,sterck2005PRL95}. In the
meantime, several variants of this scheme have been studied, both
theoretically and experimentally, including fluxon ratchets of
various designs
\cite{zapata1998PRL80,Carapella2001PRL87,berger2004PRB70,falo2002ApplPhysA75,lee2003APL83,shalom2005PRL94,beck2005PRL95};
cf. also Sec. \ref{superconductingdevice}. Moreover, transport of
fluxons in an extended, annular Josephson junction has been
demonstrated by \textcite{ustinov2004PRL93} also as a harmonic
mixing effect.

\subsubsection{Quantum regime}

\begin{figure}
\includegraphics[angle=90,width=8.0truecm]{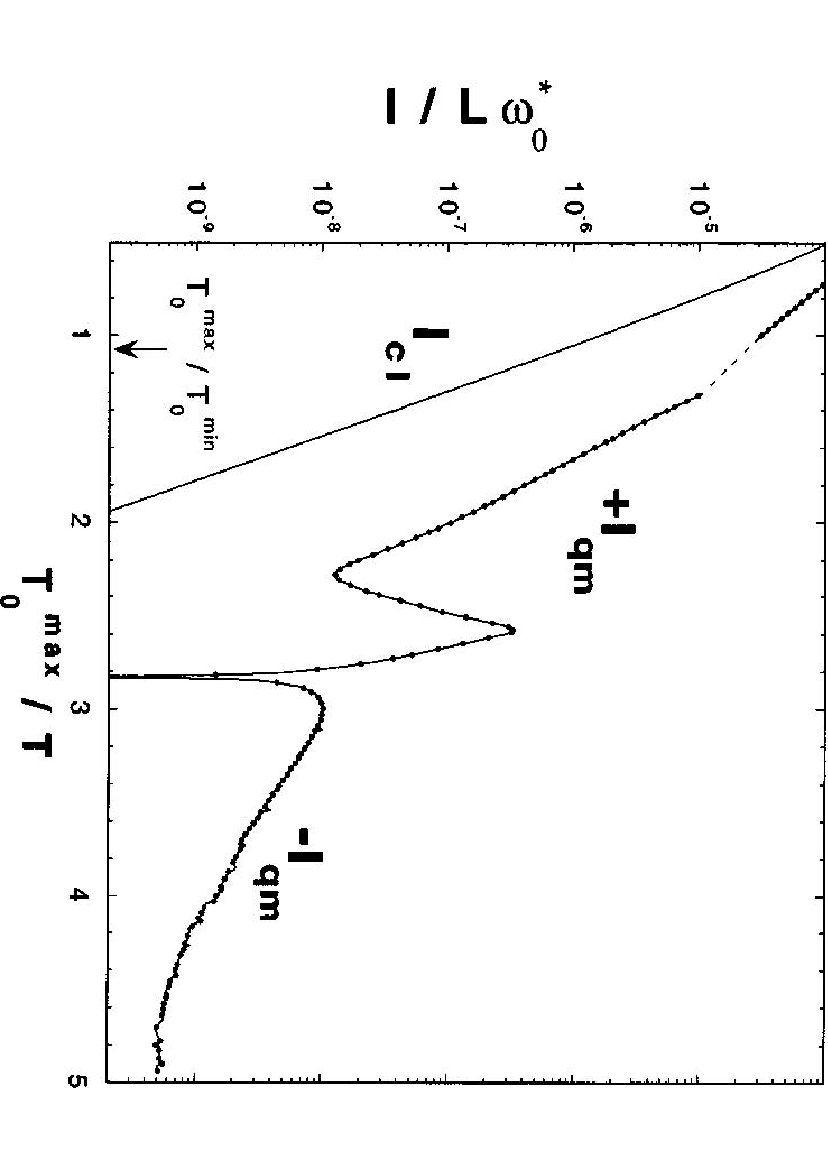}
\caption {The classical Brownian motor current $I_{cl}$ occurring in
a dichotomously rocked ratchet potential, Eq. (\ref{doublesine}), of
period $L$, is compared with the corresponding quantum Brownian
motor current $I_{qm}$ versus dimensionless inverse temperature $T$.
The current is measured in units of $ [L\omega_0^{*}]$ with
$\omega_0^{*}= [4\pi^2 V_0/(L^2m)]^{\frac{1}{2}}$  and temperature
is measured in units of the (maximal) crossover temperature
$T_{0}^{max}$ occurring in the rocked potential landscape. Notably,
in clear contrast to an adiabatically rocked classical Brownian
motor, the quantum current changes sign near $T_0^{max}/T = 2.8$
which is a manifestation of true quantum tunneling.
For more details we refer to the original work by
\textcite{reimann1998Chaos8}.} \label{FQR}
\end{figure}

To tackle  true quantum effects, including quantum tunneling
phenomena, we must resort to the theoretical scheme outlined in
Sect. \ref{quantumdissipation}. A first theoretical study of quantum
ratchets has been pioneered by \textcite{reimann1997PRL79} (see also
\cite{reimann1998Chaos8}). For a regime with several (quasi)-bound
states below a potential barrier one can evaluate the ratchet
tunneling dynamics in terms of an effective action for the extremal
bounce-solution (in combination with a fluctuation analysis around
this bounce solution) to obtain the corresponding tunneling rates.
In Fig. \ref{FQR} the result of such a quantum calculation is
compared with the classical result for an adiabatically rocked
quantum particle of mass $m$ and coordinate $q$ dwelling in a
ratchet potential $V_R(q)$. Note that within a quantum Brownian
motor scheme the role of mass $m$, i.e. the inertia, does enter the
analysis  explicitly. The most distinctive quantum signature is a
striking current reversal which emerges solely as a consequence of
quantum tunneling under adiabatic rocking conditions. Moreover, in
contrast to the classical result, the directed current no longer
vanishes as $T$ tends to zero and additional resonance-like features
show up.

\begin{figure}
\includegraphics[angle=270,width=9.0truecm]{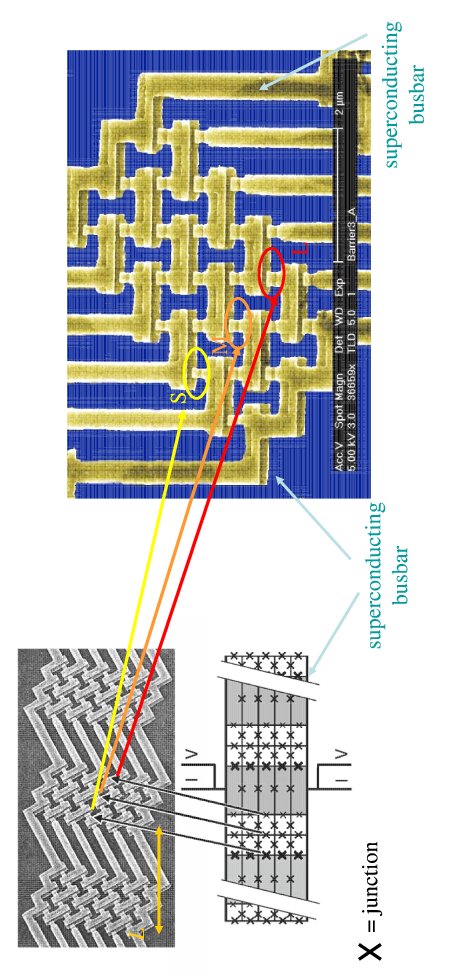}
\caption {(Color online) Top left: Scanning electron picture (with
enlargement on the right) of a strongly asymmetric array of a long,
narrow network of Josephson junctions arranged in a rectangular
lattice. This arrangement determines the potential shape felt by
vortices. Bottom left: Schematic layout. The Josephson junctions are
represented by crosses, each network cell being delimited by four
junctions. All arrays have a length of $303$ cells and a width of
$5$ cells, placed between solid superconducting electrodes
(busbars). The vortices assume a lower energy in cells with larger
area and smaller junctions. Figure provided by Milena Grifoni,
University of Regensburg, Germany.} \label{Fquantummajer}
\end{figure}

Such a dependence of the current versus decreasing temperature in a
quantum Brownian motor has been corroborated experimentally for
vortices moving in a quasi-1D Josephson junction array with ratchet
potential profile specially tailored so as to allow several bands
below the barrier \cite{majer2003PRL90}. The experimental setup of
this quantum ratchet device is displayed in Fig.
\ref{Fquantummajer}. The experimental findings are in good agreement
with the theoretical analysis reported by
\textcite{grifoni2002PRL89}.

\subsection{Quantum dot ratchets} \label{QuantumdotQR}

\begin{figure}[htb]
\includegraphics[width=6.0truecm]{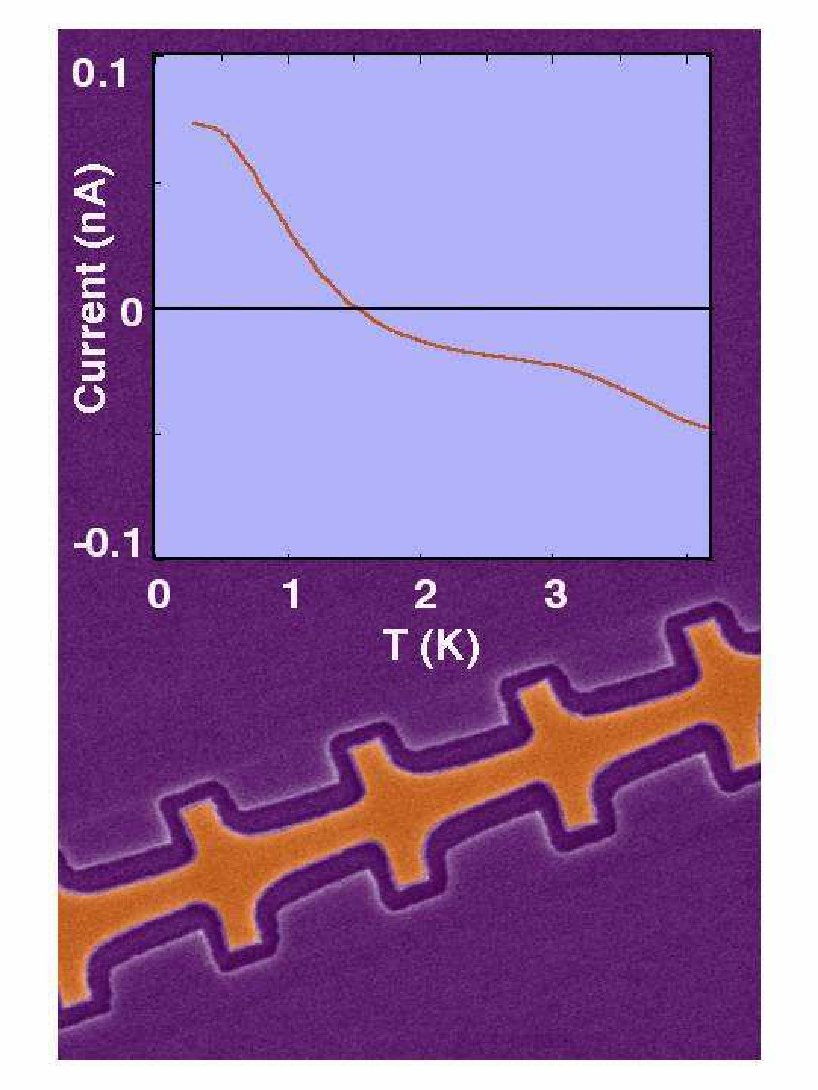}
\centering \hfil \caption{(Color online) In an experimental {\it
quantum Brownian motor} driven by an adiabatic square-wave zero-mean
rocking voltage, quantum tunneling can contribute to the electron
current. Due to the underlying asymmetric potential structure, the
two opposite components to the time-averaged driven current differ
in magnitude, yielding  a net quantum ratchet current
\cite{reimann1997PRL79,linke1999Science286}. The magnitudes can be
individually controlled by tuning the temperature. This in turn
causes a tunneling induced current reversal (occurring near $1.5$\/K
in the top graph) that can be exploited to direct electrons along
{\em a priori} designed routes \cite{linke1999Science286}. Below the
current versus T graph is a scanning electron micrograph of the
relevant quantum device. Figure provided by Heiner Linke, University
of Oregon.} \label{Fquantumlinke}
\end{figure}

Another ideal resource to observe the interplay of (i) quantum
Brownian motion, (ii) quantum dissipation, and (iii) non-equilibrium
driving are semiconductor engineered quantum rectifiers. Composed of
arrays of asymmetric quantum rectifiers \cite{linke1998EPL44}, these
devices operate on geometric and dynamical length scales ranging
between a few nanometers up to micrometers. This class of devices
allowed the first experimental validation of the distinctive
features of quantum rectifiers, namely, tunneling enhanced rachet
current and tunneling induced current reversals, by Linke and
collaborators \cite{linke1999Science286,linke2002ApplPhysA75b}.
Their central results are illustrated in Fig. \ref{Fquantumlinke}.

Their quantum Brownian motor consisted of a 2D gas of electrons
moving at the interface of a fabricated, ratchet-tailored
GaAs/AlGaAs heterostructure. Its operating regime was achieved by
adiabatically switching on and off a square wave source-drain
periodic voltage of frequency $190$Hz and amplitude of about $1$mV.
Recalling that for an adiabatically rocked classical ratchet of Sec.
\ref{rockedratchet}, noise induced transport exhibits no current
reversals, things change drastically when quantum tunnelling enters
into the dynamics. A true benchmark for the quantum behavior of an
adiabatically rocked Brownian motor is then the occurrence of a
tunneling induced reversal at low temperatures
\cite{reimann1997PRL79}. This characteristic feature has been first
experimentally verified with an electron quantum rocked ratchet by
\textcite{linke1999Science286}. Moreover, the current reversal
reported in the top panel of Fig. \ref{Fquantumlinke} indicates the
existence of parameter configurations where the current of a quantum
Brownian motor vanishes. In correspondence to such configurations,
one can imagine to operate the device as a quantum refrigerator to
separate ``cold'' from ``hot'' electrons in the absence of currents
\cite{linke2002ApplPhysA75b}. At higher temperatures this and other
asymmetric quantum-dot arrays, when subjected to unbiased ac drives,
exhibit incoherent quantum ratchet currents. Experimental such
realizations using two-dimensional electron systems with broken
spatial inversion symmetry are reported in Refs.
\cite{lorke1998PhysicaB251,vidan2004APL85,sassine2008PRB78}. An even
richer behavior of the quantum current, including for example
multiple current reversals, emerges when this class of devices is
operated in the {\it non-adiabatic} ac regime, as revealed by recent
theoretical studies \cite{
goychuk1998PRL81,goychuk1998PRL81b,goychuk1998EPL43,grifoni2002PRL89,kohler2005PhysRep406,goychuk2005AdvPhys54}.

\begin{figure}[htb]
\includegraphics[width=7.0truecm]{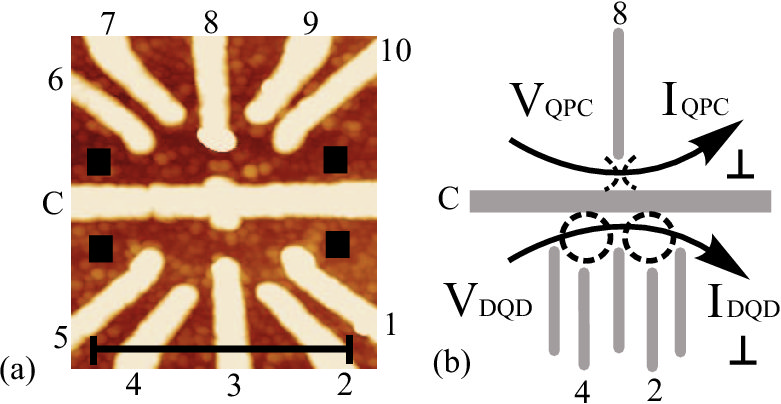}
\centering \hfil \caption{(Color online) (a) AFM micrograph of a
double quantum dot device. Metal surface gates have a light color,
black squares mark source and drain regions. The black scale bar
marks the length of $1\mu m$. (b) Schematic diagram: A biased
quantum point contact (QPC) provides the nonequilibrium fluctuation
source for driving a tunneling current $I_{\rm DQD}$ in the
asymmetric double quantum dot (DQD) (dashed two circles)
\cite{khrapai2006PRL97}. The asymmetry is induced via gate voltages
at the plunger gates $2$ and $4$. Figure provided by  Vadim S.
Khrapai, Ludwig-Maximilian University, Munich, Germany.}
\label{Fquantumkotthaus1}
\end{figure}

\begin{figure}[htb]
\includegraphics[width=7.0truecm]{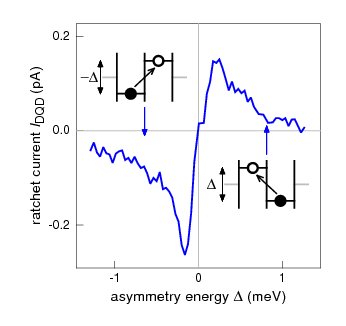}
\centering \hfil \caption{(Color online) Experimental quantum
ratchet current $I_{\rm DQD}$ measured through the double quantum
dot as a function of its asymmetry energy $\Delta$. The energy
source is a nearby quantum point contact biased with $V_{\rm QPC} =
-1.55\,$mV, see in Fig. \ref{Fquantumkotthaus1}. An elastic
contribution to $I_{\rm DQD}$ is subtracted \cite{khrapai2006PRL97}.
The two insets sketch the corresponding inelastic tunneling
processes which drive the ratchet current. Notably, this quantum
ratchet current vanishes when no finite zero asymmetry is present.
Figure provided by Stefan Ludwig, Ludwig-Maximilians-University,
Munich, Germany. } \label{Fquantumkotthaus2}
\end{figure}

As a second example of an artificial quantum dot Brownian motor, we
consider an experimental double quantum dot device where both dots
are independently coupled to a nonequilibrium energy source which is
given by a biased quantum point contact. Moreover, both quantum dots
are embedded in independent electric circuits via a common central
gate, marked by a ``C'' in the sketch of Fig.
\ref{Fquantumkotthaus1}. On breaking its internal symmetry by tuning
the dot gate voltages, this double quantum dot acts as a quantum
ratchet device \cite{khrapai2006PRL97}. For weak interdot tunneling,
detuning of the quantum dot energy levels results in the
localization of an electron in one dot, so that elastic electron
transfer to the other dot is energetically forbidden. To operate as
a quantum ratchet, the device still requires a nonequilibrium energy
source. To this purpose, a strong tunable bias on the quantum point
contact induces nonequilibrium energy fluctuations across the
dividing central gate, thus promoting inelastic interdot tunneling
inside the Coulomb blockaded double quantum dot. This in turn leads
to a net quantum ratchet current flow, as plotted in Fig.
\ref{Fquantumkotthaus2}. In the insets of the same figure, the
interdot tunneling process is sketched for the right-to-left
transition $[m,n+1] \longrightarrow [m+1,n]$, with asymmetry energy
$\Delta \equiv E_{m+1,n} - E_{m,n+1}$, and for the opposite
left-to-right transition with energy $-\Delta$. Notably, a finite
ratchet current is only observed
if the electron energy states in the dots are  detuned
asymmetrically, i.e. when $\Delta \neq 0$. In contrast, a likely
inelastic ionization of one dot towards its adjacent lead, followed
by recharging from the same lead, does not result in a net current.
The nonequilibrium energy fluctuations, carried by the quantum point
contact electrons and absorbed by the electrons in the double
quantum dot, could either consist of acoustic phonons, long
wavelength photons or plasmons \cite{khrapai2006PRL97}.

The technology available to generate 2D electron gases can be
generalized to ratchet not only charge but also spin carriers. The
interplay of spatially periodic potentials, lateral confinement,
spin-orbit or Zeeman-type interactions and ac driving then gives
rise to directed {\it spin ratchet} currents, as theoretically
proposed in recent studies \cite{scheid2007PRB76,scheid2007NJP9}. Of
particular experimental relevance is the recent theoretically
predicted phenomenon of a spin current which emerges via an unbiased
ac charge current driving a dc spin current in a non-magnetic,
dissipative spin quantum ratchet which is composed of an asymmetric
periodic structure with Rashba spin-orbit interactions
\cite{smirnov2008PRL100,flatte2008NaturePhys4}. Remarkably, this
spin current occurs although {\it no}  magnetic fields are present.

Likewise, substantial rectification of a spin current can also be
achieved  by coupling impurities to spatially asymmetric Luttinger
liquids under ac voltage rocking \cite{braunecker2007PRB76}. The
transport mechanism in those schemes is governed by coherent
tunneling processes, which will be addressed in more detail in the
following subsection.

\subsection{Coherent quantum ratchets}

The study of quantum Brownian motors is far from being complete.
Actually, there is an urgent need for further developments. In many
cases the role of incoherent tunneling and coherent transport
channels can not be clearly separated. For quantum devices  built
bottom-up with engineered molecules or artificial molecules such as
quantum dots, the role of coherent quantum transport typically plays
a dominant role, as other interactions like electron-phonon
processes turn out to be negligible. The key transport process  for
quantum ratcheting is then driven coherent quantum tunneling
\cite{grifoni1998PR304,kohler2005PhysRep406}.

\subsubsection{Quantum ratchets from molecular wires} 
\begin{figure}[btp]
\includegraphics[width=7.0truecm]{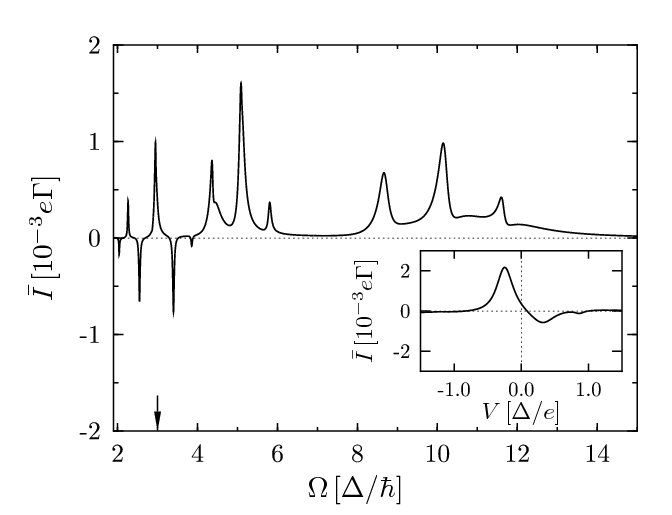}
\centering \hfil \caption{Time-averaged current $\overline I$
(measured in units of the lead coupling strength $ e\Gamma$) as a
function of the infrared drive angular frequency $\Omega$ (measured
in units of the tunneling matrix element $\Delta= 0.1 eV$ between
sites for a laser amplitude $A=\Delta$ \cite{lehmann2002PRL88}. The
inset displays the dependence of $\overline I$ on the static voltage
drop, $V$, applied along the molecule. Note that the current at zero
voltage is {\it finite}, thereby indicating a ratchet effect. The
driving frequency here corresponds to the vertical arrow in the main
panel.} \label{Fquantumlehmann}
\end{figure}

This is the
case of the quantum Brownian motors consisting of nanowires formed
by asymmetric molecular groups and subjected to infrared light
sources \cite{lehmann2002PRL88}, or symmetric molecular wires with
symmetry breaking now provided by irradiating the wire with harmonic
mixing signals \cite{lehmann2003JCP118}. The theoretically predicted
quantum ratchet current through an asymmetric molecular wire
irradiated by far-infrared light is displayed in Fig.
\ref{Fquantumlehmann}.

In these class of quantum ratchets transport proceeds coherently
between two or more fermionic leads that ensure the dissipative
mechanism for the transported electrons to relax towards equilibrium
Fermi distributions.
%
Such a ratchet current can as well be used to sensitively probe the
role of the electron correlations in the leads, as it is the case
with a ratchet device coupled to Luttinger liquids rather than to
Fermi liquids \cite{komnik2003PRB68,feldman2005PRL94}.

\subsubsection{Hamiltonian quantum ratchet for cold atoms}

Much to one's surprise, systems governed by strictly Hamiltonian
dynamics are able to yield directed transport, as well. Directed
transport may then take place even though {\it no} dissipation
mechanisms are present. Trivial such examples are provided by
integrable regular dynamics subjected to unbiased ac-drives with
{\it fixed initial phase}, i.e., not to be averaged over
\cite{yevtushenko2000PRE61,goychuk2001JPCB105}. Much more intriguing
are time-dependent driven Hamiltonian quantum Brownian motors
exhibiting a nontrivial, mixed classical phase space structure
\cite{flach2000PRL84,denisov2001PRE64,schanz2001PRL87,
schanz2005PRE71}. In these systems, the onset of a directed flow
requires, apart from the necessity of  breaking  time reversal
symmetry, also an additional dynamical symmetry breaking. As it
turns out, in the semi-classical limit, the corresponding directed
flow then obeys a remarkable sum rule: Directed currents occurring
in regular regimes of the underlying phase space dynamics are
counter-balanced by directed flows occurring in chaotic regimes in
phase space \cite{schanz2001PRL87, schanz2005PRE71}. As a result, if
those individual directed currents are summed up over all disjoint
regimes in (semi)-classical mixed phase space, no net transport
emerges.

The concept of Hamiltonian quantum Brownian motors  extends as well
to fully quantum systems governed by a unitary time-evolution
\cite{goychuk2000LNP557,schanz2001PRL87,goychuk2001JPCB105,
schanz2005PRE71,denisov2006EPL74,denisov2007PRA75,gong2007PRA75}. In
fact, the issue of pure quantum coherence in the directed transport
of chaotic Hamiltonian systems is presently a topic of active
research. This in particular holds true  for cold atoms loaded in
optical lattices: If properly detuned, the intrinsic quantum
dynamics is then practically dissipation-free, thus providing a
paradigm for Hamiltonian quantum ratchet transport. Various
experimental cold atom ratchets have been realized (Sec.
\ref{coldatom}). Relevant to the topic of this section is also the
demonstration of sawtooth-like, asymmetric cold atom potentials by
\textcite{salger2007PRL99}. All these systems are currently being
employed to effectively shuttle cold matter on a quantum scale.

In the context of quantum ratchets for cold atoms one must
distinguish the common case of ``rectification of velocity'',
implying that the mean position of particles is growing linearly in
time, from the case with a ``rectification of force'', i.e. with
mean momentum growing linearly in time. The latter class is better
classified as ``quantum ratchet accelerators''
\cite{gong2004PRE70,gong2006PRL97}. In the context of cold atom
ratchets the first situation is obtained by rocking a cold atom gas
with temporally asymmetric driving forces or temporally
asymmetrically flashing, in combination with spatially asymmetric
potential kicks \cite{denisov2007PRA75}. The accelerator case
originates from the physics of quantum $\delta$-kicked  cold atoms
displaying quantum chaos features such as the quantum suppression of
classical chaotic diffusion (dynamical localization) and the
diametrically opposite phenomenon of {\it quantum resonance}
(occurring when the kicking period is commensurate with the inverse
recoil frequency). Under  quantum resonance condition  a linear
(quadratic) increase of momentum (kinetic energy) takes place. These
ratchet acceleration features, theoretically predicted in suitably
modified kicked rotor models
\cite{lundh2005PRL94,poletti2007PRE75,kenfack2008PRL100}, requires
tuning to {\it exact} resonance. In contrast, the accelerator models
obtained from either a generalization of a quantum kicked rotor or a
generalization of a kicked Harper model
\cite{gong2004PRE70,gong2006PRL97,wang2008PRE78} are {\it generic}
``rectifiers of force'' in the sense that here no need for tuning to
exact resonance is necessary. Interestingly, these fully quantum
ratchet accelerators display unbounded linear growth of mean
momentum, while the underlying classical dynamics is fully chaotic,
a situation where classical quantum Brownian motor transport
necessarily vanishes according to the above mentioned classical sum
rule \cite{schanz2001PRL87, schanz2005PRE71}. Thus, within this full
quantum regime, which carries no clear relationship with the
dynamics in the semi-classical regime, the quantum accelerator work
by \textcite{gong2006PRL97}, see also \cite{wang2008PRE78}, presents
an intriguing and generic quantum mechanical exception to the
classical sum rule.

Early quantum resonance experiments in quantum ratchet have already
been successfully carried out: For a $\delta$-kicked rotor model
with time-symmetry broken by a $2$-period kicking cycle (asymmetric
temporal drive), directed growth of momentum has been detected by
\textcite{jones2007PRL98}. For a phase-dependent initial preparation
of a Bose-Einstein condensate kicked at resonance, a momentum
acceleration has been observed by \textcite{sadgrove2007PRL99} at
zero quasimomentum, while for an arbitrary quasimomentum directed
quantum Brownian transport has been realized in
\textcite{dana2008PRL100}. The latter experiment also evidenced that
an intrinsic experimentally non-avoidable finite width in
quasimomentum causes a suppression of the acceleration eventually
leading to a saturation effect after short times.

This field of Hamiltonian quantum ratchets is presently undergoing a
racy development. For example, it is possible to apply control
schemes to relative phases for resulting single-particle quantum
Brownian motor currents by harvesting Landau-Zener transitions
\cite{moralesmolina2008EPL83}. Of particular theoretical and
experimental interest is also the study of the role of nonlinearity
on the size of directed quantum currents in interacting cold gases,
as described within mean field theory by a nonlinear quantum ratchet
evolution of the Gross-Pitaevskii-type. There, the interplay of
time-dependent driving and {\it nonlinear} Floquet states yields new
features, such as lifting of accidental symmetries
\cite{poletti2007PRA76} and a resonant enhancement of directed
ratchet currents \cite{moralesmolina2008NJP10}.


\section{SUNDRY TOPICS} \label{sundrytopics}
In the following we discuss some classes of nanosystems and devices
which feature directed transport in the spirit of Brownian motors.
In these systems, however, rectification of Brownian motion does not
constitute the main element for directed transport. We recall that
an artificial Brownian motor is mainly noise-controlled, meaning
that such a motor operates
in a hardly predictable manner. In contrast, we discuss next systems
that exhibit directed transport 
predominantly as the result of strong coupling schemes. Typical
examples are adiabatic pump scenarios of the peristaltic type, or
nanosystems which are driven by unbiased, but asymmetric mechanical
or chemical causes that are tightly coupled to resulting motion.


\subsection{Pumping of charge, spin and heat} \label{pumps}
A first class of physical systems that comes to mind in relation to
the working principles of artificial Brownian motors are nanoscale
pump devices. Pumping is characterized by the occurrence of a net
flux of particles, charges, spins and alike, in response to time
dependent external manipulations of an otherwise unbiased system.
This mechanism is well studied and peristaltic pumps are being
widely exploited in technological applications. These systems do not
require a periodic arrangement of components nor is thermal Brownian
motion an issue for their operation. In particular, adiabatic
turnstiles and pumps for charge and other degrees of freedom have
attracted considerable interest both experimentally
\cite{Kouwenhoven1991PRL67,pothier1992EPL17,switkes1999Science283,
hohberger2001APL78} and theoretically
\cite{thouless1983PRB27,spivak1995PRB51,brouwer1998PRB58,zhou1999PRL82,shutenko2000PRB61,
vavilov2001PRB63,brandes2002PRB66,aono2003PRB67,moskalets2004PRB70,sinitsyn2007PRL99}.

Realizations of artificial pumps on the nanoscale often involve
coupled quantum dots or superlattices. Most notably, in such
peristaltic devices the number of transferred charges, or more
generally, the number of transporting units per cycle is directly
linked to the cycle period. As a result the output current thus
becomes proportional to the driving frequency. This observation
leads to the conclusion that high frequency nonadiabatic pumping
might  become more effective. Indeed, nonadiabatic pumping of
charge, spin
\cite{scheid2007NJP9,smirnov2008PRL100,flatte2008NaturePhys4} or
also heat \cite{nianbei2008arxiv} does  exhibit a rich
phenomenology, including resonances
\cite{stafford1996PRL76,moskalets2002PRB66,platero2004PhysRep395,cota2005PRL94,
kohler2005PhysRep406,rey2007PRB76,arrachea2007PRB75} and other
potentially useful noise-induced features
\cite{strass2005PRL95,sanchez2008PhysicaE40}; hence yielding a close
interrelation between {\it nonadiabatic} pumping in the presence of
noise and the physics of Brownian motors.

\begin{figure}[htb]
\includegraphics[width=7.0truecm]{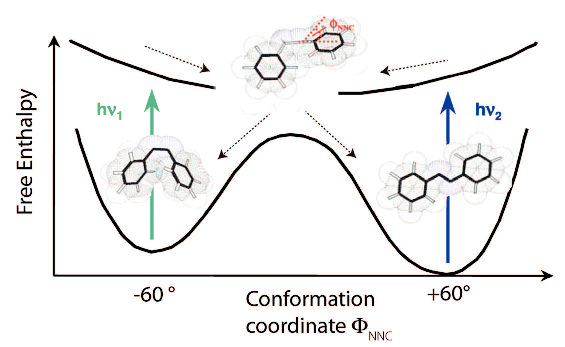}
\caption{(Color online) Free energy landscape of a single azobenzene
molecule along the reaction coordinate of the conformation
coordinate given by the bond angle $\Phi_{NNC}$, which varies from
about $-60^{\circ}$ to $\sim +60^{\circ}$. Transitions can be
induced optically ($\lambda_1 \sim 420$ nm for the short {\it
cis}-form ($0.6$ nm) and $\lambda_2 \sim 365$ nm for the extended
{\it trans}-form ($0.9$ nm), respectively) from the singlet ground
state $S_0$ into the excited singlet state $S_1$, from which the
molecule subsequently relaxes fluctuation-driven into either the
{\it cis}-form or the {\it trans}-form. Notably, the {\it
trans}-form is thermally favored. The insets depict the
corresponding conformations with the transition state located near
$\Phi_{NNC} \sim 0^{\circ}$. Figure provided by Thorsten Hugel from
TU-Munich, Germany} \label{FsundryHugelA}
\end{figure}

\subsection{Synthetic molecular motors and machines}
\label{syntheticmotors} As already emphasized in our introduction,
the field of Brownian motors has its roots in  the study and
applications of intracelluar transport in terms of molecular motors
\cite{julicher1997RMP69,reimann2002PhysRep361,
reimann2002ApplPhysA75,astumian2002PhysToday55,wang2002ApplPhysA75,lipowsky2005PhysicaA352}.
These molecular motors function in view of structure and   motility
by use of specialized proteins in living systems.
These biological motor enzymes are fueled by ATP hydrolysis and are
able to efficiently perform mechanical work on the nanoscale inside
biological cell structures
\cite{howard2001Sinauer,schliwa2002Wiley-VCH}.

Closely related to synthetic motors are the intriguing possibilities
of devising DNA-fueled artificial motors: several settings  render
possible to biologically engineer nanomachines which move along {\it
a priori} designed tracks
\cite{yurke2000Nature406,turberfield2003PRL90,yin2004AngChemie43}.

An offspring of this topic is the engineering of nanomachines based
on interlocked molecular species. This has spurred a flurry of new
investigations within the physical biology and the organic and
physical chemistry community, aimed at building bottom-up synthetic
molecular systems which carry out such diverse functions as
molecular switches, molecular rotors and any other kind of molecular
gears.

\begin{figure*}
\includegraphics[width=15.0truecm]{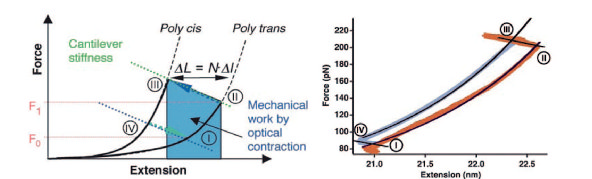}
\caption{(Color online) Operation of a light powered molecular
motor: The panel on the left depicts the schematic force extension
cycle for the opto-mechanical energy conversion cycle of a single
poly-azobenzene. In analogy to a thermodynamic Stirling-cycle, the
polymer is first stretched by the AFM-tip acting as the piston (I
$\longrightarrow$ II). Then, the application of the first optical
excitation with $\lambda_2 \simeq 365$nm {\it shortens} the polymer
by inducing a (poly){\it trans}- (poly){\it cis} transition. This
yields a first part of work (blue area) in bending the AFM
cantilever, which is clearly not as stiff as a piston in a common
Stirling motor. In analogy to the Stirling cycle, another amount of
work is done during the relaxation of the polymer (III $
\longrightarrow$ IV). Finally, a second optical excitation with
$\lambda_1 \simeq 420$nm is needed to reset the molecule into its
starting (poly){\it trans}-state (IV $\longrightarrow $ I). The
total work output of the system is the mechanical energy
corresponding to the contraction ($\Delta L=  N \Delta l$) of the
entire polymer chain of $N$ azobenzenes against the external load.
The experimental realization of a full single molecule operating
cycle \cite{hugel2002Science296} is depicted in the panel on the
right. Figure provided by Thorsten Hugel from TU-Munich, Germany}
\label{FsundryHugelB}
\end{figure*}

A beautiful example of this class of molecular motor is the light
powered molecular machine developed by
\textcite{hugel2002Science296}. Bistable photosensitive azobenzene
molecules can be synthesized into long chains, each containing
molecules in either their {\it trans}- or {\it cis}-form. The free
energy landscape corresponding to different molecule conformational
states is sketched in Fig. \ref{FsundryHugelA} versus the reaction
coordinate. Polymeric chains have the advantage of scaling up the
length changes corresponding to the two different conformations of
its constituents. The length changes of an azobenzene polymeric
chains can then be transformed into mechanical energy by means of
the lever arm of a atom force microscope (AFM) which, in combination
with the light sources, forms the core of the light driven molecular
motor of Fig. \ref{FsundryHugelB}.

Such a molecular motor can be operated very much according to an
idealized thermodynamic cycle of the Stirling type: Individual
azobenzene polymers are made stretch and contract by inducing
optical {\it trans}-{\it cis} transitions of their constituents.
This machine thus demonstrates opto-mechanical energy conversion in
a single-molecule motor device. The analogy with a thermodynamic
cycle should though not be taken too seriously. A thermal Stirling
engine operates between baths of differing temperature; here, the
``expansion'' and ``compression'' of the molecular motor are
performed at a fixed temperature and, most notably, by means of
nonequilibrium light sources. This laser operated molecular motor
can as well be used as a building block to devise a bioinspired
molecular locomotive which can be guided forth and back on a
preassigned track via a laser assisted protocol
\cite{wang2004PRE70}.

In the operation of these synthetic nanoscale devices, especially
when powered by light or chemical reactive additives, thermal noise
does not necessarily play a major role, i.e. the function of these
synthetic molecular motors is ruled predominantly by deterministic
forces, forces that depend on the mechanical and chemical properties
of the molecules \cite{neuert2006Macromolecules39}. Yet, this area
of research is fascinating and we therefore encourage the interested
reader to read some recent items and tutorials
\cite{porto2000PRL84,browne2006NatNanotechn1,
balzani2006CSR35,astumian2007Physchemchemphys9} and to consult the
timely and most comprehensive reviews by
\textcite{kottas2005ChemRev105} and \textcite{kay2007AngewChem46}
for more details.


\section{CONCLUDING REMARKS} \label{conclusions}

With this review we have taken the reader on a tour of horizon
through the many intriguing and multi-facetted applications that
Brownian motion can offer in the most diverse areas of
nanotechnology, when combined with spatio-temporal symmetry
breaking, nonlinearity and, possibly, collective interaction
effects.

The physics of classical and quantum Brownian motion is by now well
established, also in consideration of the breadth of the theoretical
modeling and experimental realizations produced over the last
century. Many research activities are still spawning in
interdisciplinary fields encompassing chemistry, biological
research, information sciences and even are extending into social
sciences and economics. The main lesson to be learned from Robert
Brown's and Albert Einstein's work is therefore: Rather than
fighting thermal motion, we should put it to work to our advantage.
Brownian motors thrive from these ceaseless noise source to
efficiently direct, separate, pump and shuttle degrees of freedom of
differing nature reliably and effectively.

In writing this overview we spared no efforts in covering a wide
range of interesting developments and potential achievements. In
doing so we nevertheless had to make some selective choices of
topics and applications, which to some extent reflect the authors'
preferences and prejudices. Closely related topics of ongoing
research were not reviewed for space limits. For example, in Sec.
\ref{syntheticmotors} we did not cover in sufficient detail the
fascinating topic of ATP-driven molecular motors and DNA-fueled
motors, for which we refer to earlier comprehensive reviews and
books, like in \textcite{julicher1997RMP69},
\textcite{howard2001Sinauer}, \textcite{schliwa2002Wiley-VCH}, and
\textcite{lipowsky2005PhysicaA352}. This surely is the case of the
bottom-up design and operation of synthetic Brownian molecular
devices, extensively covered by the excellent review of
\textcite{kay2007AngewChem46}. Another such topic is the question of
the so-called absolute negative mobility, which occurs via quantum
tunneling events in quantum systems
\cite{keay1995PRL75,aguado1997PRB55,grifoni1998PR304,platero2004PhysRep395}
and through mere nonequilibrium driven diffusive dynamics in
classical systems
\cite{eichhorn2002PRL88,eichhorn2002PRE66,ros2005Nature436,machura2007PRL98,kostur2008PRB77}.

Our main focus was on noise-assisted directional transport,
shuttling and pumping of individual or collective particle-, charge-
or matter-degrees of freedom. The concept, however, extends as well
to the transport of other degrees of freedom such as energy (heat)
and spin modes (see also in Sec.(\ref{pumps}). Both topics are
experiencing a surge of interest with new exciting achievements
being reported in the current literature. The concept carries
potential for yet other applications. Examples that come to mind are
the noise-assisted directional transport and transfer of
informational degrees of freedom such as probability or entropy and,
within a quantum context, the shuttling of entanglement information.
These issues immediately relate to the energetics of artificial
Brownian motors reviewed in Sec. (\ref{efficiency}), including
measures of open and closed loop control scenarios and other
optimization schemes.

We have attempted to catch the potential of Brownian motors in
nowadays nanotechnology by putting them to work; we discussed how
such motors can be constructed and characterized, and how
directional, Brownian motion driven transport can be controlled,
measured, and optimized. Moreover, we are confident that the
interdisciplinary style of this overview shall encourage the readers
to bring in new approaches and motivations in this challenging and
fast growing research area.

\acknowledgements

This review would not have emerged without continuous support from
and insightful discussions with our close collaborators and
colleagues. We first express our gratitude to all members of our
work groups among which special credit should be given to M.
Borromeo, J. Dunkel, I. Goychuk, G.L. Ingold, S. Kohler, G. Schmid,
P. Talkner, and U. Thiele for their scientific contributions and
personal encouragement. A special thanks goes to our many colleagues
and friends who also engaged in the field of Brownian motors and, in
particular, to R.D. Astumian, M. Bier, C. Van den Broeck, J. Casado
Pascual, D. Hennig, W. Ebeling, J.A. Freund, L. Gammaitoni, H.E.
Gaub, E. Goldobin, M. Grifoni, P. Jung, J. K\"arger, I. Kosinska,
J.P. Kotthaus, M. Kostur, H. Linke, B.-W. Li, J. Luczka, S. Ludwig, L. Machura,
V.R. Misko, M. Morillo, F. Nori, J. Prost, P. Reimann, F. Renzoni,
K. Richter, M. Rub\'i, S. Savel'ev, L. Schimansky-Geier, Z. Siwy, B.
Spagnolo, and A. Vulpiani. In addition, we are indebted to M.
Borromeo, M. Grifoni, T. Hugel, H. Linke, S. Ludwig, V.S. Khrapai,
F. M\"uller, S. Savelev, U. Thiele and C. Van den Broeck for
providing us with original figures and unpublished material.
Finally, we wish to thank the senior editor of Reviews of Modern
Physics, Prof. Achim Richter, for the invitation to write the
present review and his most efficient handling of our submission.

F.M. wishes to thank Prof. Hunggyu Park for his kind hospitality at
the Korea Institute for Advanced Study and, likewise, P.H. for the
kind hospitality at the Physics Department of the National
University of Singapore: At these two institutions we were both
given the opportunity to efficiently continue working on a
preliminary version of this review. P.H. acknowledges the financial
support by the Deutsche Forschungsgemeinschaft via the Collaborative
Research Centre SFB-486, project A10, B 13 and by the German
Excellence Cluster {\it Nanosystems Initiative Munich} (NIM). Both
authors thank the Alexander von Humboldt Stiftung, who made this
joint project possible by granting to one of us (F.M.) a Research
Award to visit the Universit\"at Augsburg, where most of the work
was done.

\newpage

\end{document}